\documentclass[letterpaper,titlepage,11pt]{article}
\usepackage{hyperref}
\usepackage{amssymb,amsmath,amsfonts}
\usepackage{epsfig}
\usepackage{graphicx}

\def\clock{{\count0=\time
           \divide\count0 60
           \ifnum\count0<10 0\fi\the\count0
           \multiply\count0 -60 \advance\count0 \time
           :\ifnum\count0<10 0\fi \the\count0
         }}
\newcommand{\timestamp}{{\small\vbox{\hbox{\tt\jobname.tex}
\hbox{\the\day/\the\month/\the\year, \clock}}}}

\def\AA{{\cal A}}

\def\GG{{\cal G}}

\def\KK{{\cal K}}

\def\MM{{\cal M}}
\def\NN{{\cal N}}
\def\OO{{\cal O}}
\def\PP{{\cal P}}

\def\TT{{\cal T}}

\def\d{{\partial}}

\def\gym{{g_{\rm YM}}}

\newcommand{\T}{\mathbb{T}}

\newcommand{\Z}{\mathbb{Z}}

\newcommand{\R}{\mathbb{R}}

\newcommand{\grad}{\vec{\nabla}}

\newcommand{\nn}{\nonumber}

\newcommand{\spa}{\ , \ \ }

\newcommand{\ds}{\displaystyle}

\newcommand{\vecto}[2]{\left( \begin{array}{c} \displaystyle #1 \\ \displaystyle  #2 \end{array}
\right) }
\newcommand{\matrto}[4]{\left( \begin{array}{cc} #1 & #2 \\
#3 & #4 \end{array} \right) }

\newcommand{\be}{\begin{equation}} \newcommand{\ee}{\end{equation}}
\newcommand{\bea}{\begin{eqnarray}} \newcommand{\eea}{\end{eqnarray}}

\newcommand{\CO}{\mathcal{O}}
\newcommand{\CN}{\mathcal{N}}

\newcommand{\CT}{\mathcal{T}}

\newcommand{\CM}{\mathcal{M}}

\newcommand{\id}{\hbox{1\kern-.27em l}}
\newcommand{\sid}{\hbox{\scriptsize1\kern-.27em l}}

\newcommand{\we}{\kern-.1em\wedge\kern-.1em}
\newcommand{\scal}{\kern-.13em\cdot\kern-.13em}

\newcommand{\al}{\alpha}

\newcommand{\Ord}{{\cal{O}}}

\newcommand{\mt}{\mathfrak{t}}
\newcommand{\ms}{\mathfrak{s}}

\newcommand{\hmt}{\hat{\mathfrak{t}}}
\newcommand{\hms}{\hat{\mathfrak{s}}}

\newcommand{\hmf}{\hat{\mathfrak{f}}}

\newcommand{\beastar}{\begin{eqnarray*}}
\newcommand{\eeastar}{\end{eqnarray*}}
\newcommand{\rom}[1]{\mathrm{#1}}

\numberwithin{equation}{section}

\begin{document}
\begin{titlepage}

\rightline{\vbox{\small\hbox{\tt hep-th/0701022}}}
\rightline{\vbox{\small\hbox{\tt CPHT-RR114.1206}}} \vskip 1.6cm

\centerline{\Large \bf Instabilities of Black Strings and Branes}

\vskip 1.6cm \centerline{\bf Troels Harmark$^a$, Vasilis
Niarchos$^b$ and Niels A.\ Obers$^a$} \vskip 0.5cm \centerline{\sl
$^a$The Niels Bohr Institute} \centerline{\sl Blegdamsvej 17, 2100
Copenhagen \O, Denmark}

\vskip 0.5cm \centerline{\sl $^b$ Centre de Physique Th\'eorique}
\centerline{\sl Ecole Polytechnique}
\centerline{\sl 91128 Palaiseau, France}

\vskip 0.5cm

\centerline{\small\tt harmark@nbi.dk,
niarchos@cpht.polytechnique.fr, obers@nbi.dk}

\vskip 1.6cm

\centerline{\bf Abstract} \vskip 0.2cm \noindent We review recent
progress on the instabilities of black strings and branes both for
pure Einstein gravity as well as supergravity theories which are
relevant for string theory. We focus mainly on Gregory-Laflamme
instabilities. In the first part of the review we provide a detailed
discussion of the classical gravitational instability of the neutral
uniform black string in higher dimensional gravity. The uniform
black string is part of a larger phase diagram of Kaluza-Klein black
holes which will be discussed thoroughly. This phase diagram
exhibits many interesting features including new phases,
non-uniqueness and horizon-topology changing transitions. In the
second part, we turn to charged black branes in supergravity and
show how the Gregory-Laflamme instability of the neutral black
string implies via a boost/U-duality map similar instabilities for
non- and near-extremal smeared branes in string theory. We also
comment on instabilities of D-brane bound states. The connection
between classical and thermodynamic stability, known as the
correlated stability conjecture, is also reviewed and illustrated
with examples. Finally, we examine the holographic implications of
the Gregory-Laflamme instability for a number of non-gravitational
theories including Yang-Mills theories and Little String Theory.

\vskip 1cm
\centerline{\sl Invited review for Classical and Quantum Gravity}


\end{titlepage}

\tableofcontents


\section{Introduction}

Gravitational instabilities were already considered around a hundred
years ago by Sir James Jeans \cite{Jeans:1902} in the context of
Newtonian gravity, where it was shown that a static homogeneous
non-relativistic fluid is unstable against long wavelength
gravitational perturbations. Subsequently, with the advent of
Einstein's theory of general relativity, the perturbative stability
of gravitational backgrounds such as black holes became an important
issue. More recently, string theory has provided a plethora of
higher dimensional black brane solutions which are captured at low
energies by higher-dimensional (super)gravity theories. The
question of perturbative stability in these backgrounds was examined
by Gregory and Laflamme in 1993 in the seminal work
\cite{Gregory:1993vy,Gregory:1994bj}, where it was shown that
neutral black strings in more than four dimensions suffer from a
long wavelength instability, in close analogy to the Jeans
instability. The main focus of this review will be the important
progress that has been achieved over the last few years in
understanding various aspects of the Gregory-Laflamme (GL)
instability.

The classical stability of black holes%
\footnote{With the term ``black hole'' we refer in this review to
any black object irrespective of its horizon topology.} is related
closely to the central question of uniqueness in gravity. Black
holes in four dimensions are known to be classically stable
\cite{Regge:1957td,Zerilli:1971wd}. This fits nicely with the fact
that in four-dimensional vacuum gravity, a black hole in an
asymptotically flat space-time is uniquely specified by the ADM mass
and angular momentum measured at infinity
\cite{Israel:1967wq,Carter:1971,Hawking:1972vc,Robinson:1975}.
Uniqueness theorems \cite{Gibbons:2002bh,Gibbons:2002av} for
$D$-dimensional ($D > 4$) asymptotically flat space-times state that
the Schwarzschild-Tangherlini black hole solution
\cite{Tangherlini:1963} is the only static black hole in pure
gravity. However, in pure gravity there are no uniqueness theorems
for non-static black holes with $D > 4$,%
\footnote{See \cite{Hollands:2006rj} for recent progress in this
direction.} or for black holes in space-times with non-flat
asymptotics. On the contrary, there are known cases of
non-uniqueness. An explicit example occurs in five dimensions for
stationary solutions in asymptotically flat space-time: for a
certain range of mass and angular momentum there exist both a
rotating black hole with $S^3$ horizon \cite{Myers:1986un} and
rotating black rings with $S^2 \times S^1$ horizons
\cite{Emparan:2001wn}.

One may speculate \cite{Kol:2002dr} that a universally valid version
of black hole uniqueness concerns only those solutions that are
perturbatively stable. If this is true, then taking into account classical
instability could restore the lost black hole uniqueness in higher dimensions.
At present, there is no more than circumstantial evidence for
this conjecture.

Beyond the question of uniqueness, instabilities in gravity are
important because they are related to time-dependent phenomena and
provide a glimpse into the full dynamics of the theory.
Understanding the full time evolution of a condensing instability of
a static or stationary solution and its end-point is an interesting
and important problem that provides information about the full
non-perturbative structure of the classical equations of motion. In
this respect, information about the full structure of the static or
stationary phases of the theory can provide important clues about
the time-dependent trajectories that interpolate between different
phases. In this review, we will see that black holes in spaces with
compact directions exhibit a rich phase structure and give strong
clues about time-dependent processes that involve exciting new
dynamics like horizon topology changing transitions. In particular,
for the Gregory-Laflamme instability of the neutral black string, a
lot of the current research is inspired by the question
\cite{Horowitz:2001cz,Horowitz:2002dc} of whether the Cosmic
Censorship Hypothesis is violated
\cite{Gregory:1993vy,Gregory:1994bj}  in the condensation process of
the Gregory-Laflamme instability, a process in which the topology of
the horizon is conjectured to change.

While the stability properties of black holes in (super)gravity are
interesting in their own right, their study is also highly relevant
for the stability of black holes and branes in string theory. At
low-energies closed string dynamics is captured by appropriate
supergravity theories admitting a large variety of interesting
solutions with event horizons that have received a lot of attention
in the past decades. A particularly interesting class of solutions
in string theory are the D-brane backgrounds. It has been
conjectured
\cite{Maldacena:1997re,Gubser:1998bc,Witten:1998qj,Itzhaki:1998dd}
that the near-horizon limit of these backgrounds is holographically
dual to a non-gravitational theory living on the brane (or
configuration of branes). This gauge/gravity (or AdS/CFT)
correspondence implies a deep connection between the thermodynamics
of the near-horizon brane backgrounds and the thermodynamics of the
dual non-gravitational theories. For example, it suggests that the
gauge theory dual of a classically unstable brane background must
have a corresponding phase transition. This hints at an interesting
connection between the classical stability of brane backgrounds and
the thermodynamic stability of the dual non-gravitational theories.

In this review, we will discuss instabilities of black holes and
branes in pure gravity and string theory.\footnote{Perturbative
instabilities can occur more generally in string theory in vacua
without supersymmetry. The instabilities are signaled by the
presence of a tachyonic mode in the perturbative spectrum with mass
that is in general of order string scale. In this review we discuss
light tachyons that are visible already in the supergravity
approximation. In recent years much progress has been made in
understanding perturbative instabilities in string theory (see
\cite{Sen:2004nf,Headrick:2004hz} for a status report on open and
closed string instabilities). In string theory there is also an
interesting relation between perturbative instabilities and the high
energy density of states of the theory
\cite{Kutasov:1990sv,Niarchos:2000kw}.} We will focus mainly on
Gregory-Laflamme-type instabilities of black strings and branes that
wrap or are smeared on a compact or non-compact direction, but along
the way we will also comment on a variety of other related cases.
The first part of this review (Sections 2-4) is a discussion that
involves exclusively pure Einstein gravity in higher dimensions, and
can be seen as separate from the second part which is more related
to string theory. In this connection, we also refer the reader to
the excellent review by Kol \cite{Kol:2004ww}, which contains
discussions on several important topics that we only briefly
mention. In the second part (Sections 5-7), we turn to
manifestations of the Gregory-Laflamme instability in supergravity
solutions that are relevant for string theory and examine the
implications for the holographically dual gauge theories and little
string theory.

In the pure gravity case, we begin the discussion with an introduction to the
Gregory-Laflamme instability of the neutral black string and its various manifestations.
The black string metric is a solution of Einstein gravity
in five or more dimensions that has a factorized form consisting of
a Schwarzschild-Tangherlini black hole and an extra flat direction. It has been
shown \cite{Gregory:1993vy,Gregory:1994bj} that this
background suffers from a long wavelength instability, known as the Gregory-Laflamme
instability, that involves perturbations with an oscillating profile along the direction
of the string. We review the precise form of this perturbation including the (time-independent)
threshold mode where the instability sets in. New features appear when the string
direction is compactified. In this case, we find that the neutral black string has a
critical mass, the Gregory-Laflamme mass, below which the solution is unstable.

The existence of the threshold mode at the critical mass strongly
suggests that a new static non-uniform black string emerges at
that point \cite{Gregory:1988nb,Gubser:2001ac}. In  recent years,
non-trivial information about the non-uniform black string
solution
\cite{Gubser:2001ac,Wiseman:2002zc,Sorkin:2004qq,Kleihaus:2006ee,Sorkin:2006wp}
has been obtained with advanced numerical methods. One remarkable
phenomenon of the non-uniform string is that it exhibits a
critical dimension \cite{Sorkin:2004qq} (see also \cite{Kol:2002xz}) at which the phase
transition of the uniform black string into the non-uniform black string changes from first
order into second order. See also Refs.~\cite{Sorkin:2004qq,Kol:2004pn,Cardoso:2006ks}
for the large dimension behavior of the threshold mode and
\cite{Kol:2002xz,Kol:2006vu} for the connection between the
Gregory-Laflamme instability and the Landau-Ginzburg theory of
phase transitions.

The unstable neutral black string is, in fact, part of a larger
phase diagram that consists of different phases of Kaluza-Klein
black holes, which are defined to be solutions with an event horizon
that asymptote to Minkowski space times a circle. This diagram
includes for instance the phase of black holes localized in the
compact direction for which much analytical
\cite{Harmark:2002tr,Harmark:2003yz,Gorbonos:2004uc,Gorbonos:2005px,Karasik:2004ds,Chu:2006ce}
and numerical \cite{Sorkin:2003ka,Kudoh:2003ki,Kudoh:2004hs}
progress has been made in recent years. All the static, uncharged
phases can be depicted in a two-dimensional phase diagram
\cite{Harmark:2003dg,Kol:2003if,Harmark:2003eg} parameterized by the
mass and tension. Despite the fact that we have only one compact
direction in this case the phase diagram exhibits a very rich
structure. For example, contrary to a neutral black hole in
Minkowski space, where black hole uniqueness restricts to just one
solution for a given mass, in the case of Kaluza-Klein black holes
one finds a continuous family of solutions \cite{Elvang:2004iz}. An
especially fascinating aspect of the phase structure of Kaluza-Klein
black holes is the occurrence of a phase transition, $i.e.$ the
occurrence of a point in a continuous line of phases where the
topology of the event horizon changes
\cite{Kol:2002xz,Wiseman:2002ti,Kol:2003ja,Sorkin:2006wp}. More
specifically, in this point, called the merger point, the localized
black hole phase meets the non-uniform black string phase.

Many of the insights obtained in this simplest case are expected to
carry over as we go to Kaluza-Klein spaces with higher-dimensional compact
spaces \cite{Kol:2004pn,Kol:2006vu}, although the degree of complexity in these cases will increase
substantially. If there are large extra dimensions in
Nature \cite{Arkani-Hamed:1998rs,Antoniadis:1998ig} the
Kaluza-Klein black hole solutions, or generalizations thereof, will
become relevant for experiments involving microscopic black holes and
observations of macroscopic black holes.

A similar structure occurs for black strings and branes in
supergravity and string theory. We analyze this aspect of the story
in the second part of this review focusing mostly on the case of
non- and near-extremal D-branes smeared on a transverse circle. In
fact, we will see that this case is directly related to the neutral
Kaluza-Klein black holes, discussed in the first part of this
review, via a boost/U-duality map
\cite{Harmark:2002tr,Bostock:2004mg,Aharony:2004ig,Harmark:2004ws}.
This powerful connection enables us to obtain many interesting
results for smeared D-branes in supergravity. The results are
interesting in their own right as statements about various phases of
string theory, but are also relevant via the gauge/gravity
correspondence for the phase structure of the dual non-gravitational
theories at finite temperature. For instance, it is possible to
obtain in this way non-trivial predictions
\cite{Aharony:2004ig,Harmark:2004ws,Harmark:2005dt} about the strong
coupling dynamics of supersymmetric Yang-Mills theories on compact
spaces and little string theory. Interesting generalizations
involving D-brane bound states
\cite{Gubser:2004dr,Ross:2005vh,Friess:2005tz} will also be
discussed briefly in various places in the main text.

Objects with an event horizon also exhibit thermodynamic properties,
in particular one can obtain from the horizon area an expression for the entropy using
 the Bekenstein-Hawking formula. It is then possible to study the
thermodynamic stability of the black holes and branes under
consideration. This raises a question: is there any connection
between thermodynamic stability and classical gravitational
stability? Holography suggests that such a relation is likely to
exist. For a certain class of branes, this connection has been
formulated more precisely and is known as the Correlated Stability
Conjecture (CSC)
\cite{Gubser:2000ec,Gubser:2000mm,Reall:2001ag,Gregory:2001bd}. The
conjecture has proven to be a rather robust statement whose validity
has been verified in a large set of examples
\cite{Gubser:2004dr,Ross:2005vh,Friess:2005tz,Harmark:2005jk} which
we will review. Recently, \cite{Friess:2005zp} presented a set of
counterexamples, which suggest that the conjecture cannot survive in
full generality in its current form. We will review the key elements
of the existing arguments in favor of the CSC, what may go wrong in
these arguments and suggested ways to refine the conjecture.

There are many topics related to this review which will not be covered fully in the main
text. For instance, besides the Gregory-Laflamme instability, solutions in (super)gravity
can exhibit a variety of other classical instabilities. We give a list of known
instabilities in the concluding section.

Although there are many exciting developments based on the material
presented here, we would like to emphasize and summarize now some of
the most prominent ones. The study of the
Gregory-Laflamme instability  of neutral black strings has revealed
a rich phase structure of black objects in Kaluza-Klein spaces, and
is a concrete manifestation of the large degree of non-uniqueness in
higher dimensional gravity, that would be very interesting to
understand better. Also, the phase diagram of Kaluza-Klein black
holes has revealed a concrete example of horizon topology changing
phase transitions in gravity. Further understanding of the
properties of the merger point, where two distinct phases meet,
would provide us with valuable new insights into the above issues.
An interesting development in this respect is the connection
\cite{Kol:2005vy,Sorkin:2005vz,Frolov:2006tc,Asnin:2006ip}
 between the merger point transition and Choptuik scaling \cite{Choptuik:1992jv} in black hole
formation.

Another important issue concerns the violation of the Cosmic Censorship Hypothesis
in the decay of the unstable uniform black string, where a naked singularity may
be formed when the horizon of the black string pinches \cite{Gregory:1993vy,Gregory:1994bj}.
 The results \cite{Wiseman:2002zc,Kudoh:2004hs,Kleihaus:2006ee,Sorkin:2006wp}
on the non-uniform black string  phase suggest that for certain
dimensions, the localized black hole is the only solution with
higher entropy than that of the uniform black string (for masses
where the black string is unstable). On the
other hand, it was argued in Ref.~\cite{Horowitz:2001cz} that the
horizon cannot pinch off in finite affine parameter, thus
suggesting that it is impossible for the black string horizon to
pinch off.
A way to reconcile these two results is if the horizon pinches off in
infinite affine parameter (see also \cite{Marolf:2005vn}).
Recently, the numerical analysis of
\cite{Choptuik:2003qd,Garfinkle:2004em,Anderson:2005zi} indicates that this
indeed is the case. If this is correct it would be interesting to examine the
implications for the Cosmic Censorship Hypothesis.

A central question regarding the material presented here
concerns the microscopic understanding of the entropy
\cite{Harmark:2006df,Chowdhury:2006qn} of the various phases of
black holes and branes on transverse circles. In particular, it
would be very interesting to pursue further a microscopic
explanation \cite{Chowdhury:2006qn} of the new phases.

Another interesting direction concerns the holographic relation
between the phase structure on the gravity side and the strongly
coupled dynamics of the dual gauge theory. By now there are many
non-trivial examples where purely gravitational phase transitions
imply via holography phase transitions in strongly coupled gauge
theories, which in some cases seem to be  continuously connected to
phase transitions that appear at weak coupling (see $e.g.$
Refs.~\cite{Hawking:1983dh,Witten:1998zw,Sundborg:1999ue,Aharony:2003sx,Aharony:2005bq}).
A case relevant to this review is that of Ref.~\cite{Aharony:2004ig}
that considered two-dimensional supersymmetric Yang-Mills on a
circle, which is dual to near-extremal smeared D0-branes on a
circle. It is remarkable to see how the phase structure of black
branes in supergravity can be matched qualitatively to the phase
structure of weakly coupled gauge theories, $c.f.$
Ref.~\cite{Aharony:2004ig} where a new phase was found in the weakly
coupled gauge theory that is also present in the holographic dual of
the strongly coupled gauge theory.

For a more detailed guide to the topics we discuss in this review,
we recommend the reader to examine the table of contents and/or read
the introductory paragraphs of each section.

\section{Classical stability analysis: black strings in pure gravity}

In this section we take a first look at the Gregory-Laflamme (GL)
instability of neutral black strings in Einstein gravity. We start
with a brief reminder of the Jeans instability, which is a long
wavelength instability already present in Newtonian gravity, and
then discuss the GL instability of the neutral black string. We will
see that there is a nice parallel between the two instabilities.
Then we go on to consider the GL instability for the neutral black
string with the string direction compactified along a circle. An
important feature of the GL instability is a critical wavelength
where the GL unstable mode becomes a marginal (threshold) mode. This
mode signals the existence of a neutral non-uniform black string.
The GL mode shows up in many different settings of black hole
physics, all connected to the question of stability. Finally, we
summarize the most prominent manifestations of the GL instability
and conclude with a brief discussion of the instability of the
Kaluza-Klein bubble. Further developments concerning the GL
instability of neutral black strings will be discussed in Section
\ref{sec:moreGL}.

\subsection{Long wavelength instabilities in gravity}

Gravitational instability due to long wavelength modes was first
discovered around a hundred years ago by Sir James Jeans
\cite{Jeans:1902} in the context of Newtonian gravity. The salient
features of this instability are as follows. Consider a
gravitational perturbation of a sample of static matter with
constant mass density $\rho=\rho_0$, constant pressure $p=p_0$ and
zero velocity field $\vec{v} = 0$. The equations of motion for
non-relativistic isotropic hydrodynamics with gravity that govern
this system are the equation of continuity
\begin{equation}
\partial_t \rho + \grad \cdot ( \rho \vec{v} ) = 0
~,
\end{equation}
the Euler equation
\begin{equation}
\partial_t \vec{v} + (\vec{v} \cdot \grad ) \vec{v} = -
\frac{1}{\rho} \grad p + \vec{g}
\end{equation}
and the equations for the gravitational field
\begin{equation}
\grad \times \vec{g} = 0 \spa \grad \cdot \vec{g} = - 4 \pi G \rho
~,
\end{equation}
where $\rho$ is the mass density, $p$ the pressure, $\vec{v}$ the
velocity field and $\vec{g}$ the external gravitational field with
$G$ being the Newton constant.

Plugging the perturbation
\begin{equation}
\rho_0 + \delta \rho \spa p_0 + \delta p \spa \delta \vec{v} \spa
\vec{g} + \delta \vec{g}
\end{equation}
into the equations of motion we get%
\footnote{In the present analysis we are neglecting the contributions due to the
self-gravitation of the sample of matter with $\rho=\rho_0$,
$p=p_0$ and $\vec{v}=0$. One can include these extra contributions
in a more careful analysis, but the end result will be the same.}
\begin{equation}
\partial_t \delta \rho + \rho_0 \grad \cdot \delta \vec{v} = 0
\spa
\partial_t \delta \vec{v}  = - \frac{1}{\rho_0} \grad \delta p + \delta \vec{g}
\spa \grad \times \delta \vec{g} = 0 \spa \grad \cdot \delta
\vec{g} = - 4\pi G \delta \rho
\end{equation}
{}From these equations we deduce that
\begin{equation}
\partial_t^2 \delta \rho = \grad^2 \delta p + 4\pi G \rho_0 \delta
\rho~.
\end{equation}
Then, demanding that the perturbation should obey the equation of
state $\delta p = v_s^2 \delta \rho$, we obtain for a particular
Fourier component $\delta \rho = A e^{i\omega t - i\vec{k} \cdot
\vec{x}}$ the dispersion relation
\begin{equation}
\omega^2 = v_s^2 k^2 - 4\pi G \rho_0
~.
\end{equation}
It is now evident that for sufficiently long wavelengths
\begin{equation}
\label{lambJ} \lambda > \lambda_J \equiv \sqrt{ \frac{\pi v_s}{G
\rho_0} }
\end{equation}
there is an unstable mode with $\omega^2 < 0$. This is the unstable
mode of the Jeans instability. Qualitatively we find that an object
becomes unstable to gravitational perturbations if its size becomes
equal to or larger than its critical Jeans wavelength $\lambda_J$.
Although originally derived in Newtonian gravity, this phenomenon
appears in a variety of contexts, among these cosmological
perturbations in the early Universe. As another example, one can
view black hole formation as a process that occurs in accordance
with the Jeans instability. Indeed, in this case we have $v_s \sim c
=1$ and by writing the mass as $M \sim \rho r^3$ and the
Schwarzschild radius as $2MG$ we find that a black hole will form
provided $r >1/\sqrt{G\rho}$, which is precisely the critical Jeans
instability wavelength.

Here we will be interested in the fact that the Jeans instability
has another pendant for black hole physics, namely the
Gregory-Laflamme instability. We will see below that the
Gregory-Laflamme instability occurs precisely when the size of the system is
larger than its critical Jeans wavelength. Another classical
analogue in which some features of the Gregory-Laflamme
instability have been successfully observed \cite{Cardoso:2006ks}
is the Rayleigh-Plateau instability of long fluid cylinders.
This is briefly reviewed in Section \ref{sec:qualitative}.

\subsection{Gregory-Laflamme instability \label{sec:GL}}

Black holes in four dimensions are known to be classically stable
\cite{Regge:1957td,Zerilli:1971wd}. When the relevance of General Relativity
in more than four dimensions emerged, it was natural to
ask whether there exist any higher-dimensional pure gravity
solutions with an event horizon exhibiting a classical
instability.

Gregory and Laflamme found in 1993 a long wavelength instability
for black strings in five or more dimensions \cite{Gregory:1993vy,Gregory:1994bj}.
The mode responsible for the instability propagates along the direction of
the string, and develops an exponentially growing time-dependent
part when its wavelength becomes sufficiently long.

The metric for a neutral black string in $D=d+1$ space-time
dimensions is
\begin{equation}
\label{ublstr} ds^2 = - f dt^2 + f^{-1} dr^2 + r^2 d\Omega_{d-2}^2
+ dz^2 \spa f=1-\frac{r^{d-3}}{r_0^{d-3}}
\end{equation}
where $d\Omega_{d-2}^2$ is the metric element of a $(d-2)$-dimensional
unit sphere. The metric \eqref{ublstr} is found by taking the $D-1$
dimensional Schwarzschild-Tangherlini static black hole%
\footnote{The classical stability of these higher-dimensional black hole solutions
was addressed in Refs.~\cite{Kodama:2003jz,Ishibashi:2003ap,Kodama:2003kk}.}
solution \cite{Tangherlini:1963} and adding a flat $z$ direction, which is the direction
parallel to the string. The event horizon is located at $r=r_0$
and has topology $S^{d-2}\times \R$.

\subsubsection{The Gregory-Laflamme mode}

\begin{figure}
\begin{center}
\resizebox{10cm}{6cm}{\includegraphics{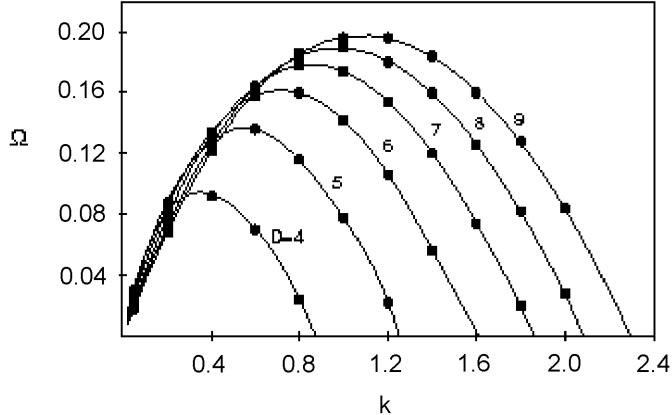}}
\end{center}
\caption{The Gregory-Laflamme $\Omega(k)$ curve in $d$
dimensions reprinted from Ref.~\cite{Gregory:1993vy}, where $D \vert_{\rm figure} =d$
used in this review.}
\label{figGL}
\end{figure}

The Gregory-Laflamme mode is a linear perturbation of the metric
\eqref{ublstr}, which we will denote as
\begin{equation}
\label{pertmet} g_{\mu\nu} + \epsilon h_{\mu\nu}
~.
\end{equation}
$g_{\mu\nu}$ stands for the components of the unperturbed
black string metric \eqref{ublstr},  $\epsilon$ is a small parameter and
$h_{\mu\nu}$ is the metric perturbation
\begin{equation}
\label{GLmode1} h_{\mu\nu} = \Re \left\{ \exp \left( \frac{\Omega
t}{r_0} + i \frac{kz}{r_0} \right) P_{\mu\nu} \right\}
\end{equation}
\begin{equation}
\label{GLmode2} P_{tt} = -f\psi , \ P_{tr} = \eta, \ P_{rr} =
f^{-1} \chi , \ P_{\rm sphere} = r^2 \kappa
\end{equation}
where $\psi$, $\eta$, $\chi$ and $\kappa$ are functions of
$x=rk/r_0$ such that the perturbed metric \eqref{pertmet} solves
the Einstein equations of motion. The symbol $\Re$ denotes the real part.
The resulting Einstein equations for the perturbation are analyzed in
Appendix \ref{app:mode} and reduce to the four independent
equations \eqref{E1}-\eqref{E4}. The gauge conditions%
\footnote{Various methods and different gauges have been employed to derive the differential
equations for the GL mode. See Ref.~\cite{Kol:2006ga} for a nice summary of these,
including a new derivation (see also \cite{Kol:2006ux}).}
on $h_{\mu\nu}$ are the tracelessness condition \eqref{traceeq} and
the transversality conditions \eqref{transeqs}. Combining the
gauge conditions with the linearized Einstein equations one can
derive a single second order differential equation for $\psi$
\begin{equation}
\label{psieq}
\psi''(x) +{\cal Q}_d(x) \psi'(x) +{\cal P}_d(x)
\psi(x) = 0 ~.
\end{equation}
The explicit form of the $d$-dependent rational functions ${\cal
Q}_d(x)$ and ${\cal P}_d(x)$ appears in Appendix \ref{app:mode}.
This equation depends only on $k$, $\Omega$ and the variable $x$.
A thorough analysis produces the curve of possible ($\Omega,k$) values
for which \eqref{psieq} has a well-behaved solution. We plot a sketch of this
curve found numerically in \cite{Gregory:1993vy,Gregory:1994bj} in
Figure \ref{figGL}.

Figure \ref{figGL} reveals that the curve of possible values
of $\Omega$ and $k$ intersects the $k$-axis in two places: $k=0$ and
$k=k_c$, where $k_c$ is a critical non-zero wave-number. The fact that the
curve does not intersect the $k$-axis at $k < 0$ follows from the
stability of the Schwarzschild-Tangherlini black hole
(otherwise we would have an unstable mode with $\Omega>0$ and
$k=0$). On the other hand, the presence of an intersection at $k=k_c$ is a signature
of a long wavelength instability: it means that there is an unstable mode for any
wavelength larger than the critical wavelength
\begin{equation}
\label{lambgl}
\lambda_{\rm GL} = \frac{2\pi r_0 }{k_c}
~.
\end{equation}

The critical wave-number $k_c$ marks the lower bound of the possible
wavelengths for which there is an unstable mode. Therefore, we
call the mode with $k=k_c$ and $\Omega=0$ the {\sl threshold mode}.
It is a time-independent mode of the form
\begin{equation}
\label{thresmode}
h_{c,\mu\nu} = \Re \left\{ \exp \left( i \frac{k_c z}{r_0} \right) P_{c,\mu\nu} \right\}
\end{equation}
As we shall see in the following, the threshold mode is important
for several reasons:
\begin{itemize}
\item[(i)] It signals the instability of the black string itself.
\item[(ii)] It can be mapped to unstable modes for other gravity solutions.
\item[(iii)] It suggests the existence of a static non-uniform black string.
\end{itemize}
Below, we go through each of these different aspects of the
threshold mode. The values of $k_c$ for $d=4,...,14$, as obtained in
\cite{Gregory:1993vy,Gubser:2001ac,Sorkin:2004qq}, are listed in
Table \ref{tabkc}.

\begin{table}[ht]
\begin{center}
\begin{tabular}{|c||c|c|c|c|c|c|c|c|c|c|c|}
\hline $d$ & $4$ & $5$ & $6$ & $7$ & $8$ & $9$ & $10$ & $11$ &
$12$ & $13$ & $14$
\\ \hline
$k_c $ & $0.88$ & $1.24$ & $1.60$ & $1.86$ & $2.08$ & $2.30$ & $2.50$ & $2.69$ & $2.87$ & $3.03$ &
$3.18$ \\
\hline
\end{tabular}
\caption{The critical wave numbers $k_c$  for the
Gregory-Laflamme instability.
\label{tabkc}}
\end{center}
\end{table}

\subsubsection{The Lichnerowitz operator and the threshold mode}

The fact that the mode \eqref{GLmode1}-\eqref{GLmode2} solves the
Einstein equations as a perturbation can be formulated in the following way.
Define the Lichnerowitz operator for a background metric $g_{\mu\nu}$ as
\begin{equation}
(\Delta_L)_{\mu \nu\rho \sigma} = - g_{\mu \rho}g_{\nu\sigma} D_\kappa
D^\kappa - 2 R_{\mu\nu\rho\sigma}
\end{equation}
Here $L$ refers to Lorentzian signature. In the ensuing it will be
convenient to use the short-hand notation
\begin{equation}
\Delta_L h_{\mu\nu} \equiv {(\Delta_L)_{\mu\nu}}^{\rho\sigma}
h_{\rho\sigma}
~.
\end{equation}
Then, the statement that the perturbation $h_{\mu\nu}$ of
$g_{\mu\nu}$ satisfies the Einstein equations of motion can be
stated as the differential operator equation
\begin{equation}
\label{licheq} \Delta_L h_{\mu\nu} = 0
\end{equation}
Hence what appears in Appendix \ref{app:mode} is a check
that the perturbation \eqref{GLmode1}-\eqref{GLmode2}
satisfies the Lichnerowitz equation \eqref{licheq}, with
\eqref{ublstr} as the background metric. We can write the relevant
equations more explicitly as
\begin{equation}
\label{Gonh1} {(\Delta_L)_{\mu\nu}}^{\rho\sigma} \left( P_{\rho\sigma}
e^{ \Omega t /r_0 } e^{ i kz/r_0 } \right) = 0
\end{equation}
In this expression, $P_{\mu\nu}$ has no explicit dependence on $t$
and $z$ and $P_{z\mu}=0$ for all $\mu$. Since the background metric
\eqref{ublstr} has $g_{zz}=1$, we deduce that we can rewrite
\eqref{Gonh1} as
\begin{equation}
\label{Gonh2} {(\hat \Delta_L)_{\mu\nu}}^{\rho\sigma} \left( P_{\rho\sigma}
e^{ \Omega t /r_0 } \right) = - \frac{k^2}{r_0^2} P_{\mu\nu} e^{
\Omega t /r_0}
\end{equation}
where $\hat \Delta_L$ is the dimensionally reduced Lichnerowitz operator
arising from the metric in one dimension less, $i.e.$ with the $z$-direction excluded.
In particular, for the threshold mode with $\Omega=0$ and $k=k_c$
\begin{equation}
\label{Goncrith} \hat \Delta_L P_{c,\mu\nu} = - \frac{k_c^2}{r_0^2}
P_{c,\mu\nu}
~.
\end{equation}
This expression has taken the form of an eigenvalue equation
for the dimensionally reduced Lichnerowitz operator  $\hat \Delta_L$.
In general, one could look for all possible solutions to the eigenvalue
equation
\begin{equation}
\hat \Delta_L U_{\mu\nu} = \alpha U_{\mu\nu}
\end{equation}
thus obtaining the full spectrum of eigenvalues $\alpha$ of the
Lichnerowitz operator. It is now clear from \eqref{Goncrith} that
{\it the existence of the threshold mode is equivalent to the existence
of a negative eigenvalue for the dimensionally reduced Lichnerowitz
operator ${\hat \Delta}_L$}. Taking the existence of the threshold mode
as a signal for the existence of the unstable Gregory-Laflamme mode, we
deduce that a negative eigenvalue of the dimensionally reduced
Lichnerowitz operator is a signal of a long wavelength instability for the solution.
This statement will be important in Section \ref{subsec:csc1}.

\subsubsection{Comparison with Jeans instability}

Is there any relation or analogy between the
Gregory-Laflamme instability of neutral black strings and
the Jeans instability of a massive object with similar
physical properties such as dimension and mass?
For example, does such an object have a Jeans Instability that sets in at
approximately the same point as the Gregory-Laflamme instability?
To see such an analogy, we note that the black string in a $d+1$
dimensional flat space-time has mass per unit length $M/L \sim
r_0^{d-3}/G$. From this we deduce that the mass density is $\rho_0
\sim M/(Lr_0^{d-1}) \sim 1/(Gr_0^2)$. Thus,
for a system with these physical parameters \eqref{lambJ}
implies that the Jeans instability should set in at wavelengths larger
than the critical wavelength
\begin{equation}
\lambda_J \sim \frac{1}{\sqrt{G \rho_0}} \sim r_0
\end{equation}
where we take the velocity of sound to be the speed of light.
So, indeed there is a Jeans instability that sets in when $\lambda/r_0$ is larger
than some number of order one. This is in good qualitative agreement
with the Gregory-Laflamme instability, which sets in at
wavelengths larger than the critical wavelength $\lambda_{\rm GL} = 2\pi r_0/k_c$
where $k_c$ is a number of order one given in Table \ref{tabkc}.

It is interesting to push this correspondence between the
Gregory-Laflamme instability and the Jeans Instability even further to the
large dimension limit $d \rightarrow \infty$, especially since
the gravitational field behaves in a Newtonian fashion in this limit. Keeping track
of the $d$-dimensional factors in the derivation above, one finds for $d \gg 1$
that $\lambda_J \sim d^{-3/4} r_0$. As we will see in Section \ref{sec:larged},
this critical wavelength does not have the right asymptotic behavior compared
to the one found for the GL mode. Interestingly, however, the (classical) Rayleigh-Plateau
instability reviewed in Section \ref{sec:qualitative} does predict the
correct asymptotic behavior \cite{Cardoso:2006ks}.

\subsubsection{Gregory-Laflamme instability for the compactified
neutral black string}

We have discussed the neutral black string solution in $D$
dimensional Minkowski space. However, for many reasons it is more
meaningful to consider instead the neutral black string in a
space-time with a compact direction parallel to the string. For one
thing, the question of the end point of the black string instability
becomes in this setting a more well-defined problem. Moreover, the
fact that the string is wrapped on a circle allows us to stabilize
it by making the circle smaller than the critical Gregory-Laflamme
wavelength, so that the Gregory-Laflamme mode cannot fit into the
geometry. In what follows, we briefly analyze the neutral black
string in a Kaluza-Klein space-time. Later in Section
\ref{sec:KKphases} we will put this into the more general context of
Kaluza-Klein black holes.

To this end, consider the neutral black string solution
\eqref{ublstr} with a periodic $z$ coordinate whose period we will
denote by $L$. This solution describes a neutral black string in a
$D=d+1$ dimensional Kaluza-Klein space-time $\CM^d \times S^1$.
Indeed, the black string solution \eqref{ublstr} asymptotes to
$\CM^d \times S^1$ as $r \rightarrow \infty$. Henceforth we shall
denote this solution as the (compactified) {\sl neutral uniform
black string} solution, in order to distinguish it from another
branch of black string solutions that will be discussed later. The
mass of the compactified neutral uniform black string is
\begin{equation}
M = \frac{\Omega_{d-2}L}{16\pi G}(d-2)r_0^{d-3}
~.
\end{equation}
It is useful to define a rescaled mass which is dimensionless,
using the circumference $L$:
\begin{equation}
\mu \equiv \frac{16\pi G}{L^{d-2}} M
~.
\end{equation}
Then for the compactified neutral uniform black string we have
\begin{equation}
\label{mustr} \mu = (d-2)\Omega_{d-2}
\left(\frac{r_0}{L}\right)^{d-3}
~.
\end{equation}
We see that the Gregory-Laflamme mode \eqref{GLmode1}-\eqref{GLmode2}
cannot obey the correct periodic boundary condition on $z$
if $L < \lambda_{\rm GL}$, with $\lambda_{\rm GL}$ given by
\eqref{lambgl}. On the other hand, for $L > \lambda_{\rm GL}$,
we can fit the Gregory-Laflamme mode into the compact direction
with the frequency and wave number $\Omega$ and $k$
in \eqref{GLmode1}-\eqref{GLmode2} determined by the ratio $r_0/L$.
Translating this in terms of the mass of the neutral black string
\eqref{mustr}, we deduce that we have a critical Gregory-Laflamme
mass given by
\begin{equation}
\label{mugl} \mu_{\rm GL} = (d-2)\Omega_{d-2} \left(
\frac{k_c}{2\pi} \right)^{d-3}
~.
\end{equation}
For $\mu < \mu_{\rm GL}$ the Gregory-Laflamme mode can be fitted
into the circle, and the compactified neutral uniform black string
is unstable. For $\mu > \mu_{\rm GL}$, on the other hand, the
Gregory-Laflamme mode is absent, and the neutral uniform black
string is stable. For $\mu=\mu_{\rm GL}$ there is a marginal mode
which, as we discuss below, signals the emergence of a new branch of
black string solutions which are non-uniformly distributed along the
circle. In Table \ref{tabmugl} we record the critical
Gregory-Laflamme mass $\mu_{\rm GL}$ that follows from the numerical
data of Table \ref{tabkc} (obtained in
\cite{Gregory:1993vy,Gubser:2001ac,Sorkin:2004qq}) using
\eqref{mugl}.

\begin{table}[ht]
\begin{center}
\begin{tabular}{|c||c|c|c|c|c|c|c|c|c|c|c|}
\hline $d$ & $4$ & $5$ & $6$ & $7$ & $8$ & $9$ & $10$ & $11$ &
$12$ & $13$ & $14$
\\ \hline
$\mu_{\rm GL}$ & $3.52$ & $2.31$ & $1.74$ & $1.19$ & $0.79$ & $0.55$ & $0.37$ & $0.26$ & $0.18$ & $0.12$ & $0.08$ \\
\hline
\end{tabular}
\caption{The critical masses $\mu_{\rm GL}$ for the
Gregory-Laflamme instability. \label{tabmugl}}
\end{center}
\end{table}

As we mentioned above, for the compactified neutral uniform black
string we can meaningfully ask about the end-point of the
instability of the string for $\mu < \mu_{\rm GL}$. We will discuss
this question further in Section \ref{sec:bhc}. One of the
problematic issues regarding the question about the end-point of the
Gregory-Laflamme instability in the uncompactified uniform black
string case is the infinite mass of the black string. If the
instability ended, for example, with a breaking up of the horizon
into disconnected pieces, there would have to be an infinite process
of localized instabilities happening one after another without end.
This process is regularized in the compactified case and therefore
it seems more sensible to consider the instability of the neutral
black string in the compactified setting.

\subsubsection{Neutral non-uniform black strings}

Let us re-consider now the threshold mode, $i.e.$ the critical
Gregory-Laflamme mode with $\Omega=0$ and $k=k_c$. This mode obeys
the equation \eqref{Goncrith}, or equivalently the equation
\begin{equation}
{(\Delta_L)_{\mu\nu}}^{\rho \sigma} \left( P_{c,\rho\sigma}
e^{ik_cz/r_0} \right) =0
~.
\end{equation}
$P_{c,\mu\nu} e^{ik_cz/r_0}$ is a marginal mode that
depends explicitly on the compact direction $z$. Therefore, as noticed in
\cite{Gregory:1988nb,Gubser:2001ac,Wiseman:2002zc},
the mode corresponds to a new static
classical solution that can be seen as a neutral black string with
a horizon that is non-uniform in the $z$ direction. In other words, the mode
signals the emergence of a new branch of black string solutions
which are non-uniformly distributed along the circle. We will call this
branch of solutions {\sl non-uniform black strings}, as opposed to
the uniform black string solutions \eqref{ublstr}. In section
\ref{sec:KKphases} we review the properties of the finitely deformed
non-uniform black string solution that has been found numerically, and
the connection with other branches in the full phase diagram
of solutions. Note that the topology of the horizon is  $S^{d-2} \times S^1$
for both the uniform as well as the non-uniform black string (in the compactified
case).

\subsubsection{Manifestations of the Gregory-Laflamme mode}

The Gregory-Laflamme mode discussed above shows up in many different
set ups in black hole physics, all connected to the question of
stability. In the following, we describe briefly various
implications and manifestations of the Gregory-Laflamme mode and the
context in which they appear:
\begin{itemize}
\item {\it Neutral non-uniform black string branch.} This is described
above in this section. Here the threshold mode corresponds to a
static black string solution which is non-uniform along the
direction of the string.
\item {\it Instability of static Kaluza-Klein bubble.} This will be discussed
below in Section \ref{sec:kkbub}. In this context the threshold mode is
double-Wick-rotated into an unstable mode for the static
Kaluza-Klein bubble.
\item {\it Semi-classical black hole instability in the canonical ensemble.}
Consider the 5D threshold mode \eqref{Goncrith}, involving
the dimensionally reduced 4D Lichnerowicz operator ${\hat \Delta}_L$. By
Wick rotating the mode, we obtain an eigenmode $u_{\mu\nu}$ for the
Lichnerowitz operator $\Delta_{\rm E}$ of the Euclidean section
of the 4D Schwarzschild black hole with eigenvalue $-k_c^2
r_0^{-2}$. Using the relation $r_0 = 2 G_4 M$, where $G_4$ is
four-dimensional Newton's constant and $M$ the four-dimensional
mass, we deduce that $u_{\mu\nu}$ obeys the equation
\begin{equation}
\label{GPYmode} \Delta_{\rm E} u_{\mu\nu} = - 0.19 (G_4M)^{-2}
u_{\mu\nu}
\end{equation}
where we used the value $k_c = 0.88$ from Table \ref{tabkc}. This is
precisely the eigenmode for the Euclidean section of the 4D
Schwarzschild black hole found by Gross, Perry, and Yaffe in
\cite{Gross:1982cv}%
\footnote{It is possible to extend this matching to the cases where
the unstable mode of the Schwarzschild-Tangherlini metric has been
calculated (see \cite{Reall:2001ag} for details).}. In
\cite{Gross:1982cv} it was argued as a consequence of the mode
\eqref{GPYmode} that the Euclidean flat space $\R^3\times S^1$ (hot
flat space) is semi-classically unstable to nucleation of
Schwarzschild black holes. In other words, for any non-zero
temperature it is thermodynamically preferred for a gas of gravitons
in Minkowski space to form a black hole. It is easily seen using the
higher-dimensional threshold modes that this can be extended to
higher dimensions, $i.e.$ that hot flat space $\R^d \times S^1$ is
unstable to nucleation of $(d+1)$-dimensional
Schwarzschild-Tangherlini black holes.
\item {\it Local thermodynamic instability.} In \cite{Reall:2001ag} it was
shown, following the conjectures of \cite{Gubser:2000ec,Gubser:2000mm},
that a Euclidean negative mode implies a local thermodynamic
instability, and vice versa. For the Schwarzschild black hole this
is clearly the case since the heat capacity is negative. If there is
in addition a flat non-compact direction one can Wick rotate the
Euclidean mode to a threshold Gregory-Laflamme mode, implying that
classical instability for black branes with non-compact directions
is in correspondence with local thermodynamic instability. We
explain and review this topic in Section \ref{sec:CSC}.
\item {\it Gregory-Laflamme modes for charged branes.}
Finally, one can also map the neutral Gregory-Laflamme mode to Gregory-Laflamme
modes for charged branes \cite{Aharony:2004ig,Harmark:2005jk}. Moreover, in
\cite{Harmark:2005jk} it is shown that the Gregory-Laflamme modes in the
limit $(k,\Omega) \rightarrow (0,0)$ become marginal modes for
extremal branes uniformly smeared along a transverse direction. We
review and discuss these issues in Section \ref{sec:sugra}.
\end{itemize}

\subsection{Instability of the static Kaluza-Klein bubble}
\label{sec:kkbub}

Kaluza-Klein bubbles were discovered in \cite{Witten:1982gj} by
Witten. In \cite{Witten:1982gj} it was explained that the
Kaluza-Klein vacuum $\CM^4 \times S^1$ is unstable semi-classically,
at least in the absence of fundamental fermions. The semi-classical
instability of $\CM^4 \times S^1$ proceeds through the spontaneous
creation of expanding Kaluza-Klein bubbles, which are Wick rotated
5D Schwarzschild-Tangherlini black hole solutions. The Kaluza-Klein
bubble is essentially a minimal $S^2$ somewhere in the space-time,
$i.e.$ a ``bubble of nothing''. In the expanding Kaluza-Klein bubble
solution the $S^2$ bubble expands until all of the space-time is
gone. However, apart from these time-dependent bubble solutions
there are also solutions with static bubbles, as we now review.

To construct the static Kaluza-Klein bubble in $d+1$ dimensions we
take the metric \eqref{ublstr} of the neutral uniform black string
and make a double Wick-rotation in the $t$ and $z$ directions.
After a convenient relabeling, we obtain the metric
\begin{equation}
\label{statbubmet} ds^2 = - dt^2 + \left( 1-
\frac{R^{d-3}}{r^{d-3}} \right) dz^2 + \frac{1}{1-
\frac{R^{d-3}}{r^{d-3}}} dr^2 + r^2 d \Omega_{d-2}^2 \ .
\end{equation}
We see that there is a minimal $(d-2)$-sphere of radius $R$
located at $r=R$. To avoid a conical singularity we need to make $z$
a periodic coordinate with period
\begin{equation}
\label{Lbub} L = \frac{4\pi R}{d-3} \ .
\end{equation}
Clearly, the solution asymptotes to $\CM^d \times S^1$ for $r
\rightarrow \infty$. Notice that the only free parameter in the
solution is the circumference of the $S^1$.

The static Kaluza-Klein bubble is classically unstable. This can be
seen using the threshold Gregory-Laflamme mode \eqref{Goncrith}.
After a double Wick-rotation $t \rightarrow iz$, $z \rightarrow it$
of the threshold mode, we obtain a mode $U_{\mu\nu}$ obeying $\Delta
U_{\mu\nu} = - k_c^2 R^{-2} U_{\mu\nu}$, where $\Delta$ is the
Lichnerowitz operator for the static Kaluza-Klein bubble
\eqref{statbubmet}. Using that $g_{tt} = -1$ in \eqref{statbubmet}
we deduce that
\begin{equation}
\label{bubmode} {\Delta_{\mu\nu}}^{\rho\sigma} \left(
U_{\rho\sigma} e^{ \pm k_c R^{-1} t  } \right) = 0
\end{equation}
Hence, $U_{\rho\sigma} e^{\pm k_c R^{-1} t  }$ corresponds
to an unstable mode for the static Kaluza-Klein bubble
\eqref{statbubmet}, implying that the static Kaluza-Klein bubble
is classically unstable to small s-wave perturbations.

The classical instability of the static Kaluza-Klein bubble causes
the bubble to either expand or collapse exponentially fast.%
\footnote{In Ref.~\cite{Sarbach:2004rm} the linear stability of
static bubble solutions of Einstein-Maxwell theory was examined. A
unique unstable mode was found and shown to be related, by double
analytic continuation, to marginally stable stationary modes of
perturbed black strings.} For five-dimensional Kaluza-Klein
space-times, there exist initial data \cite{Brill:1991qe} for
massive bubbles that are initially expanding or collapsing
\cite{Corley:1994mc}, and numerical studies \cite{Sarbach:2003dz}
show that there exist massive expanding bubbles and furthermore
indicate that contracting massive bubbles collapse to a black
hole with an event horizon. In the latter case, it is noteworthy that
the value of $\mu$ in \eqref{munstat} is smaller than $\mu_{\rm
GL}$, the Gregory-Laflamme mass, for $4 \leq d \leq 9$ (as
can be seen by comparing $\mu_{\rm b}$ in   \eqref{munstat} to
$\mu_{\rm GL}$ in  Table \ref{tabnonuni}). This implies that the
static Kaluza-Klein bubble does not decay to the uniform black
string. It is therefore likely that the Kaluza-Klein bubble in
that case decays to whatever is the endstate of the uniform black
string (see end of Section \ref{sec:bhc}).

Finally, we note that in Ref.~\cite{Horowitz:2005vp} charged bubble
solutions were constructed that are perturbatively stable, though
it was argued that these are non-perturbatively unstable.
These bubble solutions play a role in the end state of tachyon
condensation of charged black strings.

\section{Phases of  Kaluza-Klein black holes}
\label{sec:KKphases}

In the previous section we focused primarily on the uniform black string
and its GL instability. In this section we turn to a more general description of the phases
of Kaluza-Klein black holes. A $(d+1)$-dimensional
Kaluza-Klein black hole will be defined here as a pure gravity
solution with at least one event horizon that asymptotes to
$d$-dimensional Minkowski space times a circle ($\CM^d \times S^1$)
at infinity. We will discuss only static and neutral solutions,
$i.e.$ solutions without charges and angular momenta. Obviously, the
uniform black string is an example of a Kaluza-Klein black hole, but
many more phases are known to exist. In this section, we present
their properties as well as the possible relation of these phases to
the GL instability (see also the shorter review
\cite{Harmark:2005pp}).

\subsection{Physical parameters and definition of the phase diagram}
\label{sec:phase}

In this subsection we present a general method of computing the mass
$\mu$ and relative tension $n$ of a Kaluza-Klein black hole, which
will be later used to plot each phase in a $(\mu,n)$ phase diagram.
We review the main features of the split-up of this phase diagram
into two regions. Finally, we discuss some general results on the
thermodynamics of Kaluza-Klein black holes.

\subsubsection{Computing the mass and tension}

For any space-time which asymptotes to $\CM^d \times S^1$ we can
define the mass $M$ and the tension $\CT$. These quantities can be
used to parameterize the various phases of Kaluza-Klein black holes,
as we review below.

Let us define the Cartesian coordinates for $d$-dimensional
Minkowski space $\CM^d$ as $t,x^1,...,x^{d-1}$ and the radius $r
=\sqrt{\sum_i (x^i)^2 }$. In addition we use a coordinate $z$ of
period $L$ to label the $S^1$. Hence, the total space-time dimension
is $D=d+1$. In this notation, we have for any localized static
object the asymptotics \cite{Harmark:2003dg}
\begin{equation}
\label{gttzz} g_{tt} \simeq - 1 + \frac{c_t}{r^{d-3}} \spa g_{zz} \simeq 1 +
\frac{c_z}{r^{d-3}} \ ,
\end{equation}
as $r \rightarrow \infty$. The mass $M$ and tension $\CT$ are then
given by \cite{Harmark:2003dg,Kol:2003if}
\begin{equation}
\label{MT} M = \frac{\Omega_{d-2} L}{16 \pi G_{\rm N}} \left[
(d-2) c_t - c_z \right] \spa \CT = \frac{\Omega_{d-2} }{16 \pi
G_{\rm N}} \left[ c_t - (d-2) c_z \right] \ .
\end{equation}
The tension in \eqref{MT} can also be obtained from the general
formula for tension in terms of the extrinsic curvature
\cite{Harmark:2004ch}  analogous to the Hawking-Horowitz mass
formula \cite{Hawking:1996fd}. The mass and tension formulas have
been generalized to non-extremal and near-extremal branes in
\cite{Harmark:2004ws}. Gravitational tension for black branes has
also been considered in
\cite{Myers:1999ps,Traschen:2001pb,Townsend:2001rg}.

In what follows, it will be convenient to define the {\sl relative tension}
(also called the {\sl relative binding energy}) as \cite{Harmark:2003dg}
\begin{equation}
\label{then} n = \frac{\CT L}{M} = \frac{c_t - (d-2) c_z}{(d-2)
c_t - c_z } \ .
\end{equation}
This ratio provides a measure of how large the tension (or binding energy) is
relative to the mass. It is a useful quantity because it is
dimensionless and bounded as \cite{Harmark:2003dg}
\begin{equation}
\label{nbound} 0 \leq n \leq d-2 \ .
\end{equation}
The upper bound is due to the Strong Energy Condition whereas the
lower bound was found in \cite{Traschen:2003jm,Shiromizu:2003gc}.
The upper bound can also be understood physically in a more direct
way from the fact that we expect gravity to be an attractive force.
For a test particle at infinity it is easy to see that the
gravitational force on the particle is attractive when $n < d-2$ but
repulsive when $n > d-2$.

It is also useful to define a rescaled dimensionless quantity from
the mass as
\begin{equation}
\label{themu} \mu = \frac{16\pi G_{\rm N}}{L^{d-2}} M \ .
\end{equation}
The program set forth in \cite{Harmark:2003dg,Harmark:2003eg} is to
plot all phases of Kaluza-Klein black holes in a $(\mu,n)$ diagram.
We shall turn to this in the following.

\subsubsection{The split-up of the phase diagram}

According to the present knowledge of the phases of static and
neutral Kaluza-Klein black holes, the $(\mu,n)$ phase diagram
appears to be divided into two separate regions
\cite{Elvang:2004iz}:
\begin{itemize}
\item The region $0 \leq n \leq 1/(d-2)$ contains solutions
without Kaluza-Klein bubbles, and the solutions have a local
$SO(d-1)$ symmetry. We review what is known about solutions in this
part of the phase diagram in Section \ref{sec:bhc}. Because of the
$SO(d-1)$ symmetry there are only two types of event horizon
topologies: $S^{d-1}$ for the black hole on a cylinder branch and
$S^{d-2} \times S^1$ for the black string.
\item The region $1/(d-2) < n \leq d-2$ contains solutions with
Kaluza-Klein bubbles. We review this class of solutions in Section
\ref{sec:bub}. This part of the phase diagram, which is much more densely
populated with solutions compared to the lower part, is the subject of
\cite{Elvang:2004iz}.
\end{itemize}

\subsubsection{Thermodynamics, first law and the Smarr formula}

For a neutral Kaluza-Klein black hole with a single connected
horizon, we can find the temperature $T$ and entropy $S$ directly
from the metric. Together with the mass $M$ and  tension
${\cal{T}}$, these quantities obey the Smarr formula
\cite{Harmark:2003dg,Kol:2003if}
\begin{equation}
\label{Smarr1}
(d-1) TS = (d-2)M -  L {\cal{T}}
\end{equation}
and the first law of thermodynamics \cite{Townsend:2001rg,Kol:2003if,Harmark:2003eg}
\begin{equation}
\delta  M = T \delta S + {\cal{T}} \delta L
~.
\end{equation}
This equation includes a ``work'' term (analogous to $p \delta V$)
for variations with respect to the size of the circle at infinity.
See $e.g.$ Ref.~\cite{Harmark:2003eg} for a proof of the first law
based on the Smarr formula \eqref{Smarr1} using an ansatz
(see \eqref{ansatz} below) for black holes/strings on cylinders and
a class of static Ricci-flat perturbations. See also Ref.~\cite{Kastor:2006ti}
for a more general proof based on Hamiltonian methods.

It is useful to define the rescaled temperature
$\mt$ and entropy $\ms$ as
\begin{equation}
\label{tsneut} \mt = L T \spa \ms = \frac{16 \pi G_{\rm
N}}{L^{d-1}} S \ .
\end{equation}
In terms of these quantities, the Smarr formula for Kaluza-Klein
black holes and the first law of thermodynamics take the form
\begin{equation}
(d-1) \mt \ms = (d-2-n) \mu  \spa \delta \mu = \mt \, \delta \ms \
.
\end{equation}
Combining the Smarr formula and the first law, we get
the relation
\begin{equation}
\label{neutfirst2} \frac{\delta \log \ms}{\delta \log \mu} =
\frac{d-1}{d-2-n} \ ,
\end{equation}
so that, given a curve $n(\mu)$ in the phase diagram, the entire
thermodynamics can be obtained.

\begin{figure}[ht]
\centerline{\epsfig{file=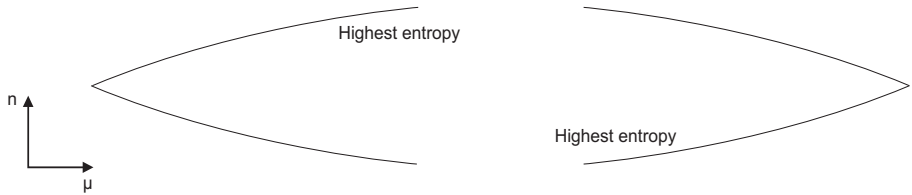,width=12
cm,height=2.5cm}} \caption{Illustration of the Intersection Rule.}
 \label{fig_intrule} \end{figure}

From \eqref{neutfirst2} it is also possible to derive the {\sl
Intersection Rule} of \cite{Harmark:2003dg}: for two branches that
intersect at the same solution, the branch with the lower
relative tension has the highest entropy for masses
below the intersection point, whereas the branch with the
higher relative tension has the highest entropy for masses
above the intersection point. An illustration of the
Intersection Rule appears in Figure \ref{fig_intrule}.

It is important to note that there are also examples of Kaluza-Klein
black hole solutions with more than one connected event horizon
\cite{Harmark:2003eg,Elvang:2004iz,Dias:2007hg}. The Smarr formula and
first law of thermodynamics generalize also to these cases.

\subsection{Black holes and strings on cylinders}
\label{sec:bhc}

In this subsection we review the properties of neutral and static
black objects without Kaluza-Klein bubbles. These turn out to have
local $SO(d-1)$ symmetry, which implies that the solutions fall into
two categories, black holes with event horizon topology $S^{d-1}$
and black strings with event horizon topology $  S^{d-2}\times S^1$.
First we discuss the ansatz that follows from the local $SO(d-1)$
symmetry. Then we review the uniform black string, non-uniform black
string and localized black hole phases. These phases are drawn in
the $(\mu,n)$ phase diagram for the five- and six-dimensional cases.
Finally we consider various related topics, including the existence
of copies of solutions, the endpoint of the decay of the uniform
black string and an observation related to the large $d$ behavior.

\subsubsection{The ansatz}

As we mentioned above, all solutions with $0 \leq n \leq 1/(d-2)$
have, to our present knowledge, a local $SO(d-1)$ symmetry. Using
this symmetry it has been shown \cite{Wiseman:2002ti,Harmark:2003eg}
that the metric of these solutions can be written in the form
\begin{equation}
\label{ansatz} ds^2 = - f dt^2 + \frac{L^2}{(2\pi)^2} \left[
\frac{A}{f} dR^2 + \frac{A}{K^{d-2}} dv^2 + K R^2 d\Omega_{d-2}^2
\right] \spa f = 1 - \frac{R_0^{d-3}}{R^{d-3}} \ ,
\end{equation}
where $R_0$ is a dimensionless parameter, $R$ and $v$ are
dimensionless coordinates and the metric is determined by the two
functions $A=A(R,v)$ and $K=K(R,v)$. The form \eqref{ansatz} was
originally proposed in Refs.~\cite{Harmark:2002tr,Harmark:2003fz} as
an ansatz for the metric of black holes on cylinders.

The properties of the ansatz \eqref{ansatz} were extensively
considered in \cite{Harmark:2002tr}. It was found that the
function $A=A(R,v)$ can be written explicitly in terms of the
function $K(R,v)$ thus reducing the number of free unknown
functions to one. The functions $A(R,v)$ and $K(R,v)$ are periodic
in $v$ with the period $2\pi$. Note that $R = R_0$ is the location
of the horizon in \eqref{ansatz}.

As already stated, all phases without Kaluza-Klein bubbles have,
to our present knowledge, $0 \leq n \leq 1/(d-2)$ and can be
described by the ansatz \eqref{ansatz}.
In what follows we review the three known phases
in this class and describe their properties.

\subsubsection{Uniform string branch}

The uniform string branch consists of neutral black strings which
are translationally invariant along the circle direction. The
metric of a uniform string in $d+1$ dimensions is
\cite{Tangherlini:1963}%
\footnote{The metric \eqref{unmet} corresponds to
$A(R,v)=K(R,v)=1$ in the ansatz \eqref{ansatz}.}
\begin{equation}
\label{unmet} ds^2 = - \left( 1 - \frac{r_0^{d-3}}{r^{d-3}}
\right) dt^2 + \left( 1 - \frac{r_0^{d-3}}{r^{d-3}} \right)^{-1}
dr^2 + r^2 d\Omega_{d-2}^2 + dz^2 \ .
\end{equation}
As we can see easily from the metric \eqref{unmet} using \eqref{gttzz} and
\eqref{then}, the relative binding energy is $n=1/(d-2)$ for the whole uniform
string branch. Hence, the uniform string branch is a horizontal line
in the $(\mu,n)$ diagram. The rescaled entropy is given by the expression
\begin{equation}
\label{su}
\ms_{\rm u} (\mu) = C_1^{(d-1)} \mu^{\frac{d-2}{d-3}} \ ,
\end{equation}
where the constant $C_1^{(q)}$ is defined as
\begin{equation}
\label{Cdef}
C_1^{(q)} \equiv 4\pi (\Omega_{q-1})^{-\frac{1}{q-2}} (q-1)^{-\frac{q-1}{q-2}} \ .
\end{equation}
The horizon topology of the uniform black string is clearly
$S^1 \times S^{d-2}$, where the $S^1$ is along the circle-direction.

The most important physical feature of the neutral uniform black
string solution is the GL instability reviewed in Section
\ref{sec:GL}. As stated there, a compactified neutral uniform black
string is classically unstable for $\mu <\mu_{\rm GL}$, $i.e.$ it is
unstable for sufficiently small masses. See Table \ref{tabmugl} for
the numerical values of the critical mass $\mu_{\rm GL}$ when $d
\leq 14$. For $\mu> \mu_{\rm GL}$ the string is believed to be
classically stable. We comment on the endpoint of the classical
instability of the uniform black string below.

\subsubsection{Non-uniform string branch}

It was realized in \cite{Gubser:2001ac}%
\footnote{See also \cite{Gregory:1988nb}.} that the classical
instability of the uniform black string for $\mu < \mu_{\rm GL}$
implies the existence of a marginal mode at $\mu =\mu_{\rm GL}$,
which again suggests the existence of a new branch of solutions.
Since the new branch of solutions is continuously connected to the
uniform black string it is expected to have the same horizon
topology, at least when the deformation away from the uniform black
string is sufficiently small. Moreover, the solution is expected to
be non-uniformly distributed in the circle-direction $z$ since there
is an explicit dependence in the marginal mode on this direction.

The new branch, which we call here the non-uniform string branch, has
been found numerically in \cite{Gubser:2001ac,Wiseman:2002zc,
Sorkin:2004qq}. Its most prominent features are:
\begin{itemize}
\item The horizon topology is $S^1 \times S^{d-2}$ with the $S^1$
being the circle of the Kaluza-Klein space-time $\CM^d \times S^1$.
\item The solutions are non-uniformly distributed along $z$.
\item The non-uniform black strings have a local $SO(d-1)$
symmetry and therefore can be written in terms of the ansatz
\eqref{ansatz} \cite{Wiseman:2002ti,Harmark:2003eg}.
\item The non-uniform string branch meets the uniform string branch
at $\mu=\mu_{\rm GL}$ on the line with $n=1/(d-2)$.
\item The branch has relative tension $n< 1/(d-2)$.
\end{itemize}
More concretely, considering the non-uniform black string branch for
$|\mu-\mu_{\rm GL}| \ll 1$ one obtains
\begin{equation}
\label{nofmu}
n(\mu) = \frac{1}{d-2} - \gamma ( \mu - \mu_{\rm GL}) + \CO ( (
\mu - \mu_{\rm GL})^2 ) \ .
\end{equation}
Here $\gamma$ is a number representing the slope of the curve
that describes the non-uniform string branch near $\mu=\mu_{\rm GL}$.
The numerical data for $\mu_{\rm GL}$ when $4 \leq d \leq 14$
are given in Table \ref{tabmugl} and the corresponding results
for $\gamma$ in Table \ref{tabnonuni}. These data were computed
in \cite{Gregory:1993vy,Gregory:1994bj,Gubser:2001ac,Wiseman:2002zc,
Sorkin:2004qq}.\footnote{$\gamma$ in Table \ref{tabnonuni} has been found
in terms of $\eta_1$ and $\sigma_2$ given in Figure 2 of \cite{Sorkin:2004qq}
by the formula
\[
\gamma = - \frac{2(d-1)(d-3)^2 }{(d-2)^2}
\frac{\sigma_2}{(\eta_1)^2} \frac{1}{\mu_{\rm GL}} \ .
\]
$\eta_1$ and $\sigma_2$ are also determined in \cite{Gubser:2001ac,
Wiseman:2002zc} for $d=4,5$.}

\begin{table}[ht]
\begin{center}
\begin{tabular}{|c||c|c|c|c|c|c|c|c|c|c|c|}
\hline $d$ & $4$ & $5$ & $6$ & $7$ & $8$ & $9$ & $10$ & $11$ &
$12$ & $13$ & $14$
\\ \hline
$\gamma$  & $0.14$ & $0.17$ & $0.21$ & $0.31$ & $0.47$ & $0.74$ & $1.4$ & $2.8$ & $7.9$ & $-40$ & $-9.2$ \\
\hline
\end{tabular}
\caption{The constant $\gamma$ which, together with $\mu_{\rm GL}$
of Table \ref{tabmugl}, determines the behavior of the non-uniform
branch for $| \mu - \mu_{\rm GL}| \ll 1$ . \label{tabnonuni}}
\end{center}
\end{table}

The large $d$ behavior of $\mu_{\rm GL}$ was examined numerically in
 \cite{Sorkin:2004qq} and analytically in \cite{Kol:2004pn}. The latter
 will be discussed in Section \ref{sec:larged}.

The qualitative behavior of the non-uniform string branch depends on
the sign of $\gamma$. If $\gamma$ is positive, then the branch
emerges at the mass $\mu=\mu_{\rm GL}$ with increasing $\mu$ and
decreasing $n$. If instead $\gamma$ is negative the branch emerges
at $\mu=\mu_{\rm GL}$ with decreasing $\mu$ and decreasing $n$. One
can use the Intersection Rule of \cite{Harmark:2003dg} (see Section
\ref{sec:phase}) to see that if $\gamma$ is positive (negative) then
the entropy of the uniform string branch for a given mass is higher
(lower) than the entropy of the non-uniform string branch for that
mass. This result can also be derived directly from \eqref{nofmu}
using \eqref{neutfirst2}. We find
\begin{equation}
\frac{\ms_{\rm nu} ( \mu )}{\ms_{\rm u}  ( \mu )}
= 1 - \frac{(d-2)^2}{2(d-1)(d-3)^2} \frac{\gamma}{\mu_{\rm GL}}
(\mu-\mu_{\rm GL})^2 + \CO ( (\mu - \mu_{\rm GL})^3 ) \ ,
\end{equation}
from which we clearly see the significance of the sign of the
parameter $\gamma$. In this expression $\ms_{\rm u} ( \mu )$
($\ms_{\rm nu} ( \mu )$) refers to the rescaled entropy of the
uniform (non-uniform) black string branch. From the data of Table
\ref{tabnonuni} we see that $\gamma$ is positive for $d \leq 12$ and
negative for $d \geq 13$ \cite{Sorkin:2004qq}. Therefore, as
discovered in \cite{Sorkin:2004qq}, the non-uniform black string
branch has a qualitatively different behavior for small $d$ and
large $d$,
$i.e.$ the system exhibits a critical dimension $D=14$.%
\footnote{The first occurrence of a critical dimension in this
system was given in \cite{Kol:2002xz}, where evidence was given
that the merger point between the black hole and the string
depends on a critical dimension $D=10$, in such a way that
for $D<10$ there are local tachyonic modes around the tip of the
cone (the conjectured local geometry close to the thin ``waist'' of the
string) which are absent for $D>10$.}

\subsubsection*{\sl Numerical results}

In six dimensions, $i.e.$ for $d=5$, Wiseman found in
\cite{Wiseman:2002zc} a large body of numerical data for the
non-uniform string branch. These data are displayed in the $(\mu,n)$
phase diagram on the right side of Figure \ref{fig1}
\cite{Harmark:2003dg}. The branch starts at $\mu = \mu_{\rm GL}$ and
the data end around $\mu \simeq 2.3 \, \mu_{\rm GL} \simeq 5.3$. In
\cite{Kudoh:2004hs} numerical evidence has been found that suggests
that the non-uniform string branch more or less ends where the data
of \cite{Wiseman:2002zc} end, $i.e.$ around $\mu \simeq 5.3$ for
$d=5$, supporting the considerations of \cite{Kol:2003ja}. Recently,
numerical results extending the non-uniform branch into the strongly
non-linear regime were also obtained for the case of five
dimensions, $i.e.$ for $d=4$, in Ref.~\cite{Kleihaus:2006ee}, and
for the entire range $d \leq 5 \leq 10$ in
Ref.~\cite{Sorkin:2006wp}. All these non-uniform black string
solutions have masses above the GL mass and less entropy than the
corresponding uniform black string.

Another interesting feature of the non-uniform solutions is the local
geometry around the ``waist'' of the most non-uniform solutions. This is
cone-like to a fairly good approximation \cite{Kol:2003ja,Sorkin:2006wp},
lending support to the ideas proposed in
\cite{Kol:2002xz,Kol:2005vy} (see also Section \ref{sec:qualitative}).
As we further discuss below, all these data provide important evidence
that the non-uniform branch has a topology changing transition
into the localized black hole branch.

\subsubsection{Localized black hole branch}

On physical grounds, it is natural to expect a branch of neutral
black holes in the space-time $\CM^d \times S^1$. One defining
feature of these solutions is an event horizon of topology
$S^{d-1}$. We will call this branch the {\sl localized black hole
branch} in the following, because the $ S^{d-1}$ horizon is
localized on the $S^1$ of the Kaluza-Klein space.

Neutral black hole solutions in the space-time $\CM^3 \times S^1$
were found and studied in \cite{Myers:1987rx,Bogojevic:1991hv,
Korotkin:1994dw,Frolov:2003kd}. However, the study of black holes
in the space-time $\CM^{d} \times S^1$ for $d \geq 4$ has begun only recently.
The complexity of the problem stems from the fact that such
black holes are not algebraically special \cite{DeSmet:2002fv}
and moreover from the fact that the solution cannot be found using
a Weyl ansatz since the number of Killing vectors is too small.

\subsubsection*{\sl Analytical results}

Progress towards finding an analytical solution for the localized
black hole was made in \cite{Harmark:2002tr} where, as reviewed
above, the ansatz \eqref{ansatz} was proposed.
Subsequently it was proven in \cite{Wiseman:2002ti,Harmark:2003eg}
that the localized black hole can be put in this ansatz.

In \cite{Harmark:2003yz} the metric of small black holes, $i.e.$
black holes with mass $\mu \ll 1$, was found analytically using
the ansatz \eqref{ansatz} of \cite{Harmark:2002tr} to first order
in $\mu$. Subsequently, an equivalent expression for the first
order metric was found in
Refs.~\cite{Gorbonos:2004uc,Gorbonos:2005px} using a different
method. An important feature of the localized black hole solution
is the fact that $n \rightarrow 0$ as $\mu \rightarrow 0$. This
means that the black hole solution becomes more and more like a
$(d+1)$-dimensional Schwarzschild black hole as the mass goes to
zero. For $d=4$, the second order correction to the metric and
thermodynamics have recently been studied in
\cite{Karasik:2004ds}. More generally, the second order correction
to the thermodynamics was obtained in Ref.~\cite{Chu:2006ce} for
all $d$ using an effective field theory formalism in which the
structure of the black hole is encoded in the coefficients of
operators in an effective worldline Lagrangian.

The first order result of \cite{Harmark:2003yz} and second order result of \cite{Chu:2006ce}
can be summarized by giving the first and second order corrections
to the relative tension $n$ of the localized black hole branch
as a function of $\mu$%
\footnote{Here $\zeta(p) = \sum_{n=1}^\infty n^{-p}$ is the
Riemann zeta function.}
\begin{equation}
\label{bhslope}
n = \frac{(d-2)\zeta(d-2)}{2(d-1)\Omega_{d-1}} \mu -
\left( \frac{(d-2)\zeta(d-2)}{2(d-1)\Omega_{d-1}} \mu \right)^2 + \CO (\mu^3) \ .
\end{equation}
Plugging this expression into \eqref{neutfirst2} one can find the leading
correction to the thermodynamics as
\begin{equation}
\label{cors} \ms_{\rm bh} (\mu) = C_1^{(d)} \mu^{\frac{d-1}{d-2}} \left( 1
+ \frac{\zeta(d-2)}{2(d-2)\Omega_{d-1}} \,  \mu -
\frac{(d^2- 6d +7)}{2(d-1)} \left( \frac{\zeta(d-2)}{2(d-2)\Omega_{d-1}} \,  \mu
\right)^2
 + \CO ( \mu^3 )\right) \ ,
\end{equation}
where $C_1^{(d)}$ is defined in \eqref{Cdef}.
This constant of integration is fixed by the physical requirement
that in the limit of vanishing mass we should recover the thermodynamics
of a Schwarzschild black hole in $(d+1)$-dimensional
Minkowski space.

\subsubsection*{\sl Numerical results}

The black hole branch has been studied numerically
for $d=4$ in \cite{Sorkin:2003ka,Kudoh:2004hs}
and for $d=5$ in \cite{Kudoh:2003ki,Kudoh:2004hs}.
For small $\mu$, the impressively accurate data of
\cite{Kudoh:2004hs} are consistent with the analytical results
of \cite{Harmark:2003yz,Gorbonos:2004uc,Karasik:2004ds}.
We have displayed the results of \cite{Kudoh:2004hs} for $d=4,5$
in a $(\mu,n)$ phase diagram in Figure \ref{fig1}.

Amazingly, the work of \cite{Kudoh:2004hs} seems to give an answer
to the question: ``Where does the localized black hole branch
end?''. Several scenarios have been suggested, see
\cite{Harmark:2003eg} for a list. The scenario favored by
\cite{Kudoh:2004hs} is the scenario suggested by Kol
\cite{Kol:2002xz} in which the localized black hole branch meets
with the non-uniform black string branch in a topology changing
transition point. This is strongly implied by the $(\mu,n)$
phase diagram for $d=5$ in Figure \ref{fig1}. Further evidence arises
by examining the geometry of the two branches near the
transition point, and also by examining the thermodynamics
\cite{Kudoh:2004hs}.

Hence, it seems reasonable to expect that the localized black hole
branch is connected with the non-uniform string branch in any
dimension. This means that we can go from the uniform black string
branch to the localized black hole branch through a connected
series of static classical geometries. The point in which the two
branches are conjectured to meet is called the {\sl merger point}.
See Section \ref{sec:merger} for more on the critical behavior of
the two branches near this point.

\subsubsection{Phase diagrams for $d=4$ and $d=5$}

\begin{figure}[ht]
\centerline{ \epsfig{file=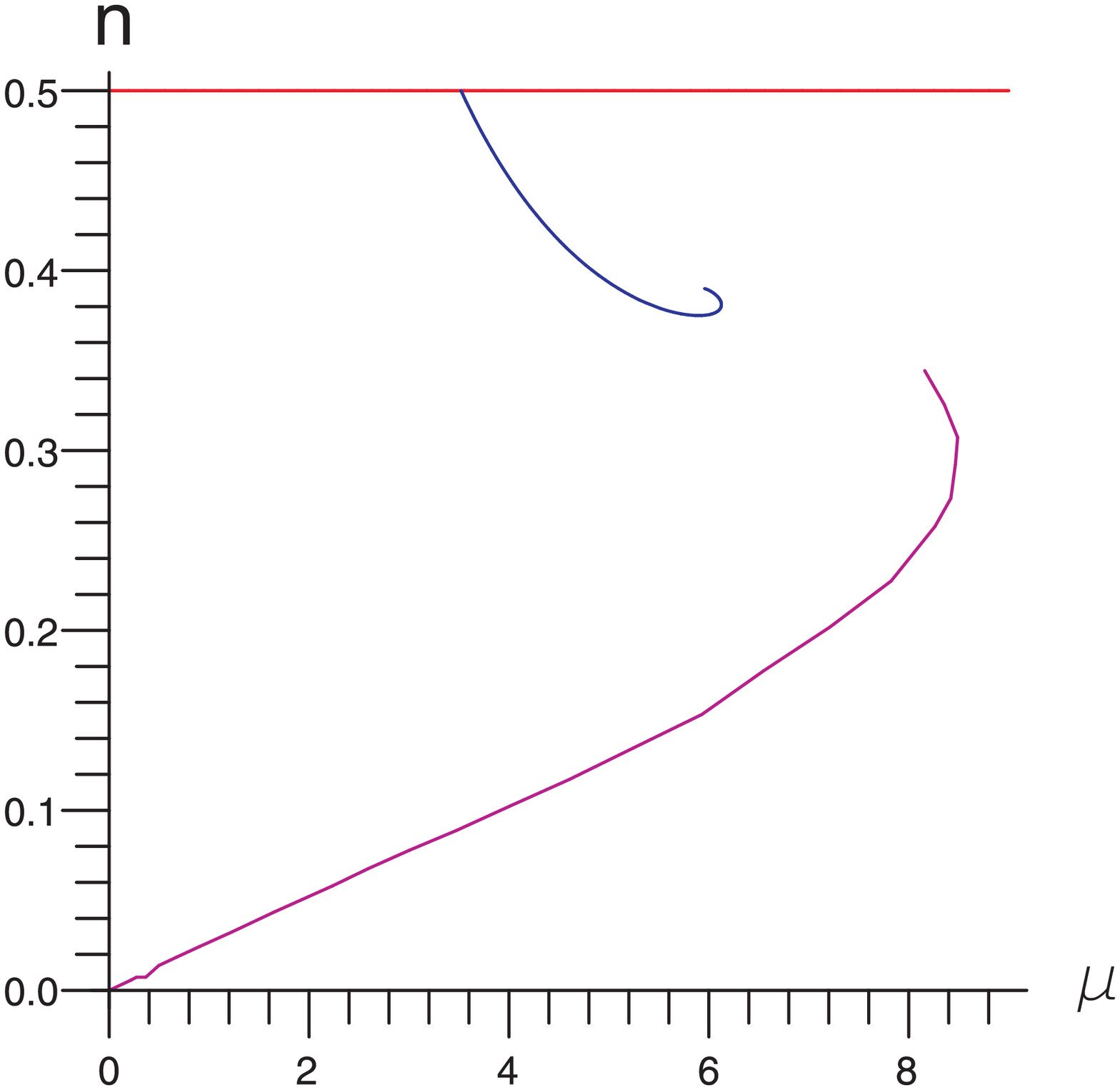,width=7 cm,height=6cm}
\hskip .5cm \epsfig{file=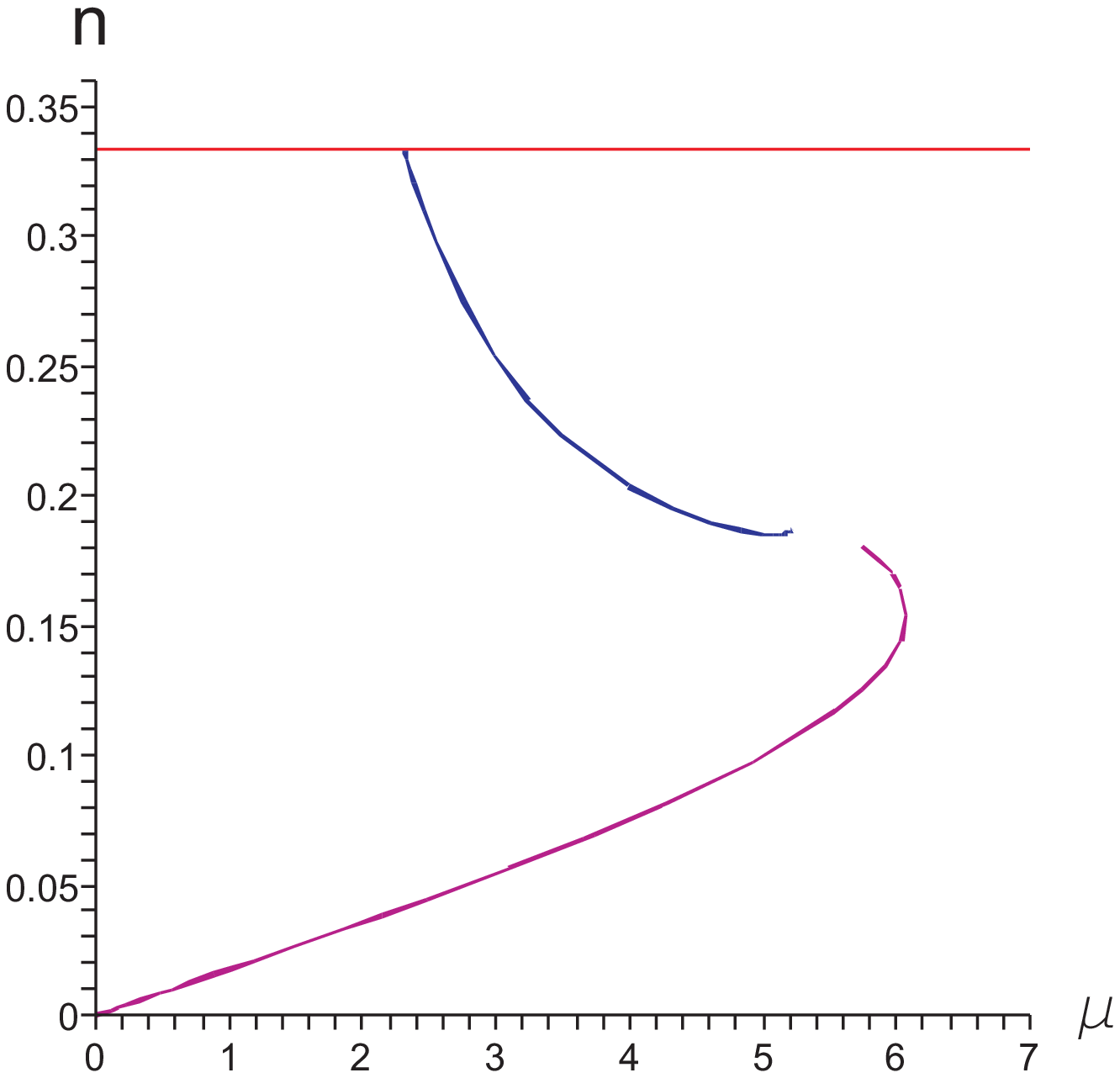,width=7 cm,height=6cm} }
\caption{Black hole and string phases for $d=4$ (left) and $d=5$
(right), drawn in the $(\mu,n)$ phase diagram. The horizontal (red)
line at $n=1/2$ and 1/3 respectively is the uniform string branch.
The (blue) branch emanating from this at the Gregory-Laflamme mass
is the non-uniform string branch. For $d=4$ and $d=5$ this branch
was obtained numerically in \cite{Kleihaus:2006ee} and
\cite{Wiseman:2002zc}. The (purple) branch starting at the point
$(\mu,n)=(0,0)$ is the black hole branch which was obtained
numerically by Kudoh and Wiseman \cite{Kudoh:2004hs}. For both
dimensions the results strongly suggest that the black hole and
non-uniform black string branches meet.} \label{fig1}
\end{figure}

In Figure \ref{fig1} we display the $(\mu,n)$ diagram for $d=4$
and $d=5$, which are the cases where the most information about
the various branches of black holes and strings on cylinders has been accumulated.
For $d=4$ we have shown the complete non-uniform branch as obtained numerically
by \cite{Kleihaus:2006ee}, which emanates at $\mu_{\rm GL} = 3.52$
from the uniform branch given by the horizontal line $n=1/2$.
For $d=5$, we have shown the complete non-uniform
branch, as obtained numerically by Wiseman \cite{Wiseman:2002zc}, which
emanates at $\mu_{\rm GL} = 2.31$ from the uniform branch that has $n=1/3$.
These data were first incorporated into the $(\mu,n)$ diagram
in Ref.~\cite{Harmark:2003dg}. For the black hole branch we have plotted the recently
obtained numerical data of Kudoh and Wiseman \cite{Kudoh:2004hs}, both
for $d=4$ and $d=5$. It is evident from the figure that this branch has an
approximate linear behavior for a fairly large range of $\mu$ close to the origin and the
numerically obtained slope agrees very well with the analytic result
\eqref{bhslope}.

These data strongly suggest that the localized and non-uniform phases meet in a
topology changing transition point. Another interesting feature to note is
the upwards bending, both for $d=4$ and $d=5$ at the end of the non-uniform
branch.

\subsubsection{Copies of solutions}

In \cite{Harmark:2003eg} it has been argued that one can generate new
solutions by copying solutions on the circle several times,
following an idea of Horowitz \cite{Horowitz:2002dc}. This works
for solutions which vary along the circle direction ($i.e.$ in the $z$
direction), so it works both for the black hole branch and the
non-uniform string branch. Let $k$ be a positive integer. Then if
we copy a solution $k$ times along the circle we get a new
solution with the following parameters:
\begin{equation}
\label{coptrans} \tilde{\mu} = \frac{\mu}{k^{d-3}} \spa \tilde{n}
= n \spa \tilde{\mt} = k \mt \spa \tilde{\ms} =
\frac{\ms}{k^{d-2}} \ .
\end{equation}
See Ref.~\cite{Harmark:2003eg} for the corresponding expression of
the metric of the copies in the ansatz \eqref{ansatz}.
Using the transformation \eqref{coptrans}, one easily sees that
the non-uniform and localized black hole branches depicted in
Figure \ref{fig1} are copied infinitely many times in the
$(\mu,n)$ phase diagrams.

Beyond these copies of solutions, more general multi black-hole
configurations with localized black hole solutions can be shown to
exist as well \cite{Dias:2007hg}. These solutions correspond to having
several localized black holes of different sizes on the transverse
circle.

\subsubsection{The endpoint of decay of the uniform black string \label{sec:end} }

As we mentioned above, the uniform black string is
classically unstable for $\mu < \mu_{\rm GL}$. It is natural
to ask: ``What is the endpoint of this classical instability?''.

The entropy of a small localized black hole is much larger
than that of a black string with the same mass,
$i.e.$ $\ms_{\rm bh} (\mu) \gg \ms_{\rm u} (\mu)$ for $\mu \ll 1$,
as can be easily seen by comparing eqs.\ \eqref{su} and \eqref{cors}.
This suggests that a light uniform string will decay to a
localized black hole. This is the global thermodynamic argument that appeared
already in the original work \cite{Gregory:1993vy,Gregory:1994bj}.
However, one can imagine other possibilities,
for example the uniform black string could decay to another static geometry,
or it could even keep decaying without reaching an endpoint.

\begin{figure}[ht]
\centerline{\epsfig{file=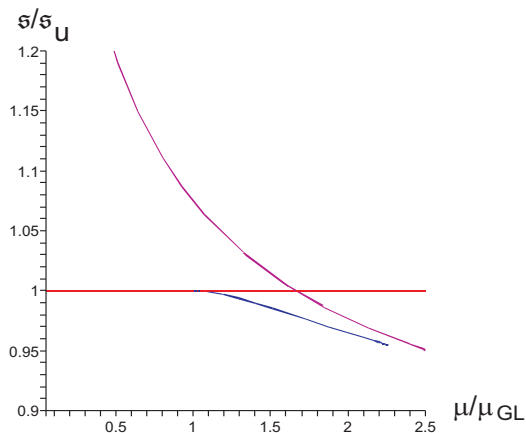,width=7 cm,height=6cm}}
\caption{Entropy $\ms/\ms_{\rm u}$ versus the mass
$\mu/\mu_{GL}$ diagram for the uniform string (red),
non-uniform string (blue) and localized black hole (purple)  branches.}
\label{fig_entr}
\end{figure}

For $d=4,5$ we can be more precise about this issue. Since both
cases show similar behavior, we focus on $d=5$ and have displayed
in Figure \ref{fig_entr} the entropy $\ms$ versus the mass $\mu$
diagram for the localized black hole, uniform string and
non-uniform string branches. We see from this figure that in six
dimensions the localized black hole has always greater entropy
than the uniform strings in the mass range where the uniform
string is classically unstable. This fact together with the
absence of the non-uniform string branch in this mass range
suggest that the unstable uniform black string decays classically
to the localized black hole branch in six dimensions ($d=5$). The
same conclusion holds for five dimensions ($d=4$), and is expected
to hold in any dimension less than fourteen (see below).

The viewpoint that an unstable uniform string decays to a
localized black hole has been challenged in
\cite{Horowitz:2001cz}. In this work it is shown that in a classical
evolution an event horizon cannot change topology, $i.e.$ it cannot
pinch off in finite affine parameter (on the event horizon).

However, this does not exclude the possibility that a classically
unstable horizon pinches off in infinite affine parameter
\cite{Marolf:2005vn}. Indeed, in
\cite{Garfinkle:2004em,Anderson:2005zi} the numerical study
\cite{Choptuik:2003qd} of the classical decay of a uniform black
string in five dimensions was reexamined, suggesting that the
horizon of the string pinches off in infinite affine parameter.

Interestingly, the classical decay of the uniform string is quite
different for $d \geq 13$. As we reviewed above, the results of
\cite{Sorkin:2004qq} show that for $d \geq 13$ the non-uniform
string has decreasing mass $\mu$ for decreasing $n$, as it
emanates from the uniform string at the Gregory-Laflamme point at
$\mu=\mu_{\rm GL}$. This means that the entropy of a non-uniform
black string is higher than the entropy of a uniform string with
same mass. Hence, for $d \geq 13$ we have a certain range of
masses where the uniform black string is unstable, and where a
non-uniform black string exists with higher entropy. This suggests
that a uniform black string in that mass range will decay
classically to a non-uniform black string. This type of decay can
occur in a finite affine parameter, according to
\cite{Horowitz:2001cz}, since the horizon topology remains fixed
during the decay.

Note that the range of masses for which we have a non-uniform
string branch with higher entropy is extended by the fact that we
have copies of the non-uniform string branch. The copies, which
have the thermodynamic quantities given by the transformation rule
\eqref{coptrans}, can easily be seen from the Intersection Rule of
\cite{Harmark:2003dg} (see Section \ref{sec:phase}) to have higher
entropy than that of a uniform string of the same mass,
since they also have decreasing $\mu$ for
decreasing $n$. Thus, for $d \geq 13$, it is even possible that
there exists a non-uniform black string for any given $\mu <
\mu_{\rm GL}$ with higher entropy than that of the uniform black
string with mass $\mu$. This will occur if the non-uniform string
branch extends to masses lower than $2^{3-d} \mu_{\rm GL}$.
Otherwise, the question of the endpoint of the decay of the
uniform black string will involve a quite complicated pattern of
ranges.

We will return to the $d$-dependence of the phase diagram in
Section \ref{sec:larged} where we consider the large $d$ behavior
of a number of relevant quantities.

\subsection{Phases with Kaluza-Klein bubbles \label{sec:bub}}

Until now all the solutions we have been discussing lie within
the range $0 \leq n \leq 1/(d-2)$. But, there are no direct physical
restrictions on $n$ preventing it from being in the range
$1/(d-2) < n \leq d-2$. Hence, the question arises about
the existence of solutions in this range. This question was answered
in \cite{Elvang:2004iz} where it was shown that pure gravity
solutions with Kaluza-Klein bubbles can realize all values of
$n$ in the range $1/(d-2) < n \leq d-2$.

In this subsection we start by reviewing the place of the static Kaluza-Klein bubble
(see Section \ref{sec:kkbub}) in the $(\mu,n)$ phase diagram. We then discuss the
main properties of bubble-black hole sequences, which are phases of
Kaluza-Klein black holes involving Kaluza-Klein bubbles. In particular,
we comment on the thermodynamics of these solutions. Subsequently, we present
the five- and six-dimensional phase diagrams as obtained by including
the simplest bubble-black hole sequences. Finally, we comment on
non-uniqueness in the phase diagram.

\subsubsection{Static Kaluza-Klein bubble}

In Section~\ref{sec:kkbub} we reviewed the static Kaluza-Klein bubble.
Since it is a static solution of pure gravity that asymptotes to
$\CM^d \times S^1$ the static Kaluza-Klein bubble belongs to the class
of solutions we are interested in. Thus, it is part of our phase diagram, and using
\eqref{gttzz}, \eqref{MT}, \eqref{then}, \eqref{themu} and \eqref{Lbub}
we can read off $\mu$ and $n$ as
\begin{equation}
\label{munstat} \mu = \mu_{\rm b} = \Omega_{d-2} \left(
\frac{d-3}{4\pi} \right)^{d-3} \spa n = d-2 \ .
\end{equation}
These expressions give the static Kaluza-Klein bubble as a specific
point in the $(\mu,n)$ phase diagram. This is consistent with the fact
that this solution does not have any free dimensionless parameters.
Notice that $n=d-2$ precisely saturates the upper bound on $n$ in
\eqref{nbound}. In fact, a test particle at infinity will not
experience any force from the bubble (in the Newtonian limit).

\subsubsection{Bubble-black hole sequences}

We now have a solution at $n=d-2$ with no event horizon,
and hence no entropy or temperature.
But so far in this review we have not mentioned any solutions lying in the
range $1/(d-2) < n < d-2$. It was argued in \cite{Elvang:2004iz} that such
solutions exist and contain both Kaluza-Klein bubbles and black hole event
horizons with various topologies.

For $d=4,5$ Emparan and Reall constructed in \cite{Emparan:2001wk}
exact solutions describing a black hole attached to a Kaluza-Klein
bubble using a generalized Weyl ansatz, describing axisymmetric static space-times
in higher-dimensional gravity.%
\footnote{See Ref.~\cite{Harmark:2004rm} for the generalization of
this class to stationary solutions.} For $d=4$ this was
generalized in \cite{Elvang:2002br} to two black holes plus one
bubble or two bubbles plus one black string. There, it was also
argued that the bubble balances the gravitational attraction
between the two black holes, thus keeping the configuration in
static equilibrium.

In \cite{Elvang:2004iz} these solutions were generalized to a
family of exact metrics for configurations with $p$ bubbles and
$q=p,p\pm 1$ black holes in $D=5,6$ dimensions. These are regular
and static solutions of the vacuum Einstein equations, describing
alternating sequences of Kaluza-Klein bubbles and black holes,
$e.g.$~for $(p,q)=(2,3)$ there is a sequence of the form: \beastar
  \rom{black~hole}~-~\rom{bubble}~-~\rom{black~hole}
  ~-~\rom{bubble}~-~\rom{black~hole} \, .
\eeastar
These solutions will be called \emph{bubble-black
hole sequences} and we will refer to particular elements in this class as
$(p,q)$ solutions. This large class of solutions, which was
anticipated in Ref.~\cite{Emparan:2001wk}, includes
the $(1,1)$, $(1,2)$ and $(2,1)$ solutions obtained and
analyzed in \cite{Emparan:2001wk,Elvang:2002br} as special
cases.

We refer the reader to \cite{Elvang:2004iz} for the explicit
construction of these bubble-black hole sequences and a comprehensive
analysis of their properties. Here we list a number of essential
features:
\begin{itemize}
\item All values of $n$ in the range $1/(d-2) < n < d-2$ are
realized. \item The mass $\mu$ can become arbitrarily large, and
for $\mu \rightarrow \infty$ we have $n \rightarrow 1/(d-2)$.
\item The solutions contain bubbles and black holes of various
topologies. In the five-dimensional case we find black holes with
spherical $S^3$ and ring $S^2 \times S^1$ topology, depending on
whether the black hole is at the end of the sequence or not.
Similarly, in the six-dimensional case we find black holes with
ring topology $S^3 \times S^1$ and tuboid topology $S^2 \times S^1
\times S^1$, depending on whether the black hole is at the end of
the sequence or not. The bubbles support the $S^1$'s of the
horizons against gravitational collapse. \item The  $(p,q)$
solutions are subject to constraints enforcing regularity, but
this leaves $q$ independent dimensionless parameters allowing for
instance the relative sizes of the black holes to vary.  The
existence of $q$ independent parameters in each $(p,q)$ solution
is the reason for the large degree of non-uniqueness in the
$(\mu,n)$ phase diagram, when considering bubble-black hole
sequences.
\end{itemize}

Another interesting feature of the bubble-black hole sequences is that
there exists a  map between five- and six-dimensional
solutions \cite{Elvang:2004iz}. As a consequence there is a
corresponding map between the physical parameters which reads
\begin{equation}
\label{mun5to6} \mu^{(6D)} = \frac{2\pi}{3L^{(5D)}} \mu^{(5D)}
\left( 5 - n^{(5D)} \right) \spa n^{(6D)} =
\frac{5n^{(5D)}-1}{5-n^{(5D)}} \ ,
\end{equation}
\begin{equation}
\label{ts5to6} \mathfrak{t}^{(6D)}_k = \mathfrak{t}^{(5D)}_k \spa
\mathfrak{s}^{(6D)}_k = \frac{4\pi}{L^{(5D)}}
\mathfrak{s}^{(5D)}_k \ ,
\end{equation}
where the superscripts $5D$ and $6D$ label the five- and six-dimensional
quantities respectively. This form of the map assumes a certain normalization of the parameters
of the solution, or equivalently, a choice of units, as further
explained in Ref.~\cite{Elvang:2004iz}.

\subsubsection{Thermodynamics}

For static space-times with more than one black hole horizon we
can associate a temperature to each black hole by analytically
continuing the solution to Euclidean space and performing the
proper identifications needed to make the Euclidean solution
regular near the locations of the Wick rotated event horizons. The
temperatures of the black holes need not be equal, and one can
derive a generalized Smarr formula that involves the temperature
of each black hole. The Euclidean solution is regular everywhere
only when all the temperatures are equal. It is always possible to
choose the $q$ free parameters of the $(p,q)$ solution to give a
one-parameter family of regular equal temperature solutions, which
we denote $(p,q)_{\mathfrak{t}}$.

The equal temperature $(p,q)_{\mathfrak{t}}$ solutions  are of
special interest for two reasons. First, the two solutions,
$(p,q)_{\mathfrak{t}}$ and $(q,p)_{\mathfrak{t}}$, are directly
related by a double Wick rotation which effectively interchanges
the time coordinate and the coordinate parameterizing the
Kaluza-Klein circle. This provides a duality map under which
bubbles and black holes are interchanged, giving rise to the
following explicit map between the physical quantities of the
solutions
\begin{equation}
\label{dW2} \mu' = n  \mt^{d-3} \mu \spa n' = \frac{1}{n} \spa
\mathfrak{t}' = \frac{1}{\mathfrak{t}} \spa \mathfrak{s}' =
\frac{(d-2)n-1}{d-2-n} \mt^{d-1} \mathfrak{s} \ .
\end{equation}
Secondly, for a given family of $(p,q)$ solutions,
the equal temperature solution extremizes the entropy for fixed
mass $\mu$ and fixed size of the Kaluza-Klein circle at infinity.
For all explicit cases considered we find that the entropy is
minimized for equal temperatures.%
\footnote{This is a feature that is particular to black holes,
independently of the presence of bubbles. As an analog, consider
two Schwarzschild black holes very far apart. It is
straightforward to see that for fixed total mass, the entropy of
such a configuration is minimized when the black holes have the
same radius (hence same temperature), while the maximal entropy
configuration is the one where all the mass is located in a single
black hole.}

Furthermore, the entropy of the $(1,1)$ solution is always lower
than the entropy of the uniform black string of the same mass
$\mu$. We expect that all other bubble-black hole sequences
$(p,q)$ have entropy lower than the $(1,1)$ solution, and this
is  confirmed for all explicitly studied examples.
The physical reason behind the expectation that
all bubble-black hole sequences have lower entropy than
a uniform string of the same mass, is that
some of the mass has gone into the bubble rather than the black
holes, giving a smaller horizon area.

\subsubsection{Phase diagrams for $d=4$ and $d=5$}

The general $(\mu,n)$ phase diagrams for $d=4,5$ can in principle
be drawn with all possible values $(p,q)$, though the explicit solution
of the constraints becomes increasingly complicated for high $p,q$.
However, the richness of the phase structure and the non-uniqueness in the
$(\mu,n)$ phase diagram, is already illustrated by considering some
particular examples of the general class of solutions, as was
done in \cite{Elvang:2004iz}.
As an illustration, we give here the exact form of the curve
for the (1,1) solution, corresponding to a bubble on a black hole,
\begin{equation}
\label{nmu11} d=4: \quad n_{(1,1)}(\mu )
= \frac{1}{4} \left[ -1 + 3 \sqrt{1 + \frac{8}{\mu^2}} \right] \qquad
; \qquad d=5: \quad
n_{(1,1)} (\mu ) =\frac{1}{3}  + \frac{4}{3\mu} \, .
\end{equation}
in five and six dimensions respectively. These two solutions
are related by the map in \eqref{mun5to6} and one may also check
that these curves are correctly self-dual under the duality
map \eqref{dW2}.

\begin{figure}[ht]
\centerline{\epsfig{file=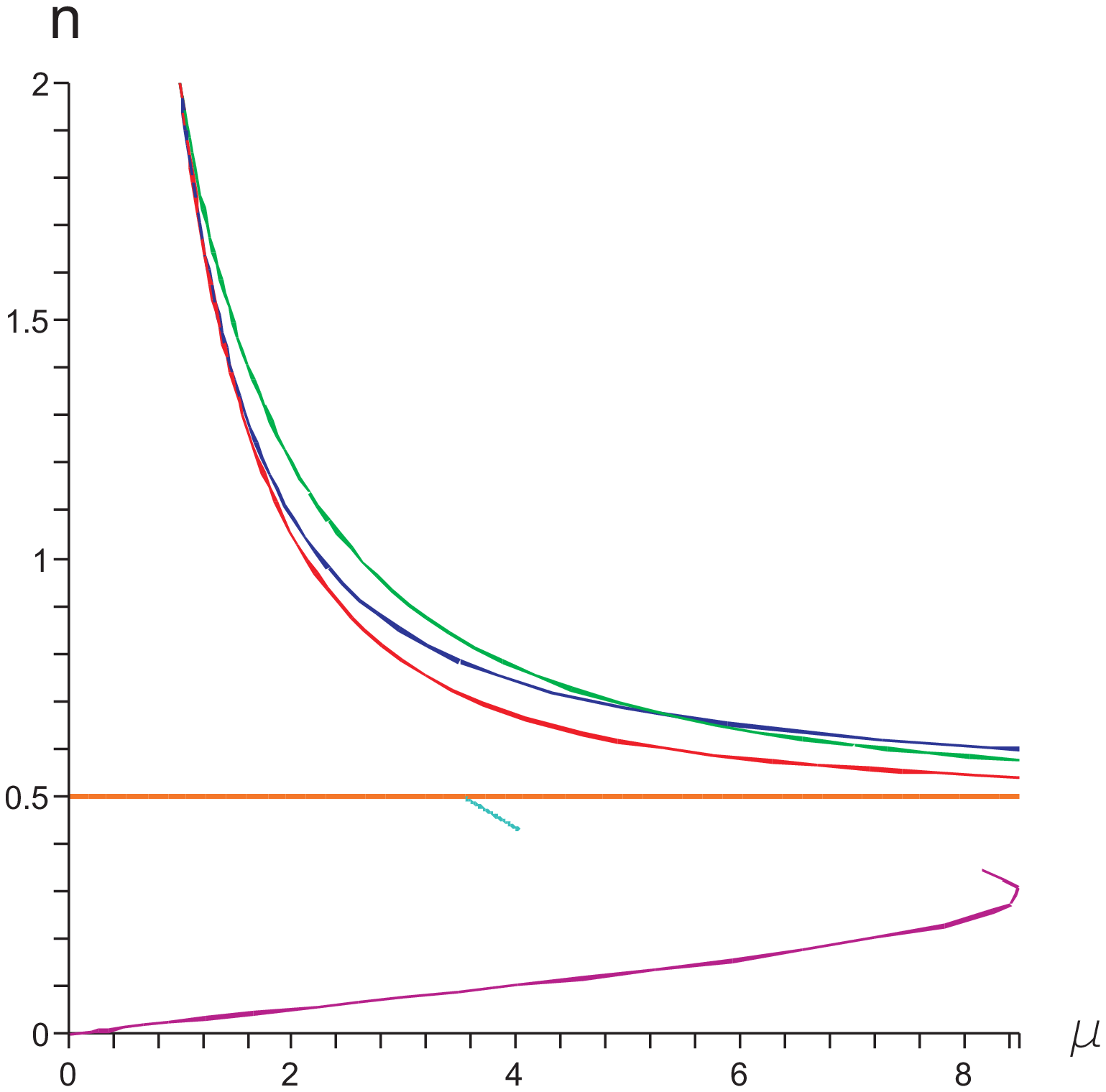,width=7cm,height=7cm}
\hskip .5cm \epsfig{file=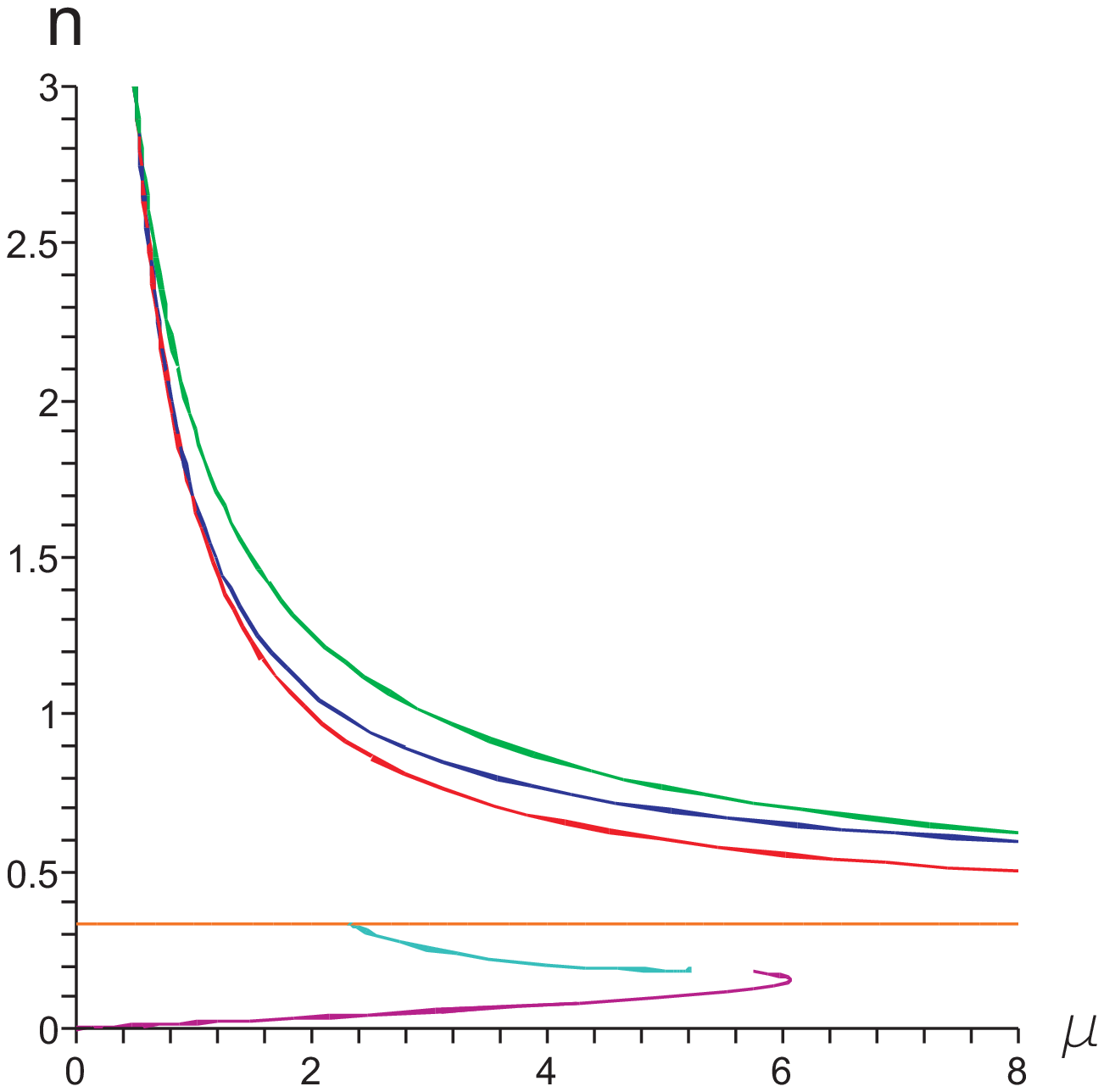,width=7 cm,height=7cm}}
\caption{$(\mu,n)$ phase diagrams for five (left figure) and six
(right figure) dimensions. We have drawn the $(p,q)=(1,1)$,
$(1,2){}_{\mathfrak{t}}$ and $(2,1)$ solutions. These curves lie
in the region $1/2 < n \leq 2$ for the five-dimensional case and
$1/3 < n \leq 3$ for the six-dimensional case. The lowest (red)
curve corresponds to the $(1,1)$ solution. The (blue) curve that
has highest $n$ for high values of $\mu$ is the equal temperature
$(1,2){}_{\mathfrak{t}}$ solution. The (green) curve that has
highest $n$ for small values of $\mu$ is the $(2,1)$ solution. The
entire phase space of the $(1,2)$ configuration is the wedge
bounded by the equal temperature $(1,2){}_{\mathfrak{t}}$ curve
and the $(1,1)$ curve. For completeness we have also included the
uniform (orange) and non-uniform (cyan) black string branch, and
the small black hole branch (magenta) displayed in
Figure \ref{fig1}.} \label{fig2}
\end{figure}

In Figure \ref{fig2} we have drawn for $d=4$ and $d=5$ the curves in
the $(\mu,n)$ phase diagram for the
$(p,q)=(1,1)$, $(1,2){}_{\mathfrak{t}}$ and $(2,1)$ solutions.
These correspond to the configurations
\begin{equation}
\label{11config}
 \begin{array}{lccc}
 (1,1): \qquad &  \rom{black~hole}&-&\rom{bubble} \\
 D=5 & S^3 &&  D \\
 D=6 &  S^3 \times S^1 &&  D \times S^1
  \end{array}
\end{equation}
\begin{equation}
\label{12config}
 \begin{array}{lccccc}
(1,2) : \qquad  & \rom{black~hole}&-&\rom{bubble} &-& \rom{black~hole} \\
 D=5 & S^3 &&  S^1 \times I   & & S^3 \\
 D=6  & S^3 \times S^1 &&   T^2 \times I && S^3 \times S^1
  \end{array}
\end{equation}
\bea
\label{21config}
   \begin{array}{lccccc}
 (2,1): \qquad & \rom{bubble}&-&\rom{black~ring}&-&\rom{bubble}  \\
 D=5 & D  &&   S^2 \times S^1 && D \\
 D=6 &  D \times S^1 &&   S^2 \times T^2 && D \times S^1
  \end{array}
\eea
where the first/second line in each configuration
corresponds to the topology in five/six dimensions. Here $D$ denotes the
disc and $I$ the line interval.

\subsubsection{Non-uniqueness in the phase diagram}

Finally we remark on the non-uniqueness of Kaluza-Klein black
holes in the $(\mu,n)$ phase diagram.%
\footnote{Non-uniqueness in higher dimensional pure gravity has
also been found for stationary black hole solutions in
asymptotically flat space-time. In five dimensions for a certain
range of parameters there exists both a rotating black hole with
$S^3$ horizon \cite{Myers:1986un} and rotating black rings with
$S^2 \times S^1$ horizons \cite{Emparan:2001wn}.} Clearly for a
given mass there is a high degree of non-uniqueness. The
non-uniqueness is not lifted by taking into account the relative
tension $n$, as there are explicitly known cases of physically
distinct solutions with the same mass and tension. For example the
$(1,2){}_{\mathfrak{t}}$ solution and the $(2,1)$ solution
intersect each other in the phase diagram. This means that we have
two physically different solutions in the same point of the
$(\mu,n)$ phase diagram. Moreover, there is in fact a continuously infinite non-uniqueness%
\footnote{Infinite non-uniqueness has also been found in
\cite{Emparan:2004wy} for black rings with dipole charges in
asymptotically flat space.} for certain points in the $(\mu,n)$
phase diagram. This is due to the fact that the $(p,q)$ solution
has $q$ free parameters \cite{Elvang:2004iz}. Hence, for given $p
\geq 2$ and $q \geq 3$, we have $q-2$ free continuous parameters
labeling physically different $(p,q)$ solutions, for certain
points in the $(\mu,n)$ phase diagram.

\section{More on the Gregory-Laflamme instability}
\label{sec:moreGL}

In this section we present further results on the Gregory-Laflamme instability.
First we discuss the large dimension limit of the GL critical mass and
the phase diagram that emerges. The case of the GL instability for
boosted neutral black strings, which has various interesting applications, will be
discussed next. We also summarize some of the main results connecting the
GL analysis to an LG (Landau-Ginzburg) analysis of the thermodynamics
and what is known about the GL instability for neutral black
branes with higher-dimensional compact spaces. Finally, we comment on an
interesting connection between the GL instability and a classical membrane instability,
known as the Rayleigh-Plateau instability.

\subsection{Large $d$ limit \label{sec:larged}}

In Section \ref{sec:GL} we reviewed the GL instability of the neutral black string
in $d+1$ dimensions, and in particular presented the differential equations
satisfied by the threshold mode. Unfortunately, no analytic solution to these
equations is known and one has to solve them numerically for a given value of $d$
on a case by case basis.
This was done for $d$ up to 50 in Ref.~\cite{Sorkin:2004qq}. It is
interesting to regard $d$ as a parameter in this system and study the behavior
of the GL mode as $d$ becomes very large.

In Ref.~\cite{Sorkin:2004qq} it was observed from the data that the critical
mass exhibits the exponential behavior
\begin{equation}
\label{eq:muGLnumer}
\mu_{\rm GL} \simeq 16.21 \cdot (0.686)^d
\end{equation}
at large $d$. One can verify this result by solving analytically for
the threshold mode in the large $d$ limit, as was done in
Ref.~\cite{Kol:2004pn}. Combined with the behavior of small
localized black holes for large dimensions, this result enables us
to conjecture the way the localized and non-uniform phases appear in
the $(\mu,n)$ phase diagram when the dimension is large.

To derive the large $d$ behavior of the GL mode, it is convenient to start with the linear
second order ODE  that appears when considering the negative modes
of a static and spherically symmetric perturbation of the
Schwarzschild black hole in $d$ dimensions (see \eqref{GPYmode}).
By taking the large $d$ limit, this ODE simplifies considerably
and by dropping the exponentially growing solution, one finds%
\footnote{Equivalently, one can take the large $d$ limit of the second order
differential equation \eqref{E5}.}
\begin{equation}
\chi (r) \sim \KK_{\frac{d-1}{2}} (r) r^{-\frac{d-1}{2}}
\end{equation}
where $\KK_s$ is the modified Bessel function of the second kind (we
remind the reader that the function $\chi (r)$ is related to the
$h_{rr}$ component of the perturbation, $c.f.$ eqs.\ \eqref{GLmode1}, \eqref{GLmode2}).
The GL wave number $k_{\rm GL} \simeq \sqrt{d}$ can be determined
by imposing the appropriate boundary conditions and the resulting
GL mass takes the form \cite{Kol:2004pn}
\begin{equation}
\label{eq:muGLlarged}
\log \mu_{\rm GL} \simeq d \log  \sqrt{\frac{e}{2\pi}} \spa d \gg 1
\end{equation}
This analytical result meshes very nicely with the earlier numerically obtained
result \eqref{eq:muGLnumer}.

\begin{figure}
\begin{center}
\includegraphics[width=7cm,height=4cm]{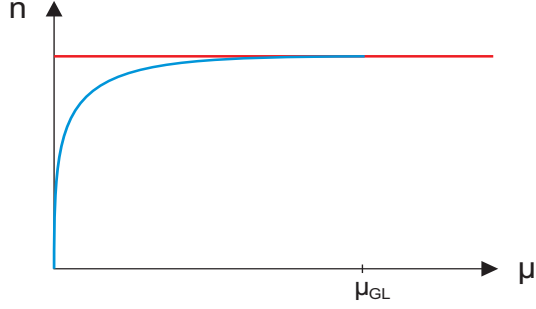}
\caption{Sketch of the $(\mu,n)$ phase diagram for large $d$.}
\label{fig_sketch}
\end{center}
\end{figure}

Moreover, it is interesting to consider at large $d$ the slope of the localized
black hole branch \eqref{bhslope}, measured
in units normalized with respect to the Gregory-Laflamme
point $(\mu_{\rm GL},n_{\rm GL}= 1/(d-2))$,
with $\mu_{\rm GL}$ given by the large $d$ formula
\eqref{eq:muGLlarged}. One finds
\begin{equation}
\frac{n}{n_{\rm GL}} \simeq  d^{d/2}
\frac{\mu}{\mu_{\rm GL}} \spa d \gg 1
~.
\end{equation}
From this expression we see that the slope becomes infinitely steep
as $d \rightarrow \infty$. This suggests that the curve
describing the localized black hole and non-uniform black string
branches should behave as sketched in Figure \ref{fig_sketch}. In particular,
the large $d$ behavior of the phase structure is expected to be such
that the non-uniform string gets closer and closer to having
$n=n_{\rm GL}= 1/(d-2)$.

Further evidence for the above picture stems from the realization that
the behavior \eqref{eq:muGLlarged} is equivalent to the critical
Schwarzschild radius being equal to $r_0 \simeq L/(2\pi) \sqrt{d} $.
This shows that the large $d$ critical string is ``fat'', and that most probably
it cannot decay into a black hole since the black hole cannot fit into
the compact dimension at the GL point \cite{Kol:2004pn}.
Hence, the uniform black string should decay towards a different end-state,
which presumably is the non-uniform black string.

Finally, from the comparison of the GL instability with
the Jeans instability in Section \ref{sec:GL} we recall that the Jeans instability
predicts that the large $d$ behavior of the critical wave number is
$k_J \sim d^{3/4}$. Now we see that this is different from the result
$k_c \sim \sqrt{d}$ above. The classical Rayleigh-Plateau instability, on
the other hand, does predict the correct large $d$ behavior as we
will review in Section \ref{sec:qualitative}.

\subsection{Boosted black strings \label{sec:booststr}}

Another interesting direction that has been pursued in \cite{Hovdebo:2006jy}
is the study of the GL instability for boosted black strings.
As we briefly review below it was found in \cite{Hovdebo:2006jy}
that the critical dimension (see Section \ref{sec:bhc}) depends on
the internal velocity and disappears for sufficiently
large boosts. Moreover, the five-dimensional black ring \cite{Emparan:2001wn}
reduces in the large radius limit to a boosted neutral black string (in five dimensions),
so these instabilities are also relevant to the stability analysis of black rings (we will
briefly return to those issues in Section \ref{sec:otherins}).

Boosted black strings can be obtained by boosting the static black string
\eqref{ublstr} along the $z$-direction. The resulting solution is
\begin{eqnarray}
\label{eq:boostedbs}
 ds^2 &=& -\left( 1 - \cosh^2 \alpha\frac{r_0^{d-3}}{r^{d-3}} \right) dt^2 +
 2 \frac{r_0^{d-3}}{r^{d-3}}
 \cosh \alpha \sinh \alpha dt dz + \left( 1 + \sinh^2 \alpha
\frac{r_0^{d-3}}{r^{d-3}} \right) dz^2 \nn \\ &&
 + \left( 1- \frac{r_0^{d-3}}{r^{d-3}} \right)^{-1} dr^2 + r^2
(d\theta^2 + \sin^2 \theta d\Omega_{d-3}^2 )  ~.
\end{eqnarray}

The mass and tension of the boosted black string can be computed from
the general expressions \eqref{MT}, \eqref{then}, \eqref{themu}, giving
\begin{equation}
\label{muboost}
\mu = \Omega_{d-2} \left( \frac{r_0}{L} \right)^{d-3} [ (d-3) \cosh^2 \alpha +1]
\spa n =  \frac{(d-3) \sinh^2 \al -1}{ (d-3) \cosh^2 \alpha +1 }
~.
\end{equation}
In this case the string also carries the dimensionless momentum%
\footnote{The momentum is given by $P  = T_{tz} =
 \frac{\Omega_{d-2} L}{16 \pi G_{\rm N}} c_{tz}$ with $c_{tz}$ the first
 correction around flat space in $g_{tz} \simeq 1 + c_{tz}/r^{d-3}$.
 We define, in analogy with \eqref{themu}, the dimensionless momentum
 as $ p =\frac{16\pi G_{\rm N}}{L^{d-2}} P$.}
\begin{equation}
p =  \Omega_{d-2} (d-3) \cosh \al \sinh \al
~.
\end{equation}

One can now extend the original global thermodynamic argument of Gregory and
Laflamme to this case, by comparing the entropy of the boosted black string
(with horizon $r_0$ and boost $\al$) to the entropy of a boosted spherical
black hole (with horizon $r_0'$ and boost $\al'$)
with the same energy and momentum. Because we compare the solutions
at the same energy
and momentum, one can express $(r_0',\al')$ in terms of $(r_0,\al)$
of the boosted black string. Then, equating the two entropies determines
a minimal circumference \cite{Hovdebo:2006jy}
\begin{equation}
\label{eq:Lminboost}
L_{\rm min} = \frac{r_0}{\cosh \al} \frac{(d-1)^{d-1} }{ [
(d-3) (d-1) + \cosh^{-2} \al ]^{(d-1)/2 }}  \frac{\Omega_{d-1}}{\Omega_{d-2}}
\end{equation}
and hence from \eqref{muboost} one can define a maximum mass
$\mu_{\rm max} \equiv \mu_{L = L_{\rm min}}$. This mass can be
seen as a rough estimate of the point where the boosted
black string becomes unstable. The instability appears for
$\mu < \mu_{\rm max}$ (or $L > L_{\rm min}$). Note that $L_{\rm min}$
scales like $1/\cosh \al$, so this naive analysis suggests that the
instability will persist for large boosts.

One can further analyze the instability of the boosted black string
by appropriately boosting the time-dependent GL mode \cite{Hovdebo:2006jy}.
In this way, one finds that the instabilities of the boosted string
can be related to instabilities of the static black string with complex frequencies.
The relation is even more direct for the static threshold mode
\eqref{thresmode}, on which one can simply apply the same boost that
gave \eqref{eq:boostedbs}. As a result one finds
the mode $\exp ( i \omega t + i k_{c,{\rm boost}} z)$ with
\begin{equation}
\label{eq:kmaxboost}
k_{c,{\rm boost}} = \cosh \al k_{c} \spa \omega  = \sinh \al k_{c}
\end{equation}
where $k_{c}$ is the threshold wave number for the static black
string (see Table \ref{tabkc}). So the threshold modes are waves
that travel  in the $z$-direction with the same speed as the
boosted string. Note that the boost of the GL mode in a transverse
direction, including the time-dependent one, was considered in
Refs.~\cite{Aharony:2004ig,Harmark:2005jk} and will be reviewed in
Section \ref{sec:nonGL}.

Given the above results one can show that the critical dimension
(see Section \ref{sec:bhc}) depends on the boost parameter, and
in fact disappears (for any given $d$) for large enough boost
\cite{Hovdebo:2006jy}. To see this observe from \eqref{eq:kmaxboost} that
\begin{equation}
L_{\rm GL,boost} = \frac{r_0}{\cosh \al} \left[ \frac{ (d-2) \Omega_{d-2}}{\mu_{GL} }
\right]^{1/(d-3)}
\end{equation}
and hence from \eqref{muboost},
\begin{equation}
\mu_{\rm GL,boost} = \frac{\cosh^{d-3} \alpha}{d-2} [ (d-3) \cosh^2 \alpha +1] \mu_{GL}
\end{equation}
where $\mu_{GL}$ is the neutral GL mass.  Comparing this to the $\mu_{\rm max}$
obtained above (see below \eqref{eq:Lminboost}) one finds that for small
boosts $\mu_{\rm max} > \mu_{\rm GL,boost}$ so that there is a regime where the
boosted black string is stable, but the localized black hole has higher entropy,
suggesting the existence of an unstable non-uniform black string in between.
For zero boost this occurs for dimensions $D \leq 13$. On the other hand,
for sufficiently large boosts one has $\mu_{\rm max} < \mu_{\rm GL,boost}$.
So in this case, there is a regime, $ \mu_{\rm max} < \mu <\mu_{\rm GL,boost}$,
where the boosted black string is unstable but the localized black hole does not
have a higher entropy, suggesting that the end-state is a stable non-uniform
(boosted) black string in this range. In other words, for sufficiently large
boosts the critical dimension disappears and we have a phase diagram
that is analogous to   that found for large $d$ in the neutral case
in Fig.~\ref{fig_sketch}.

We will return to some other aspects of boosted black strings in
Section \ref{sec:otherins}

\subsection{Higher-dimensional compact spaces}
\label{sec:highdspaces}

Most of the work on the stability and phases of black objects in
spaces with compact transverse directions has been performed so
far for the simplest case of one compact direction. Already for
this case one observes a very rich phase structure. This phase
structure is expected to become even richer as we go to higher
dimensional compact spaces, such as $T^n$, $S^n$, $K3$ and $CY_3$
manifolds. As the number of moduli increases, the structure of the
moduli space and hence the phase structure is clearly going to
become more involved.

\subsubsection{Instability of neutral black $p$-branes}

The one higher-dimensional case that has been studied \cite{Kol:2004pn,Kol:2006vu} in
more detail is that of the torus $\T^p$, which is the simplest generalization of the
circle $S^1$. We summarize some of the main results here.

Consider neutral solutions of Einstein gravity with an event horizon, that asymptote
to $\CM^{d-1,1} \times \T^p$. Here we have the uniform black $p$-brane phase
which is the obvious generalization of the black string, and is just a
Schwarzschild-Tangherlini black hole in $d$ dimensions, with $p$ flat directions added
$ds^2 = \sum_{i=1}^p dz^i dz^i$ for the $p$-dimensional torus.
The torus is defined by $p$ period vectors $\vec{e}_i$ such
that any function $Y(\vec{z} + \vec{e} ) = Y(\vec{z})$ with $\vec{e}$
belonging to the integer lattice spanned by the period vectors.
Performing a similar stability analysis as in Section \ref{sec:GL},
one finds that the threshold mode $P_{c,\mu \nu}(r) Y(\vec{z}) $ satisfies \cite{Kol:2004pn}
\begin{equation}
\label{eq:toteig}
 \lambda_{\rm Sch} + \lambda_{\T^p} = 0
 \end{equation}
 where $\lambda_{\rm Sch}$ and $\lambda_{\T^p}$ are the eigenvalues in
 \begin{equation}
\hat \Delta_L P_{c,\mu \nu} (r) =\lambda_{\rm Sch} P_{c,\mu \nu} (r) \spa
 - \Delta Y (\vec{z}) = \lambda_{\T^p} Y (\vec{z})
\end{equation}
The only way to satisfy \eqref{eq:toteig} is to take the GPY mode for the
Schwarzschild black hole, and then it suffices to consider modes which
are tensor on the Schwarzschild geometry and scalar on the torus.
Consequently, we find that the instability of the black brane sets in when
\begin{equation}
r_0^2 = \frac{k_c^2}{\lambda_{\rm min}}
\end{equation}
where $k_c$ is the eigenvalue in \eqref{Goncrith} and $\lambda_{\rm min}$
is the mininimal (non-zero) eigenvalue of the Laplacian on the torus $\T^p$.
In fact, this argument is general and applies to any stable compact manifold.
For the case of $\T^p$ this eigenvalue is given explicitly by
\begin{equation}
\lambda_{{\rm min},\T^p} = | \vec{k}_{\rm min}  |^2
\end{equation}
where $\vec{k}_{\rm min}$ is the shortest vector in the reciprocal lattice of $\T^p$
($i.e.$ the lattice spanned by the vectors $\vec{k}$ with
$\vec{k} \cdot \vec{e}  = 2\pi n $, $n \in \Z$).

It is also physically clear that for a highly asymmetric torus,
with $e.g.$ one large direction and the other ones small, the
instability will set in for the mode along the large direction. In
this sense the stability properties of black branes on such
asymmetric tori are closely related to those of the black string.
Hence, it seems more plausible that new effects will show
up in the study of the other extreme, namely a square torus, with
all directions of equal length. Moreover, understanding these two
extremes may lead to a qualitative understanding of the
intermediate region.

Recently, the Landau-Ginzburg method (see Section
\ref{sec:qualitative}), was applied to the case of black
$p$-branes on a square torus $\T^p$. In this case it is easy to
see that $p$ modes (corresponding to each of the directions)
become marginally tachyonic at the same GL point.
Ref.~\cite{Kol:2006vu} then studied the order of the phase
transition, $i.e.$ the critical dimension, for such tori. It was
shown that for the phase transition to be first order, it is
enough to find a single direction in tachyon space for which this
is the case. It was furthermore shown that in order to study the $\T^p$
case one needs to combine the known results for the $S^1$ case
together with results for the $\T^2$. As a result it was
concluded that the transition order for square tori depends only
on the number of extended dimensions $d$, and not on $p$. In other
words, the critical dimension discussed in Section \ref{sec:bhc}
for $p=1$ is unchanged when considering higher-dimensional square
tori. Note that in an earlier work \cite{Kol:2004pn} (see also \cite{Park:2004zr})
a more naive analysis based on the comparison of the entropy of black holes
localized in the compact directions and the entropy of black branes
seemed to indicate that the critical dimension could become lower
as the number of compact directions increases. As explained in
\cite{Kol:2006vu}, this conclusion is misleading because the corrections to
the entropy of localized black holes can be shown to have an
important effect in this case, and these were ignored in the
analysis of  \cite{Kol:2004pn}. More generally, this shows that
one should be cautious when using the Schwarzschild black hole
entropy as an approximation to make qualitative statements,
$i.e.$ one should carefully study the effects of corrections due to
the presence of the compact manifold.

\subsubsection{Phase structure for higher tori}

So far we have just considered the aspect of instability of uniform black branes
with compact transverse directions, but if we think about possible
phases it is not difficult to see that this will be very rich.
Consider, for example, the case of the two-torus. First of all, for
any phase of Kaluza-Klein black holes on $\CM^{1,d-1} \times S^1$ we trivially
get a phase for the case of  $\CM^{1,d-1} \times \T^2$ by adding a flat compact
direction. This includes the uniform black two-brane phase discussed above,
but also $e.g.$ semi-localized black holes of horizon topology $ S^{d-1} \times S^1$
(these are of black string type) as well as bubble-black hole sequences
with an extra circle direction. There are also going to be entirely new
phases each time we add a compact direction.
For example, in the case discussed above, there is also a phase of black holes that
are localized in two of the $\T^2$-directions, $i.e.$ a phase with horizon topology
$S^d$. These solutions reduce to the $(d+2)$-dimensional Schwarzschild black hole
in the limit of zero mass, and it would be interesting to obtain the first order correction
away from the zero mass limit using the methods of \cite{Harmark:2002tr,Harmark:2003yz}
or \cite{Gorbonos:2004uc}. There are also new bubble-black hole sequences
for the case of $d=5,6$ (see for example the Conclusion
of \cite{Elvang:2004iz}) that are in principle not difficult to construct
explicitly. All these will be part of some complicated higher dimensional
phase structure with a complicated pattern of (in)stabilities.

\subsubsection{Comments on general compact Ricci flat manifolds}

More generally we may consider KK black hole solutions in
$(d+p)$-dimensional pure gravity asymptoting to $\R^{d-1,1} \times
\NN_p$, with $\NN_p$ being a generic $p$-dimensional compact Ricci
flat manifold, for example a compact Calabi-Yau manifold. In this
case, the phase diagram of neutral, static black hole solutions
will be parameterized by the mass $\MM$, the components of a
generalized tension tensor $\TT$ and the moduli that characterize
the manifold $\NN_p$ at infinity (minus one for the overall volume
of $\NN_p$ that can be scaled away). The components of the tension
tensor $\TT$ are straightforward generalizations of the single
tension $n$ appearing above and can be thought of as the chemical
potentials corresponding to the moduli of $\NN_p$. See Ref.~\cite{Kastor:2006ti} for
a treatment of the two-torus case, including a derivation of the first law of
thermodynamics.

To obtain the uniform black membrane solution we simply tensor
the $d$-dimensional Schwarzschild solution with the Ricci flat metric of
$\NN_p$
\begin{equation}
\label{blackuniform}
ds^2=-\left(1-\frac{r_0^{d-3}}{r^{d-3}}\right)
dt^2+\left(1-\frac{r_0^{d-3}}{r^{d-3}}\right)^{-1} dr^2
+r^2 d\Omega^2_{d-2}+ds^2_{\NN_p}
~.
\end{equation}
An obvious generalization of the argument appearing around eq.\
\eqref{eq:toteig} suggests again the presence of a
Gregory-Laflamme instability. The discussion of the higher tori
case above indicates the presence of many more stable or unstable
phases in the phase diagram with a complicated pattern of
(in)stabilities. Beyond the phases of localized black holes,
non-uniform black membranes and bubble-black hole sequences one
can imagine new ``uniform'' phases, where the $\NN_p$ moduli are
radially varying, the tension tensor $\TT^{ab}$ has values other
than those of the uniform black brane solution
\eqref{blackuniform} and the metric components have no dependence
on the coordinates of $\NN_p$. Black hole solutions of a similar
spirit are abundant in extended supergravity, for example in
$d=4$, $\NN=2$ supergravity \cite{Ferrara:1995ih,Kastor:1997wi}
(see \cite{Pioline:2006ni} for a recent review).

In general, many features of the phase diagram are expected to be strongly
correlated with the topology of the compact manifold $\NN_p$.
It would be interesting to explore this connection further.

\subsection{Other developments}
\label{sec:qualitative}

We conclude this section by briefly discussing several other
developments related to the GL instability and the Kaluza-Klein
black hole phase diagram. These include an application of the
Landau-Ginzburg theory of phase transitions, the use of Morse
theory and the study of the merger point, where the black string
and black hole branches meet. While these developments are
important and very interesting, we choose to be brief here since
many of these are very nicely explained in the review
\cite{Kol:2004ww} by Kol. We also comment on a recent connection
observed in Ref.~\cite{Cardoso:2006ks} between the (classical)
Rayleigh-Plateau instability of membranes and the GL instability.

\subsubsection{LG in GL}

Some of the features of the black hole-black string transition, in
particular the order of the phase transition, can be nicely
studied using the Landau-Ginzburg (LG) theory for the
thermodynamics, as was done in
Refs.~\cite{Kol:2002xz,Kol:2004ww,Kol:2006vu} (see also
\cite{Gubser:2001ac,Wiseman:2002zc,Sorkin:2004qq}). We briefly
summarize some of the main results here, referring the reader to
the review \cite{Kol:2004ww} for a useful expos\'e of the general
idea and further details, as well as the recent article
\cite{Kol:2006vu} where a new method for determining the order of
the phase transition was developed based on the LG perspective.

\begin{figure}
\begin{center}
\includegraphics[width=7cm,height=5cm]{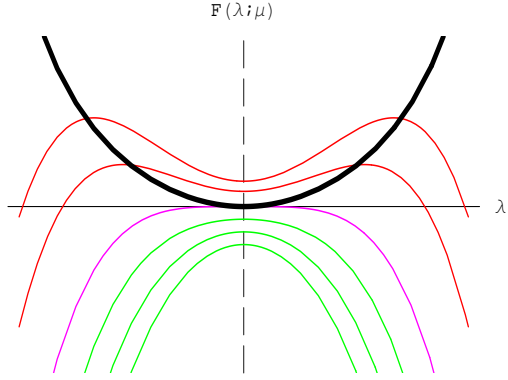}
\caption{The typical behavior of the free energy during a first order phase transition.
The red lines are at constant values of $\mu>\mu_c$ and the green lines at
constant $\mu<\mu_c$. The thick black line represents a set of unstable asymmetric
($\lambda \neq 0$) vacua at $\mu>\mu_c$. The dashed horizontal line represents
the symmetric $(\lambda=0)$ phase of the system, which is stable at $\mu>\mu_c$ and
unstable at $\mu\leq \mu_c$.}
\label{firstorder}
\end{center}
\end{figure}

\begin{figure}
\begin{center}
\includegraphics[width=7cm,height=5cm]{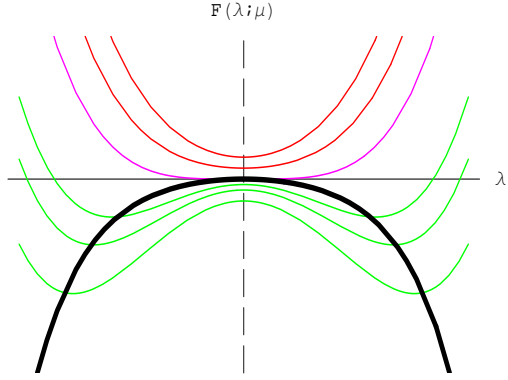}
\caption{The typical behavior of the free energy during a second (or higher) order phase transition.
Again, the red lines are at constant values of $\mu>\mu_c$ and the green lines at
constant $\mu<\mu_c$. The thick black line represents a set of stable asymmetric
($\lambda \neq 0$) vacua at $\mu<\mu_c$. The dashed horizontal line represents
the symmetric $(\lambda=0)$ phase of the system, which is stable at $\mu>\mu_c$ and
unstable at $\mu\leq \mu_c$.}
\label{secondorder}
\end{center}
\end{figure}

In statistical physics the Landau-Ginzburg theory \cite{Landau}
determines the order of a phase transition from the local behavior
of the free energy $F$ near a critical point. For a simple system
parameterized by a single variable $\lambda$ and a control
parameter $\mu$ the free energy is a function $F=F(\lambda;\mu)$.
We are interested in systems where a certain parity symmetry
forces the free energy to be an even function of $\lambda$, $i.e.$
$F=F(\lambda^2;\mu)$. For such systems the existence of a
marginally tachyonic mode at some critical value $\mu_c$ of the
control parameter implies that we can expand $F$ near $\lambda=0$,
$\mu=\mu_c$ in a power series in $\lambda$
\begin{equation}
\label{F1}
F=F_1(\mu-\mu_c)\lambda^2+F_2 \lambda^4+\OO(\lambda^6)
~,
\end{equation}
where without loss of generality we can take the constant $F_1$ to be positive.
The two possible behaviors of $F$ are summarized in Figures \ref{firstorder},
\ref{secondorder} and are distinguished by the sign of the quartic
coefficient $F_2$.

For negative $F_2$ (see Fig.\ \ref{firstorder}) the system
undergoes a first order transition. As $\mu$ is varied the system passes
from a symmetric stable phase ($\lambda=0$ and $\mu>\mu_c$)
to an unstable symmetric phase. Since $F$ is bounded from below
there is a new stable asymmetric vacuum at finite $\lambda$ and the system
discontinuously jumps to this new stable vacuum when $\mu<\mu_c$.

For positive $F_2$ (see Fig.\ \ref{secondorder}) the system undergoes a second (or higher)
order phase transition. As we vary $\mu$ from $\mu>\mu_c$ to $\mu<\mu_c$
the system passes continuously from the stable symmetric vacuum at $\lambda=0$
to a new asymmetric vacuum at $\lambda\neq 0$.

The non-uniform black string exhibits similar behavior with
details that depend crucially on the space-time dimension. In the
language of the above one-dimensional example, we can think of
$\lambda$ as a parameter that measures how far the black string is
from the uniform solution. This role can be played by the
amplitude of the marginally tachyonic GL mode \eqref{Goncrith},
$i.e.$ $\lambda$ can be taken as the coefficient of the threshold
mode metric perturbation
\begin{equation}
h \sim \lambda ~ e^{ik_c z} ~P_{c}
~.
\end{equation}
The control parameter $\mu$ can be taken as the mass of the black brane solution,
or better yet in the canonical ensemble as the dimensionless ratio $\beta/L$ of the
inverse temperature over the $z$ radius.

When dealing with gravity one has to consider the free energy\footnote{The
r.h.s.\ of this equation is the standard Gibbons-Hawking expression for the gravity
action. $R$ is the Ricci scalar and $K$ the extrinsic curvature
on the boundary $\d \MM$. $K^0$ is the same quantity for a reference
geometry.}
\begin{equation}
\label{freeenergy}
-\beta F[g_{\mu \nu}]=I_{GH}[g_{\mu \nu}]=
\frac{1}{16\pi G_N}\left(\int_{\MM} R+ 2 \int_{\d \MM}(K-K^0) \right)
~,
\end{equation}
as a function on the infinite dimensional configuration space of all metrics
with given boundary conditions. For black branes smeared on a transverse
$\T^p$ space the corresponding Landau-Ginzburg analysis around the
uniform solution has been performed in \cite{Kol:2006vu}.
In this system one finds a generalized multi-dimensional version of
\eqref{F1}, which arises by analyzing the geometry of the non-uniform
black brane up to second order in perturbation theory (for more details
we refer the reader to \cite{Kol:2006vu}). The order of the transition
is determined by the sign of an appropriate set of quartic coefficients in
the free energy expansion. One finds that the result depends crucially
on the dimension $d=D-p$ of the non-compact part of the space time
(see also comments in Section \ref{sec:highdspaces}).
For $D\leq 12$ the transition is first order and the phase diagram
looks like Fig.\ \ref{fig1}, whereas for $D\geq 13$ the transition is second order
and the phase diagram becomes as in Fig.\ \ref{fig_sketch}.
Hence, there is an interesting critical dimension ``$D^*=12.5$" where the
behavior of the system changes dramatically.

The order of the phase transition for a black string
on $\R^{3,1}\times S^1$ was first determined with a different method by Gubser in
\cite{Gubser:2001ac} and was improved upon by Wiseman in
\cite{Wiseman:2002zc}. This computation was generalized to
a more general background $\CM^{d} \times S^1$ in \cite{Sorkin:2004qq}
where the critical dimension $D^*$ was first observed.
In addition, \cite{Kudoh:2005hf} pointed out that the critical dimension
depends also on the ensemble; in the micro-canonical ensemble the
critical dimension is one unit higher than that in the canonical ensemble.

\subsubsection{Morse theory}

Another useful method that has been applied to address the issue of
the endpoint of the decay of the unstable black string is Morse theory \cite{Milnor},
as originally proposed in \cite{Kol:2002xz} and reviewed in \cite{Kol:2004ww}.
While Landau-Ginzburg theory can be employed to study the local properties of
the phase diagram, Morse theory can be used to obtain information about the
global (topological) structure of the phase diagram.

In general, when we try to determine the pattern of static
solutions in the phase diagram of Kaluza-Klein black holes on $\CM^{d}
\times S^1$, we are in essence studying the extrema of the
gravitational action in the space of metrics with given boundary
conditions. Morse theory can be useful in this study, since it is,
by definition, the topological theory of the extrema of functions
on a given manifold. For instance, it can be used to put
constraints on how the extrema of a function behave under
continuous deformations. The latter aspect of Morse theory was
employed in \cite{Kol:2002xz} to argue that a natural completion
of the black string phase diagram involves a point, called the
"merger point", where the non-uniform black string branch meets
the localized black hole branch.

The intuitive argument behind this analysis is based on the fact
that we know that there is a localized black hole phase as well as a
non-uniform black string phase, but a priori not where these phases
end in the phase diagram. Since it is unphysical for a phase to end in
nothing, it follows by ``phase conservation'' that the simplest
possibility is that these two phases meet in a topology changing
phase transition, the merger point.

\subsubsection{Merger point}
\label{sec:merger}

The prediction of a merger point in the phase diagram of black strings
implies an intriguing topology changing transition where the black string horizon
with topology $S^{d-2}\times S^1$ pinches off and becomes a black hole
horizon with topology $S^{d-1}$. This transition reminds of the Gross-Witten
transition \cite{Gross:1980he} in two dimensional large $N$ gauge theories.
Also certain features of the topology change near the merger point resemble
the conifold transition in the moduli space of Calabi-Yau threefolds \cite{Greene:1996cy}.

To this effect, it was argued in Ref~\cite{Kol:2002xz} that not
only does the horizon topology change, but also the local topology
in the Euclidean section. This topology change can be modelled in
analogy to the conifold transition as a pyramid, with the singular
topology being the cone over $S^{d-2} \times S^2$,
\begin{equation}
ds^2 = d \rho^2 + \frac{\rho^2}{d-1} [ d\Omega_{2}^2 + (d-3) d \Omega_{d-2}^2]
\end{equation}
which was termed the ``double-cone''. Here, $\rho$ measures the
distance from the tip of the cone and the constant prefactors
ensure Ricci-flatness. Refs.~\cite{Kol:2003ja,Sorkin:2006wp} show
that the local geometry around the ``waist'' of the most
nonuniform solutions is indeed cone-like to a good approximation.

Evidence for the merger point transition has been found with a numerical
study in \cite{Kol:2003ja,Sorkin:2006wp}.
Interesting parallels between the merger point transition
and the Choptuik scaling in black hole formation \cite{Choptuik:1992jv}
have been observed in
\cite{Kol:2005vy,Sorkin:2005vz,Frolov:2006tc,Asnin:2006ip}. More details about the merger point
can be found in the original work \cite{Kol:2002xz} and the review \cite{Kol:2004ww}.

\subsubsection{Rayleigh-Plateau instability}

The first law of thermodynamics
\begin{equation}
\label{firstlaw}
dM={\cal{T}} dA~,
\end{equation}
$A$ being the horizon area, suggests a natural parallel between
the mechanics of black holes and hydrodynamics. Indeed, by
viewing \eqref{firstlaw} as a law for fluids where ${\cal{T}}$ is interpreted as
an effective surface tension one is led to a simple picture of black holes,
where the event horizon is treated as a kind of membrane with well-defined
mechanical, electrical and magnetic properties \cite{Thorne:1986iy}.
It has been noticed that the membrane approach works surprisingly
well in reproducing various aspects of black hole physics in different contexts
(see for instance \cite{Kovtun:2003wp,Kovtun:2004de} for an application
in the context of AdS/CFT).

The membrane analogy was applied to black branes in
\cite{Cardoso:2006ks,Cardoso:2006sj}, where it was observed that there is an
impressive correspondence between the Rayleigh-Plateau instability
of long fluid cylinders and the Gregory-Laflamme instability of
black strings \cite{Cardoso:2006ks}. Both instabilities are long-wavelength
instabilities involving only the axisymmetric $s$-modes (for the
black  string this statement has been verified in
\cite{Kudoh:2006bp}). The critical wave numbers $k_{RP}$ and
$k_{c}$ agree exactly at large dimension $d$ - in particular they
both scale as $\sqrt d$ for $d \gg 1$ - and there is qualitative
agreement between the behavior of the instability time scales.
Hence, this classical system appears to provide a better
description of the Gregory-Laflamme instability as compared to the
Jeans instability (see Section \ref{sec:GL}).

\section{Instabilities in supergravity}
\label{sec:sugra}

Our discussion of the GL instability has so far been entirely within
the context of Einstein gravity. A natural question is whether
GL-like instabilities can also arise in the presence of matter fields,
such as scalar fields and gauge potentials. This question is especially
relevant in supergravity theories which appear in the low-energy limit of string theory.
In this context one encounters a variety of black brane solutions
whose stability properties are not only important in their own
right, but can also have non-trivial consequences for their dual
non-gravitational theories via the gauge/gravity correspondence
(see Section \ref{sec:holo}).

Gregory and Laflamme already considered instabilities of charged branes in
\cite{Gregory:1994bj,Gregory:1994tw}. Further work in this direction appeared in
\cite{Gregory:2000gf,Gubser:2000ec,Gubser:2000mm,Reall:2001ag,Gregory:2001bd,%
 Hirayama:2001bi,Hubeny:2002xn,Hirayama:2002hn,Gubser:2002yi,%
Hartnoll:2003as,Kang:2004hm,Kang:2004ys,Bostock:2004mg,Aharony:2004ig,%
Harmark:2004ws,Gubser:2004dr,Ross:2005vh,Friess:2005tz,Kang:2005is,Kudoh:2005hf,%
Friess:2005zp,Harmark:2005jk,Miyamoto:2006nd}.

The investigation into the classical stability of charged branes
began with the study of non-extremal singly-charged black $p$-branes
\cite{Gregory:1994bj,Gregory:1994tw,Reall:2001ag,Hirayama:2002hn,Gubser:2002yi,Kang:2004hm,Miyamoto:2006nd}.
In this case, there is a GL instability when the charge is
sufficiently small compared to the total mass of the brane, whereas
the brane is stable for large charge, $i.e.$ close to extremality.
This has been shown using numerical methods. In Section
\ref{sec:CSC} we comment on how this is related to the
thermodynamics of these branes. In this review we consider instead
the case of smeared non-extremal singly-charged black $p$-branes.
The reason is that in this case we can be more precise, since we can
explicitly find  the GL instability by mapping it from the GL
instability of the neutral black string
\cite{Aharony:2004ig,Harmark:2005jk}.

Before turning to the GL instability of non- and near-extremal
smeared branes, we review the boost/U-duality map between phases of
neutral Kaluza-Klein black holes and this class of branes
\cite{Harmark:2004ws} (see also
\cite{Bostock:2004mg,Aharony:2004ig,Kudoh:2005hf}). This map
suggests the presence of a GL instability in the case of non- and
near-extremal smeared branes, which we then proceed to verify more
explicitly. We also review recent work on GL instabilities of
D-brane bound states that appeared in
Refs.~\cite{Gubser:2004dr,Ross:2005vh,Friess:2005tz}.

\subsection{New phases of non- and near-extremal branes}
\label{sec:nonphases}

The starting point of our discussion is
the set of $1/2$ BPS branes in type IIA/B string theory and M-theory
with a circle in the transverse space. We would like to consider
the class of solutions that arise as thermal excitations of these branes.
We will refer to the branes in this class as {\sl non-extremal branes on a circle}.
The near-extremal limit of these solutions will be defined in a way
that keeps the non-trivial physics related to the presence
of the circle. We will refer to this particular class of near-extremal branes as
{\sl near-extremal branes on a circle}. The dual non-gravitational theories
living on these branes will be reviewed briefly in Section \ref{sec:holo}.

In analogy with the neutral Kaluza-Klein black holes of Section \ref{sec:KKphases},
one can define phase diagrams for the non- and near-extremal branes on a circle. See
Ref.~\cite{Harmark:2004ws} for details and conventions. For non-extremal branes on a circle
the dimensionless asymptotic quantities are the mass $\bar \mu$, the relative tension in
the circle direction $\bar n$ and the charge $q$. For
near-extremal branes on a circle there are two quantities,
the energy above extremality $\epsilon$, rescaled to be dimensionless,
and the {\sl relative tension} $r$ which is the ratio between the tension of the brane along the
transverse circle and the energy. The tension is measured using the general
tension formula found in \cite{Harmark:2004ch} that holds for non-asymptotically flat spaces.
Hence, the resulting phases can be plotted in a two-dimensional
$(\epsilon,r)$ phase diagram.

In this section we review the following:
\begin{itemize}
\item One can transform any static and neutral Kaluza-Klein black hole
to a non-extremal brane on a circle with a combined boost
and U-duality%
\footnote{See $e.g.$\ Ref.~\cite{Obers:1998fb} for a review on U-duality.}
 transformation. For a given charge $q$, this gives a map from points in the
$(\mu,n)$ phase diagram to points in the non-extremal $(\bar \mu, \bar n)$ phase diagram.
\item By taking a particular near-extremal limit of the map to
non-extremal branes, one can transform any static and neutral
Kaluza-Klein black hole to a near-extremal brane on a circle.
In particular, this provides a map relating points in the $(\mu,n)$
phase diagram to points in the $(\epsilon,r)$ phase diagram.
\item The above maps can be applied to the known phases
of neutral Kaluza-Klein black holes, from which we generate
(new) phases of non- and near-extremal branes on a circle.
\end{itemize}

\subsubsection{Extremal $p$-branes}

We consider here a class of branes that are thermal excitations of
the $1/2$ BPS branes of type IIA and IIB string theory and
M-theory. The singly-charged dilatonic $p$-branes in a
$D$-dimensional space-time are solutions to the equations of
motion of the action
\begin{equation}
\label{IDact}
I_D = \frac{1}{16 \pi G_D} \int d^D x \sqrt{-g} \left(
R - \frac{1}{2} \partial_\mu \phi \partial^\mu \phi
- \frac{1}{2(p+2)!} e^{a\phi} (F_{(p+2)})^2 \right) \ ,
\end{equation}
where $\phi$ is the dilaton field and $F_{(p+2)}$ is a $(p+2)$-form
field strength with $F_{(p+2)} = d A_{(p+1)}$ and
$A_{(p+1)}$ the corresponding $(p+1)$-form gauge field.
For $D=10$ the action \eqref{IDact} is the bosonic part of the low energy
action of type IIA and IIB string theory (in the Einstein-frame)
when only one of the gauge-fields is present.
For $D=11$ it is the low energy action
of M-theory for suitable choices of $a$ and $p$.

The extremal $1/2$ BPS $p$-brane solutions in string and M-theory
take the well-known form
\begin{equation}
\label{extr1}
ds^2 = H^{-\frac{d-2}{D-2}} \left[ - dt^2 + \sum_{i=1}^p (du^i)^2
+ H ds_d^2 \right] \ ,
\end{equation}
\begin{equation}
\label{extr2}
e^{2\phi} = H^a \spa
A_{(p+1)}= ( H^{-1} - 1 ) dt \wedge du^1 \wedge \cdots \wedge
du^p \ ,
\end{equation}
with $\nabla^2 H = 0$ (away from the sources).
$ds_d^2$ denotes the metric and $\nabla^2$ the Laplacian of
the $d$-dimensional transverse space. The solutions \eqref{extr1},
\eqref{extr2} represent the $1/2$ BPS extremal $p$-branes of String/M-theory
when $D=10,11$ and
\begin{equation}
\label{asq}
a^2 = 4 - 2 \frac{(p+1)(d-2)}{D-2} \ .
\end{equation}
With \eqref{asq} obeyed we get for $D=10$ the D-branes ($a=(3-p)/2$),
NS5-brane ($a=1$) and the F-string ($a=-1$) of type IIA and IIB string theory, and
for $D=11$ the M2-brane and M5-brane of M-theory ($a=0$). Note that the
string theory solutions are written here in the Einstein frame.

In the following we place the extremal solutions
\eqref{extr1}-\eqref{extr2} on a transverse space of the form
$\R^{d-1} \times S^1$. The thermal excitations of these solutions
corresponds to the non-extremal solutions that we find below.

\subsubsection{Charging up solutions via U-duality}
\label{secUdual}

Consider a static and neutral Kaluza-Klein black hole solution
(see Section \ref{sec:KKphases}), $i.e.$ a static vacuum solution
of $(d+1)$-dimensional General Relativity with at least one event
horizon that asymptotes to $\CM^d \times S^1$. We can always write
the metric of such a solution in the form
\begin{equation}
\label{startmet}
ds^2_{d+1} = - U dt^2 + \frac{L^2}{(2\pi)^2} V_{ab} dx^a dx^b \ ,
\end{equation}
where $a,b=1,...,d$ and $V_{ab}$ is a symmetric tensor.
$U$ and $V_{ab}$ are functions of $x^1,..,x^d$. In the asymptotic
region $U \rightarrow 1$ and $L^2 (2\pi)^{-2} V_{ab} dx^a dx^b$
describes the cylinder $\R^{d-1} \times S^1$. The factor $L^2/(2\pi)^2$,
where $L$ is the circumference of the circle in $\CM^d \times S^1$,
has been included in \eqref{startmet} for later convenience.

By adding $p+1$ (where $p = 9-d$) flat directions, this solution is trivially uplifted to
a solution of eleven-dimensional (super)gravity. Denoting these directions
by $z_i$ ($i=1 \ldots p$) and $y$, we perform a Lorentz boost in the $y$ direction
\begin{equation}
\label{boost}
\vecto{t_{\rm new}}{y_{\rm new}}
= \matrto{\cosh \alpha}{\sinh \alpha}{\sinh \alpha}{\cosh \alpha}
\vecto{t_{\rm old}}{y_{\rm old}} \ .
\end{equation}
The resulting boosted metric has an isometry in the $y$-direction so
we can make an S-duality to obtain a non-extremal D0-brane solution
of type IIA string theory. For $d < 9$ the D0-brane is uniformly smeared
along an $\R^{9-d}$ space and has transverse space $\R^{d-1} \times S^1$.
One can use further U-dualities to transform the solution into a D$p$-brane
solution, an F-string solution or an NS5-brane solution of
type IIA/B String theory, or to an M2-brane or M5-brane solution
of M-theory, depending on what dimension $d$ we start with. In this way we obtain non-extremal singly-charged%
\footnote{By applying further boosts it is also possible to add
two or three charges. For the case of $d=4$ this is considered in
Ref.~\cite{Harmark:2006df}, which considers the corresponding
phases of two- and three-charge brane configurations on a circle.}
$p$-branes on a transverse circle in $D$ dimensions with $D=d+p+1$
(for String/M-theory we have $D=10$/$D=11$).%
\footnote{One could also use the U-duality to get
branes that are smeared uniformly in some directions, but we
choose not to consider this option here.}
The resulting non-extremal solution takes the form
\begin{equation}
\label{gen1}
ds^2 = H^{-\frac{d-2}{D-2}} \left( - U dt^2 + \sum_{i=1}^p (du^i)^2
+ \frac{L^2}{(2\pi)^2} H \, V_{ab} dx^a dx^b \right) \ ,
\end{equation}
\begin{equation}
\label{gen2}
H = 1 + \sinh^2 \alpha \, (1 - U) \ ,
\end{equation}
\begin{equation}
\label{gen3}
e^{2\phi} = H^a \spa
A_{(p+1)} = \coth \alpha \, (H^{-1} -1) dt \wedge du^1 \wedge \cdots \wedge
du^p \ .
\end{equation}

Notice that $U \rightarrow 0$ near a event horizon in the metric
\eqref{startmet} for the neutral solution. This translates after
the map into an event horizon in the non-extremal solution
\eqref{gen1}-\eqref{gen3}. Since the harmonic function
\eqref{gen2} stays finite and non-zero for $U \rightarrow 0$, we
deduce that the source of the electric potential $A_{(p+1)}$ in
\eqref{gen3} is hidden behind the event horizon. Note also that if
there is a number of event horizons defined by $U=0$, the chemical
potential $\nu = - A_{t u^1 \cdots u^p} |_{U=0} = \tanh \alpha$ is
the same for each disconnected component of the event horizon. In
the following we assume that we have at least one event horizon
present.

Using the map from \eqref{startmet} to \eqref{gen1}-\eqref{gen3} a
Kaluza-Klein black hole solution in $d+1$ dimensions can be mapped
to a corresponding solution of supergravity describing non- or
near-extremal $p$-branes on a circle. The dimensions $d$ and $p$
are related by the equation $D=d+p+1$ with $D=10$ for string
theory and $D=11$ for M-theory.

The map from \eqref{startmet} to \eqref{gen1}-\eqref{gen3} is a
development of an earlier result in \cite{Harmark:2002tr}. In
\cite{Harmark:2002tr} it was pointed out that one can map any
neutral solution into a corresponding non- and near-extremal
solution in supergravity for the class of Kaluza-Klein black holes
that fall into the $SO(d-1)$-symmetric ansatz \eqref{ansatz} of
Ref.~\cite{Harmark:2002tr}. While this observation was made at the
level of the equations of motion, the combined boost and U-duality
transformation reviewed above reveals the physical reason behind
the existence of this map and moreover shows that it works for any
static and neutral Kaluza-Klein black hole.

Here we present a summary of the map between physical quantities
of the non-extremal branes and physical quantities of the neutral
Kaluza-Klein black holes (we refer the reader to
\cite{Harmark:2004ws} for more details on the boost/U-duality
map). In terms of the definitions of Ref.~\cite{Harmark:2004ws}
\begin{equation}
\label{mapp}
\bar \mu = q  +  \frac{(d+n)}{2(d-1)} \mu
+ \frac{ b^2 q}{1+ \sqrt{1 + b^2}}
\spa
\bar n = n  \spa b \equiv \frac{d-2-n}{2(d-1)} \frac{\mu}{q}
\end{equation}
where $(\mu,n)$ (see \eqref{themu}, \eqref{then}) are the
dimensionless mass and relative tension of the neutral
Kaluza-Klein black hole and $(\bar \mu,\bar n,q)$ the dimensionless mass, relative
tension and dimensionless charge of the non-extremal brane.
For the temperature and entropy one finds
\begin{equation}
\label{mapp2}
\bar{\mt} = \frac{\mt}{\cosh \alpha} \spa \bar{\ms} = \cosh \alpha \ms \spa
\spa
\cosh \al (\mu,n,q) \equiv \frac{1 + b + \sqrt{1 + b^2}}{2 \sqrt{b(1 + \sqrt{1+b^2})}}
\end{equation}
where $(\mt,\ms)$ and ($\bar{\mt},\bar{\ms}$) are the temperature and entropy of the
neutral
Kaluza-Klein black hole and non-extremal brane respectively.

\subsubsection{Phases of non-extremal branes}

We can now apply the above boost/U-duality map to the known phases
of Kaluza-Klein black holes reviewed in Section
\ref{sec:KKphases}. We restrict ourselves here to the lower part
of the phase diagram $n \leq 1/(d-2)$, $i.e.$ the uniform black
string, the non-uniform black string and localized black holes.
Recall at this point that all these phases have $SO(d-1)$ symmetry
and have a metric that can be taken to be of the form of the
ansatz \eqref{ansatz}. Hence the corresponding non-extremal
solutions are of the form \eqref{gen1}-\eqref{gen3} with the
ansatz \eqref{ansatz} substituted. The phases obtained from the
bubble-black hole sequences \cite{Elvang:2004iz} will be
considered elsewhere \cite{Harmark:2007}.

Applying the map to the uniform string branch we recover the
known phase of non- and near-extremal branes smeared on a circle,
which we call the {\it uniform phase}. For the sake of clarity we write
the solution explicitly here
\begin{equation}
\label{uni1}
ds^2 = H^{-\frac{d-2}{D-2}} \left( - f dt^2 + \sum_{i=1}^p (du^i)^2
+ H \frac{L^2}{(2\pi)^2} \left[ \frac{1}{f} dR^2
+ dv^2 + R^2 d\Omega_{d-2}^2 \right] \right) \ ,
\end{equation}
\begin{equation}
e^{2\phi} = H^a \spa
A_{(p+1)} = \coth \alpha \, (H^{-1} -1) dt \wedge du^1 \wedge \cdots \wedge
du^p \ ,
\end{equation}
\begin{equation}
\label{uni3}
f = 1 - \frac{R_0^{d-3}}{R^{d-3}} \spa
H = 1 + \sinh^2 \alpha \frac{R_0^{d-3}}{R^{d-3}} \ .
\end{equation}
In Section \ref{sec:nonGL} we demonstrate that this
solution suffers from a GL instability which we relate to the GL instability
of the uniform black string. The thermodynamic quantities of the solution
will be given in Section \ref{sec:examples} when
we discuss thermodynamic stability and the correlated stability conjecture.

A strong indication of a possible instability can be deduced already
by applying the map to the non-uniform black string branch from which we generate a
new phase of non-extremal $p$-branes on a circle with a connected but
not translationally invariant horizon around the circle, $i.e.$ a horizon which is
non-uniformly distributed along the circle. We denote
this phase as the {\it non-uniform phase}.%
\footnote{In \cite{Horowitz:2002ym}
a new non-uniform phase of certain near-extremal
branes on a circle was conjectured to exist for small energies.
There does not seem to be any direct connection between the branch
that we find here and the one in \cite{Horowitz:2002ym}.
We comment further on  \cite{Horowitz:2002ym}
in the conclusions in Section \ref{sec:open}.}

Using the map \eqref{mapp} from the neutral case
to the non-extremal case we see that this new branch
emerges out of the uniform phase (with $\bar{n} = 1/(d-2)$)
at the critical mass
\begin{equation}
\label{e:muc}
\bar{\mu}_{\rm c} = q + \frac{(d-1) }{2(d-2)} \mu_{\rm GL} + \frac{b_c^2}{1+\sqrt{1+b_c^2}} q
\spa b_c \equiv \frac{(d-3)\mu_{\rm GL}}{2(d-2)q} \ .
\end{equation}
We can furthermore use the map \eqref{mapp} on Eq.~\eqref{nofmu}.
This gives
\begin{equation}
\label{e:nunonex}
\bar{n} (\bar \mu;q) = \frac{1}{d-2} - \bar{\gamma}(q) (\bar{\mu} - \bar{\mu}_c)
\spa 0 \leq \bar{\mu} - \bar{\mu}_c  \ll 1 \ ,
\end{equation}
\begin{equation}
\bar{\gamma}(q) = \gamma \left[
\frac{d-1}{2(d-2)} - \frac{\gamma \mu_{\rm GL}}{2(d-1)}
+\frac{b_c}{\sqrt{1+b_c^2}}
\frac{(d-1)(d-3)+(d-2)\gamma \mu_{\rm GL}}{2(d-1)(d-2)} \right]^{-1} \ .
\end{equation}
with $\gamma$ given in Table \ref{tabnonuni}.
Eq.~\eqref{e:nunonex} captures the behavior of the non-uniform
phase of non-extremal branes on a circle for masses slightly higher than
$\bar{\mu} = \bar{\mu}_c$. Note that $\bar{\gamma}(q) \rightarrow \gamma$
for $q\rightarrow 0$, as expected. We see from \eqref{e:nunonex}
that the non-uniform phase starts at
$(\bar{\mu},\bar{n})=(\bar{\mu}_c,1/(d-2))$ and continues
with increasing $\bar{\mu}$ and decreasing $\bar{n}$
with a slope given by $\bar{\gamma}(q)$. It is not difficult to show
\cite{Harmark:2004ws} from the above results
that the uniform phase has higher entropy than the non-uniform
phase for the same mass $\bar{\mu}$, just as in the neutral case.

For those dimensions $d$ where further numerical data are
available on the neutral non-uniform branch, one can use the map to derive the
corresponding data for the non-uniform phase of non-extremal branes. In this way one can
obtain, for a given charge $q$, diagrams for any thermodynamic quantity
one may wish to consider.

In Section \ref{sec:nonGL} we show explicitly that $\bar{\mu}_c$ is
a critical GL mass for the uniform phase of non-extremal branes on a
circle. This is natural in view of the fact that the non-uniform
phase implies the presence of a marginal mode at that mass.

Finally, the map can be applied to localized black holes on a circle
to generate a phase of non-extremal $p$-branes that are localized on a circle.
This phase will be called the {\it localized phase}. Ref.~\cite{Harmark:2004ws} gives the
background to first order in $(\bar \mu -q)$, obtained from \eqref{gen1}-\eqref{gen3} and
the neutral solution of Ref.~\cite{Harmark:2003yz}. For the relative tension we have
in this case
\begin{equation}
\label{mapploc2}
\bar n (\bar \mu;q) =\frac{2(d-1)}{d} \lambda_d (\bar \mu -q)
+  \Ord \Big( (\bar \mu -q)^2\Big) \spa
\lambda_d \equiv \frac{(d-2)\zeta(d-2)}{2(d-1)\Omega_{d-1}}
\end{equation}
In the $(\bar \mu,\bar n)$ phase diagram, the localized phase
starts at the extremal point $(q,0)$ and goes upwards with a slope that can be read off
from \eqref{mapploc2}. The corresponding entropy and temperature
of the branch can also be computed from the neutral
thermodynamics. Again, for those dimensions $d$ where numerical data
are available for the neutral phase, one can use the map to
obtain corresponding diagrams for any thermodynamic quantity of the
non-extremal localized phase.

\subsubsection{Phases of near-extremal branes}

Given the boost/U-duality map from neutral and static Kaluza-Klein black holes
to non-extremal branes on a circle, we can go one step further and take the
near-extremal limit. In this way, we obtain a map that takes a
Kaluza-Klein black hole solution to a near-extremal brane solution on a circle.

The definition of the near-extremal limit of a non-extremal $p$-brane on a circle
is
\begin{equation}
\label{nelimit}
q \rightarrow \infty \spa L \rightarrow 0 \spa g
\equiv \frac{16\pi G_D}{V_p L^{d-2}}\ \, \mbox{fixed } \spa l
\equiv L \sqrt{q} \ \, \mbox{fixed } \spa
x^a\ \, \mbox{fixed } \
\end{equation}
where $V_p$ is the (spatial) volume of the $p$-brane, $L$ the circumference
of the circle, $q$ the rescaled charge\footnote{In terms of the charge $Q$ and
the ratio $g$, the rescaled charge is defined as $q=g Q$.}
and $x^a$ the transverse directions parameterizing $\R^{d-1} \times S^1$
in the asymptotic region. This definition
ensures that the energy above extremality $\bar{\mu} - q$
is finite and sensitive to the non-trivial physics associated to the compact
direction.

The near-extremal branes defined in this way can be characterized by two
dimensionless quantities $(\epsilon,r)$ which are respectively rescaled versions
of the energy above extremality and the tension in the circle direction
(see \cite{Harmark:2004ws} for the precise definitions).
Applying the limit \eqref{nelimit} to the general solution \eqref{uni1}-\eqref{uni3}
we obtain the following result for near-extremal branes on a circle
\begin{equation}
\label{gensolnh}
ds^2 = \hat H^{-\frac{d-2}{D-2}} \left( - U
dt^2 + \sum_{i=1}^p (du^i)^2 + \hat H V_{ab} d x^a d x^b
\right) \ ,
\end{equation}
\begin{equation}
\label{gensolnh2}
e^{2\phi} = \hat H^a \spa
A_{(p+1)} = \hat H^{-1}  dt \wedge du^1 \wedge \cdots \wedge du^p \ ,
\end{equation}
where
\begin{equation}
\label{Hhat}
\hat H = \hat{h}_d \frac{1-U}{\hat{c}_t} \spa
\hat h_d \equiv \frac{ l^2 (2\pi)^{d-5} }{(d-3) \Omega_{d-2}} \ .
\end{equation}
The fields in \eqref{gensolnh}-\eqref{gensolnh2} have been written
in units of $L/(2\pi)$, $i.e.$ we have rescaled the fields with
the appropriate powers of $L/(2\pi)$ to get a finite solution. As
a trivial check, one can show that the near-extremal $p$-brane
solutions generated in this way correctly asymptote to the
near-horizon limit of the extremal $p$-brane on a transverse
circle \eqref{extr1}, \eqref{extr2}, which is taken as the
reference space when calculating energy and tensions.

To summarize, we have seen that there is a direct map from any
Kaluza-Klein black hole in $d+1$ dimensions
to a near-extremal $p$-brane on a circle ($D=d+p+1$).
The corresponding map between thermodynamic quantities
gives the near-extremal quantities \cite{Harmark:2004ws}
\begin{equation}
\label{nemap}
\epsilon = \frac{d+n}{2(d-1)} \mu \spa
r = 2 \frac{(d-1)n}{d+n}
\spa
\hat{\mt} =  \mt \sqrt{\mt \ms}
\spa
\hat{\ms} = \frac{\ms}{\sqrt{ \mt \ms}}
 \ ,
\end{equation}
where $(\mu,n,\mt,\ms)$ and $(\epsilon,r,\hmt,\hms)$ are the
(rescaled) mass/energy, relative tension, temperature, entropy of the
neutral Kaluza-Klein black hole and near-extremal brane respectively.
The latter quantities are defined as
\begin{equation}
\label{epsrdef}
\epsilon = g E \spa
r = \frac{2 \pi \hat{\CT}}{E} \spa\hat{\mt} = l \, \hat{T} \spa
\hat{\ms} = \frac{g}{l} \hat{S}
\end{equation}
where $E$, ${\hat{\CT}}$, $\hat{T}$, $\hat{S}$ are the energy,
transverse tension, temperature, entropy of the near-extremal
brane, and $l$ and $g$ appear in \eqref{nelimit}. With the use of
\eqref{nemap} we can obtain the near-extremal $(\epsilon,r)$ phase
diagram from the $(\mu,n)$ phase diagram of neutral Kaluza-Klein
black holes.

In the above language the near-extremal Smarr formula and first law of thermodynamics
take the form
\begin{equation}
\hat{\mt} \hat{\ms} = 2 \frac{d-2-r}{d} \epsilon \spa
\delta \epsilon = \hmt \, \delta \hms
\end{equation}
As a consequence, knowing a curve $r(\epsilon)$ in the near-extremal phase
diagram determines the entire thermodynamics.

We can apply the above map to all the known phases of Kaluza-Klein
black holes as we did with the non-extremal branes on a circle .
Here we restrict ourselves again to the uniform, non-uniform and
localized phase. Note that since these solutions fall within the
ansatz \eqref{startmet}, we know that the corresponding
near-extremal branes will be of the form
\eqref{gensolnh}-\eqref{Hhat} with the ansatz substituted. Hence
we get the following three phases of near-extremal branes on a
circle:
\begin{itemize}
\item {\sl The uniform phase.}
In this phase a near-extremal brane is uniformly
smeared on a transverse circle. The solution can be obtained with the map
from the uniform black string and takes the well-known form
\begin{equation}
\label{uninear1}
ds ^2
= \hat{H}^{-\frac{d-2}{D-2}} \left( - f dt^2
+ \sum_{i=1}^p (du^i)^2 + \hat{H} \left[ \frac{1}{f} dR^2
+ dv^2 + R^2 d\Omega_{d-2}^2 \right] \right) \ ,
\end{equation}
\begin{equation}
e^{2\phi} = \hat{H}^a
\spa
A_{(p+1)} = \hat{H}^{-1} dt \wedge du^1 \wedge \cdots \wedge du^p \ ,
\end{equation}
\begin{equation}
\label{uninear3}
f = 1 - \frac{R_0^{d-3}}{R^{d-3}}  \spa
\hat{H} = \frac{\hat h_d}{R^{d-3}} \spa \hat h_d \equiv \frac{
l^2 (2\pi)^{d-5} }{(d-3) \Omega_{d-2}} \ .
\end{equation}
Setting $n=1/(d-2)$ in \eqref{nemap} we find the relative tension
$r = 2/(d-1)$.
The thermodynamics of the uniform phase is%
\footnote{$\hmf$ is the rescaled free energy defined
as $\hmf = \epsilon - \hmt \hms$.}
\begin{equation}
\label{unise}
\hms_u (\epsilon) = \frac{4\pi}{\sqrt{d-3}} (\Omega_{d-2})^{-\frac{1}{d-3}}
\left( \frac{2 \, \epsilon }{d-1} \right)^{\frac{d-1}{2(d-3)}} \ ,
\end{equation}
\begin{equation}
\label{unifr}
\hmf_u (\hmt) = - \frac{d-5}{2} (\Omega_{d-2})^{-\frac{2}{d-5}}
\left( \frac{4 \pi \, \hmt }{(d-3)^{3/2}} \right)^{\frac{2(d-3)}{d-5}} \ .
\end{equation}
In Section \ref{sec:nonGL} we demonstrate that the uniformly
smeared near-extremal $p$-brane solution has a GL-instability.
\item {\sl The non-uniform phase.}
This is a phase with a configuration of near-extremal branes that are
non-uniformly distributed around a circle. This phase is obtained by
applying the above near-extremal map to the non-uniform black string branch
reviewed in Section \ref{sec:KKphases}. It emerges from the uniform phase given
above at the critical energy
\begin{equation}
\label{crite}
\epsilon_c = \frac{d-1}{2(d-2)} \mu_{\rm GL}
\end{equation}
which follows from the map \eqref{nemap}. The thermodynamics of this
phase near the critical point is
\begin{equation}
\label{ssu}
\frac{\hms ( \epsilon ) }{\hms_u ( \epsilon ) }
= 1 - \frac{(d-1)^2 }{4d(d-3)^2} \frac{ \hat{\gamma}}{ \epsilon_c}
(\epsilon - \epsilon_c)^2
+ \CO ( (\epsilon - \epsilon_c)^3 ) \ .
\end{equation}\begin{equation}
\label{freeen}
\hmf ( \hmt ) = - \frac{d-5}{d-1} \epsilon_c - \hms_c ( \hmt - \hmt_c )
- \frac{c}{2\hmt_c} ( \hmt - \hmt_c )^2
+ \CO (( \hmt - \hmt_c )^3) \ ,
\end{equation}
with $\epsilon_c$ the critical energy \eqref{crite}, $\hmt_c$ and $\hms_c$
the corresponding critical temperature and entropy, and $c$ the (rescaled) specific heat
\begin{equation}
\label{heatcap}
c = \frac{d (d-1) \hms_c}{
d(d-5) + 2 (d-1) \hat{\gamma} \epsilon_c } \ ,
\end{equation}
The numerical values $\epsilon_c$, $\hat \gamma$, $\hmt_c$,
$\hms_c$ and $c$, of the  energy, slope, temperature, entropy and
specific heat at the critical point are listed in Table
\ref{tabnear} \cite{Harmark:2004ws}.
\begin{table}[ht]
\begin{center}
\begin{tabular}{|c||c|c|c|c|c|c|}
\hline
$d$ & $4$ & $5$ & $6$ & $7$ & $8$ & $9$
\\ \hline \hline
$\epsilon_c$ & $2.64$ & $1.54$ & $1.09$ & $0.71$ & $0.46$ & $0.31$ \\
\hline
$\hat{\gamma}$  & $0.25$ & $0.39$ & $0.55$ & $0.88$ & $1.42$ & $2.33$ \\
\hline
\hline
$\hmt_c$  & $0.75$ & $1$ & $1.07$ & $1.05$ & $0.97$ & $0.89$ \\
\hline
$\hms_c$  & $2.33$ & $1.54$ & $1.22$ & $0.91$ & $0.68$ & $0.53$ \\
\hline
$c$  & $-433$ & $6.46$ & $3.06$ & $1.77$ & $1.14$ & $0.79$ \\
\hline
\end{tabular}
\caption{In this table we list the critical energy $\epsilon_c$
 and the constant $\hat{\gamma}$ determining
the non-uniform phase of near-extremal branes on a circle for $0
\leq \epsilon - \epsilon_c \ll 1$. In addition we list the
critical temperature $\hmt_c$ and the critical entropy $\hms_c$,
as well as the heat capacity $c$ of the non-uniform phase at the
critical point. These numbers are obtained in
Ref.~\cite{Harmark:2004ws}. \label{tabnear}}
\end{center}
\end{table}

\item {\sl The localized phase.} This is a phase of a near-extremal
brane localized on a transverse circle. It is obtained by applying
the map to the black hole on cylinder branch.
Since the black hole on cylinder branch starts at $(\mu,n)=(0,0)$,
we deduce from the map \eqref{nemap} that the
localized phase starts at $(\epsilon,r)=(0,0)$.
The vanishing of the tension along the transverse circle
is expected since the brane becomes completely localized on the circle
in the limit $\epsilon \rightarrow 0$.
Furthermore we can use the analytical results of
Ref.~\cite{Harmark:2003yz} to explicitly compute the first correction
to the solution and thermodynamics of the non- and near-extremal
branes localized on a circle (see \cite{Harmark:2004ws} for more details).
The corresponding small black hole thermodynamics is
\begin{equation}
\label{entne}
\hms (\epsilon) = \hat{C}_1^{(d)} \epsilon^{\frac{d}{2(d-2)}}
\left( 1 + \frac{(d-1) \zeta (d-2)}{d(d-2) \Omega_{d-1}} \epsilon
 +\Ord (\epsilon^2) \right) \ ,
 \end{equation}
 \begin{equation}
 \hat{C}_1^{(d)} \equiv \frac{ 4 \pi ( \Omega_{d-1})^{-\frac{1}{d-2}}}{\sqrt{d-2}}
 \left( \frac{2}{d} \right)^{\frac{d}{2(d-2)}}
\end{equation}
and
\begin{equation}
\label{freene}
\hmf  (\hmt) = - \hat{K}_1^{(d)} \hmt^{\frac{2(d-2)}{d-4}}
\left( 1 +  \frac{2(d-1) \zeta (d-2) \hat{K}_1^{(d)}}{(d-4)^2 \Omega_{d-1}}
 \hmt^{\frac{2(d-2)}{d-4}}
+\Ord \Big( (\hmt^{\frac{2(d-2)}{d-4}} )^2 \Big)  \right) \ ,
\end{equation}
\begin{equation}
\label{K1def}
\hat{K}_1^{(d)} \equiv \frac{d-4}{2} (\Omega_{d-1})^{-\frac{2}{d-4}} \left(
\frac{ 4\pi}{(d-2)^{3/2}} \right)^{\frac{2(d-2)}{d-4}} \ .
\end{equation}
One may also compute the second order corrections to these
expressions using the second order correction in \eqref{bhslope}.
We leave this as an exercise for the reader.
\end{itemize}

From the analysis of the non-uniform phase we observe that, in
parallel with the non-extremal case, we also have a map of the
Gregory-Laflamme mass $\mu_{\rm GL}$ of the uniform black string
branch to a critical energy $\epsilon_{c}$. This is the energy
where the non-uniform phase connects to the uniform phase of
near-extremal branes on a circle. The existence of the non-uniform
phase suggests that near-extremal branes smeared on a circle have
a critical energy below which they are classically unstable. This
is indeed the case, as we show in Section \ref{sec:neGL}.

\subsection{GL-instability of smeared non- and near-extremal branes }
\label{sec:nonGL}

In the previous subsection we reviewed the boost/U-duality transformation
that takes a neutral black string to a D$p$-brane smeared on a transverse
direction. Following Ref.~\cite{Harmark:2005jk} we describe in this subsection
how to use this transformation to map the unstable
GL mode of the neutral black string to the unstable mode of the non-extremal
and near-extremal smeared D$p$-branes. As we take the extremal limit the unstable
GL mode connects smoothly to the marginal modes of the extremal smeared brane.

\subsubsection{Non-extremal smeared D$p$-branes}

The boost/U-duality transformation of the previous subsection converts the
uniform neutral black string solution \eqref{ublstr} to the smeared D$p$-brane solution
\begin{eqnarray}
\label{rzuniform} && ds^2=H^{-1/2} \left[ -f dt^2+\sum_{i=1}^p
dx_i^2\right]+ H^{1/2}\left[f^{-1}dr^2+dz^2+r^2
d\Omega^2_{7-p}\right]~,
\\
e^{2\phi}&=&H^{(3-p)/2}~, ~ A_{01\cdots p}=\coth \alpha(H^{-1}-1)~, ~ ~
 f=1-\frac{r_0^{6-p}}{r^{6-p}}~, ~ ~ H=1+\sinh^2\alpha(1-f)
~.\nonumber
\end{eqnarray}
This is the same solution as \eqref{uni1}-\eqref{uni3} now written
in the string frame without the $\frac{L}{2\pi}$ rescaling $i.e.$
with dimensionful radial coordinate $r$ and $S^1$ coordinate $z
\sim z+L$. Applying the transformation to the neutral GL mode
\eqref{GLmode1}, \eqref{GLmode2} will give (before the type II
reduction) an eleven dimensional background with a
non-normalizable exponential dependence on the M-theory circle
direction $y$ (the critical case $k=k_c$ is an exception). It was
pointed out in \cite{Aharony:2004ig} that this problem can be
avoided by considering a complex metric perturbation and finding a
complex transformation that $i)$ gets rid of the exponential $y$
dependence and $ii)$ still has the same effect on the zeroth order
part of the metric, $i.e.$ on the neutral black string metric. In
the end one is instructed to take the real part of the transformed
perturbation. This last step is allowed because the perturbed
Einstein equations of motion are linear.

The precise form of the transformation is as follows.
First define the coordinates $y'$ and $\tilde{z}$ by making a
complex rotation of the $y$ and $z$ coordinates
\begin{equation}
\label{complrot} \vecto{y'}{\tilde{z}} = M \vecto{y}{z} \spa M =
\matrto{\cosh w}{-i \sinh w }{i \sinh w}{\cosh w} \ .
\end{equation}
Recall that the $y$ and $z$ directions correspond to the M-theory circle and transverse
circle respectively.
Notice that the neutral black string metric \eqref{ublstr} embedded in eleven
dimensional gravity is invariant under \eqref{complrot} since
$(dy')^2 + (d\tilde{z})^2 = dy^2 + dz^2$. At the next step
we supplement the transformation
\eqref{complrot} with a boost that takes the $(t,y')$
coordinates to the $(\tilde{t},\tilde{y})$ coordinates given by
\begin{equation}
\vecto{\tilde{t}}{\tilde{y}} = \Lambda \vecto{t}{y'} \spa \Lambda
= \matrto{\cosh \alpha }{\sinh \alpha}{\sinh \alpha}{\cosh \alpha}
~.
\end{equation}
We choose the rotation parameters so that
\begin{equation}
\label{ktilde} \sinh w = \frac{\Omega}{k} \tanh \alpha \spa
\tilde{k} = k\cosh w \spa \tilde{\Omega} = \frac{\Omega}{\cosh
\alpha}
~.
\end{equation}
Then we can write the real part of the boosted perturbation of the
eleven dimensional metric (obtained as described around \eqref{boost}) as
\begin{equation}
\label{boostmodeI} \tilde{h}_{\mu \nu} = \mbox{Re} \left\{ \exp
\left( \frac{\tilde{\Omega} t}{r_0} + i \frac{\tilde{k}z}{r_0}
\right) \tilde{P}_{\mu \nu} \right\}
~.
\end{equation}
Using the relation $\tilde{P} = \Lambda^{-1} M^{-1} P (M^{-1})^T
(\Lambda^{-1})^T$ with $P$ given in \eqref{GLmode2} we find
\begin{equation}
\label{boostmodeII}
\begin{array}{c} \ds \tilde{P}_{tt} = - f \psi
\cosh^2 \alpha \spa \tilde{P}_{yy} = - f \psi \sinh^2 \alpha \spa
\tilde{P}_{ty} = f \psi \sinh \alpha \cosh \alpha \ ,
\\[3mm] \ds
\tilde{P}_{tr} = \eta \cosh \alpha \spa \tilde{P}_{yr} = - \eta
\sinh \alpha \spa \tilde{P}_{rr} = f^{-1} \chi \spa \tilde{P}_{\rm
sphere} = r^2 \kappa ~,
\end{array}
\end{equation}
where we made the relabelings $\tilde{t} \rightarrow t$ and
$\tilde{y} \rightarrow y$. Notice that the complex rotation
\eqref{complrot} precisely ensures that the exponential factor
does not have a $y$-dependence. We can therefore apply the same
U-duality transformations on
\eqref{boostmodeI}-\eqref{boostmodeII} as on the boosted neutral
black string. After these U-duality transformations, consisting of
one S-duality and $p$ T-dualities, we conclude that the perturbed
non-extremal D$p$-brane smeared on a transverse direction
can be written as \cite{Harmark:2005jk}%
\footnote{As usual, in the special case of D3-branes the gauge field
strength is self dual, so that we have
$F_5=(dA_4+\star dA_4)/\sqrt{2}$.}
\begin{equation}
\label{nonmode}
\begin{array}{c} \ds
\begin{array}{rcl}
ds^2 &=& \ds H_{\rm c}^{-1/2} \left[ - f_{\rm c} dt^2 +
\sum_{i=1}^p dx_i^2 + 2 \eta \cosh \alpha {\cal E} dt dr \right]
\\[5mm] && \ds + H_{\rm
c}^{1/2} \Big[ f^{-1} (1+\chi {\cal E}) dr^2 + dz^2 +    r^2
(1+\kappa {\cal E}) d\Omega_{7-p}^2 \Big] \ ,
\end{array}
\\[12mm] \ds e^{2\phi} = H_{\rm c}^{(3-p)/2} \spa A_{01\cdots p} = \coth
\alpha (H_{\rm c}^{-1} - 1) \spa A_{r1\cdots p} = - H^{-1} \eta
\sinh \alpha {\cal E} \ ,
\\[3mm] \ds f = 1 - \frac{r_0^{6-p}}{r^{6-p}} \spa
H=1 + \sinh^2 \alpha ( 1 - f ) \spa f_{\rm c} \equiv f (1 + \psi
{\cal E} ) \spa H_{\rm c} \equiv 1 + \sinh^2 \alpha ( 1 - f_{\rm
c} ) \ ,
\\[3mm] \ds
{\cal E} = \cos \left(\tilde{k} r_0^{-1} z \right) \exp \left(
\tilde{\Omega} r_0^{-1} t \right) \ .
\end{array}
\end{equation}
The perturbed D$p$-brane background appears here
in one compact expression. From this one can easily deduce the
explicit form of the perturbation by expanding to first
order. The unstable mode appearing in \cite{Aharony:2004ig}
is equivalent to \eqref{nonmode} for $p=0$, though written in a different gauge.

Note that the functions $\psi$, $\eta$, $\chi$ and $\kappa$ are
still solutions to Eqs.~\eqref{E1}-\eqref{E4} with $f$ given by
\eqref{fkx}. In particular $\psi$ is a solution of
Eq.~\eqref{psieq}. All these functions depend on the
variable $x = rk/r_0$ just as in the case of the neutral black string
perturbation.

Furthermore, we see from \eqref{ktilde} that $\tilde{k}^2 = k^2 +
\Omega^2 \tanh^2 \alpha$ and $\tilde{\Omega} = \Omega /\cosh
\alpha$. Hence, we can obtain $\tilde{\Omega}$ as a function of
$\tilde{k}$ by using the functional dependence $\Omega(k)$ for the
neutral black string (as sketched for $p=0,...,5$ in Figure
\ref{figGL}, where $p=9-d$). We note that the critical point $(k,\Omega)=(k_c,0)$
is mapped to the critical point $(\tilde{k},\tilde{\Omega})=(k_c ,
0)$, which corresponds to the marginal mode of the D$p$-brane with
a transverse direction. This marginal mode appears at the origin
of the non-uniform phase of non-extremal D$p$-branes with a
transverse direction \cite{Harmark:2004ws}. From Figure
\ref{figGL} we also deduce that $\tilde k<k_c$ for any $k<k_c$ and
the function $\tilde \Omega(\tilde k)$ never exhibits
time-dependent modes with wavelength smaller than the critical
one.

\subsubsection{Near-extremal smeared D$p$-branes}
\label{sec:neGL}

It is straightforward to take the near-extremal limit \eqref{nelimit} of the
perturbed non-extremal solution \eqref{nonmode}. In order to make
the connection with gauge theory quantities more transparent later on
we find it convenient to define here the 't Hooft and Yang-Mills couplings
as
\begin{equation}
\label{ymcouplings}
\lambda=g^2_{\rm YM} N~, ~ ~ g^2_{\rm YM}=(2\pi)^{p-1} \hat g_s l_s^{p-2}
~,
\end{equation}
where $\hat g_s$ is the T-dual string coupling.
It can be expressed in terms of the circumference
$\hat L$ of the T-dual circle (on which the D$(p+1)$-brane is wrapped) with
the use of the T-duality relations
\begin{equation}
\label{tdual}
L\hat L=(2\pi l_s)^2~,  ~ ~g_s=\hat g_s \frac{2\pi l_s}{\hat L}
~.
\end{equation}
In these parameters it is convenient to re-express the near-extremal limit \eqref{nelimit} as
\begin{equation}
\label{nelim}
l_s \rightarrow 0 \spa u = \frac{r}{l_s^2} \spa
\hat{z} = \frac{z}{l_s^2} \spa g_{\rm YM} \spa \hat{L} ~ ~
\mbox{fixed} \ .
\end{equation}
Respectively, the uniform near-extremal smeared D$p$-brane solutions take the form
\begin{equation}
\label{NEbrane}
\begin{array}{c}
\ds l_s^{-2} ds^2 = \hat{H}^{-1/2} ( - f dt^2 + \sum_{i=1}^p
dx_i^2 ) + \hat{H}^{1/2} ( f^{-1} du^2 + d\hat{z}^2 + u^2
d\Omega_{7-p}^2 ) \ ,
\\[3mm] \ds
e^{2\phi} = \hat H^{\frac{3-p}{2}}  \spa A_{01\ldots p} = \hat
H^{-1} \spa \hat{H} = \frac{K}{u^{6-p}} \spa f = 1 -
\frac{u_0^{6-p}}{u^{6-p}} \ ,
\end{array}
\end{equation}
where $K$ is defined as
\begin{equation}
\label{Kdef}
r_0^{6-p} \cosh \alpha \sinh \alpha = K l_s^{2(4-p)}
\spa K \equiv \frac{\lambda (2\pi)^{7-2p}}{(6-p)\Omega_{7-p}}
\ .
\end{equation}
$K$ is related to the near-extremal parameters $l$ and $\hat h_d$
in \eqref{uninear3} via the relation
\begin{equation}
\label{lKrelation} K=\left( \frac{L}{l_s^2}\right)^{4-p}
\frac{l^2}{(6-p)\Omega_{7-p}} ~.
\end{equation}

Applying the limit \eqref{nelim} to the background \eqref{nonmode}
we obtain the following perturbation of the near-extremal D$p$-brane
\eqref{NEbrane} \cite{Harmark:2005jk}
\begin{equation}
\label{nearmode}
\begin{array}{c}
\begin{array}{rcl}
l_s^{-2} ds^2 &=& \ds \hat{H}_{\rm c}^{-1/2} \left[ - f_{\rm c}
dt^2 + \sum_{i=1}^p dx_i^2 + 2 \eta \sqrt{K} u_0^{p/2-3} {\cal E} dt du
\right]
\\[5mm]
&& \ds + \hat{H}_{\rm c}^{1/2} \Big[ f^{-1} (1+\chi {\cal E}) du^2
+ d\hat{z}^2 + u^2 (1+\kappa {\cal E}) d\Omega_{7-p}^2 \Big] \ ,
\end{array}
\\[12mm] \ds
e^{2\phi} = \hat{H}_{\rm c}^{(3-p)/2} \spa A_{01\cdots p} =
\hat{H}_{\rm c}^{-1} \spa A_{r1\cdots p} = - u^{6-p} K^{-1/2}
u_0^{p/2-3} \eta {\cal E} \ ,
\\[3mm] \ds f = 1 - \frac{u_0^{6-p}}{u^{6-p}} \spa
f_{\rm c} = f ( 1 + \psi {\cal E} ) \spa \hat{H}_{\rm c} =
\frac{K}{u^{6-p}} \left[ 1 - \left( \frac{u^{6-p}}{u_0^{6-p}} - 1
\right) \psi {\cal E} \right] \ ,
\\[4mm] \ds
{\cal E} = \cos \left( \sqrt{k^2+\Omega^2} \frac{ \hat{z}}{u_0}
\right) \exp \left( u_0^{2-\frac{1}{2}p} K^{-1/2} \Omega t \right)
\ .
\end{array}
\end{equation}
We note that the near-extremal limit \eqref{nelim} keeps $k$ and
$\Omega$ fixed, and moreover keeps $\psi$, $\eta$, $\chi$ and
$\kappa$ fixed as functions of the variable $x$ which is now given
as
\begin{equation}
x = \frac{u k}{u_0} \ .
\end{equation}
Therefore, the functions $\psi$, $\eta$, $\chi$ and $\kappa$ are
still solutions of the Eqs.~\eqref{E1}-\eqref{E4} with $f$ given by
\eqref{fkx}. In particular $\psi$ is still a solution of
Eq.~\eqref{psieq}.

The above map shows that the near-extremal limit of the
time-dependent mode of smeared branes is well-defined and hence
that near-extremal smeared D$p$-branes are also classically unstable.

\subsubsection{Relation to the marginal modes of extremal smeared branes}

In general, the solution of extremal D$p$-branes distributed
along a single flat direction $z$ is given as
\begin{equation}
\label{extrback}
\begin{array}{c} \ds
ds^2 = H^{-1/2} \left( - dt^2 + \sum_{i=1}^p dx_i^2 \right) +
H^{1/2} \left( dr^2 + dz^2 + r^2 d\Omega_{7-p}^2 \right) \ ,
\\[5mm] \ds
e^{2\phi} = H^{(3-p)/2} \spa A_{01\cdots p} = H^{-1} - 1 \ ,
\end{array}
\end{equation}
where the harmonic function $H(r,z)$ obeys the differential equation
\begin{equation}
\left( \partial_r^2 + \frac{7-p}{r} \partial_r + \partial_z^2
\right) H (r,z) =0 \
\end{equation}
away from the source distribution. With appropriate boundary conditions
the general solution is
\begin{equation}
H(r,z) = 1 + \int_{-\infty}^{\infty} dz'
\frac{\rho(z')}{(r^2+(z-z')^2)^{(7-p)/2}} \ ,
\end{equation}
where the charge distribution function $\rho(z)$ is arbitrary.
The extremal uniformly smeared D$p$-brane case corresponds to a constant
density $\rho$. However, it is clearly possible to add to the
constant charge distribution a single mode of any wave-number $q$,
in which case the harmonic function takes the form
\begin{equation}
\label{Hm}
H = 1 + \frac{Kl_s^{8-2p}}{r^{6-p}} + m(qr) \cos( q z)
\ .
\end{equation}
$K$ is the parameter we defined in \eqref{Kdef} and
the function $m(q r)$ solves the differential equation
\begin{equation}
\label{meq} m''(y) + \frac{7-p}{y} m'(y) - m(y) = 0 \ .
\end{equation}
The solutions of Eq.~\eqref{meq} are of the form $m(y) \propto
y^{-(6-p)/2)} {\cal{K}}_{(6-p)/2} (y)$, ${\cal{K}}_s(y)$ being the
modified Bessel function of the second kind. The modes given by
\eqref{Hm} are marginal, time-independent modes.
Hence, we recover the well-known statement that the extremal
smeared branes exhibit marginal modes of any wavelength.

It is interesting to consider what happens to such marginal modes
if one perturbs the extremal brane to a non-zero temperature. From
the above study of non-extremal D$p$-branes distributed on
transverse directions one does not expect the existence of static
solutions for arbitrary charge distribution $\rho(z)$. In
particular, it is worth exploring whether there is any relation
between the extremal brane modes \eqref{Hm} and the GL modes of
the near-extremal branes. For that purpose, we need to consider
briefly the near-horizon limit \eqref{nelim} of the extremal
D$p$-brane background given by \eqref{extrback} and \eqref{Hm}.
One easily finds
\begin{equation}
\label{extrnh}
\begin{array}{c} \ds
l_s^{-2} ds^2 = \hat{H}^{-1/2} \left( - dt^2 + \sum_{i=1}^p dx_i^2
\right) + \hat{H}^{1/2} \left( du^2 + d\hat{z}^2 + u^2
d\Omega_{7-p}^2 \right) \ ,
\\[2mm] \ds
e^{2\phi} = \hat{H}^{(3-p)/2} \spa A_{01\cdots p} = \hat{H}^{-1}
\spa \hat{H} = \frac{K}{u^{6-p}} + m(qu) \cos( q \hat{z}) \ .
\end{array}
\end{equation}
Note that the function $m(qu)$ has been rescaled appropriately,
and that $m(y)$ still obeys Eq.~\eqref{meq}.

Consider now the extremal limit of the perturbed near-extremal
solution given by \eqref{nearmode}. The extremal limit corresponds
to $u_0 \rightarrow 0$ with $u$ and $K$ kept fixed. Finiteness of
the argument of the cosine factor implies that we also need to
keep $\sqrt{k^2+\Omega^2} / u_0$ finite in the limit. In terms of
the dispersion diagram of Figure \ref{figGL} we move closer and
closer to the point $(k,\Omega)=(0,0)$ on the left part of the
curve. In the process, the time-dependent exponential factor in
the function ${\cal E}$ in \eqref{nearmode} disappears. A closer
look at the solution and the equations of motion of appendix
\ref{app:mode} reveals the scaling \cite{Harmark:2005jk}
\begin{equation}
\label{extrlim} u_0 \rightarrow 0 \ \mbox{ with }\  u \ , \ K  \ ,
\ \frac{k}{u_0} \ , \ \frac{\Omega}{u_0} \ , \
\frac{\psi(x)}{u_0^{6-p}} \ , \ \frac{\eta(x)}{u_0^{6-p}} \ , \
\frac{\chi(x)}{u_0^{6-p}} \ , \ \frac{\kappa(x)}{u_0^{6-p}} \
\mbox{ kept fixed} \ .
\end{equation}
Applying this limit to the perturbed near-extremal
solution \eqref{nearmode}, we get the following background
\begin{equation}
\label{extrmode}
\begin{array}{c}
\begin{array}{rcl}
l_s^{-2} ds^2 &=& \ds \hat{H}_{\rm c}^{-1/2} \left[ dt^2 +
\sum_{i=1}^p dx_i^2  \right] + \hat{H}_{\rm c}^{1/2} \Big[  du^2 +
d\hat{z}^2 + u^2 d\Omega_{7-p}^2 \Big] \ ,
\end{array}
\\[4mm] \ds
e^{2\phi} = \hat{H}_{\rm c}^{(3-p)/2} \spa A_{01\cdots p} =
\hat{H}_{\rm c}^{-1} \spa \hat{H}_{\rm c} = \frac{K}{u^{6-p}} -
\frac{\psi}{u_0^{6-p}} {\cal E} \ ,
\\[4mm] \ds
{\cal E} = \cos \left(  \frac{\sqrt{k^2+\Omega^2}}{u_0}
\hat{z}\right) \ .
\end{array}
\end{equation}
We see that out of the four functions $\psi$, $\eta$, $\chi$ and
$\kappa$ that make up the near-extremal GL mode only $\psi$ survives.
Moreover, the equation \eqref{psieq} for $\psi(x)$ becomes
\begin{equation}
\label{extrpsieq}
\psi''(x)+\frac{7-p}{x} \psi'(x)-
\bigg(1+\frac{\Omega^2}{k^2}\bigg)\psi(x)=0
~.
\end{equation}
We deduce that the background \eqref{extrmode} corresponds
precisely to the background \eqref{extrnh} describing an extremal
smeared D$p$-brane perturbed by a marginal mode with the
wave-number
\begin{equation}
q = \frac{\sqrt{k^2+\Omega^2}}{u_0} \ .
\end{equation}
In particular, one can check explicitly that the $\psi$ equation
\eqref{extrpsieq} turns into Eq.~\eqref{meq} by identifying $m
(qu) = \psi (k u / u_0)$. This means that the extremal limit of the GL
mode becomes stable as also observed for unsmeared black branes in
Ref.~\cite{Gregory:1994tw}.

One can get marginal modes of any wave-number in this way. For
small $k$ we have $\Omega = \gamma k$, where $\gamma$ is an
appropriate number. Hence, $q =\sqrt{1+\gamma^2} k /u_0$, and by
tuning $k/u_0$ we can get any value of $q$ we want. To summarize
the picture, we see that the GL modes of near-extremal smeared
branes become marginal modes of extremal smeared branes in the
extremal limit. To put it differently, when we deform an extremal
smeared brane to make it non-BPS, the marginal modes of the
supersymmetric configuration turn into near-extremal GL modes
\eqref{nearmode}. In general, one can imagine extremal branes with
a charge distribution of the form
\begin{equation}
\label{rhorho} \rho = \rho_0 + \sum_{n \neq 0} \rho_n \cos( q_n
\hat{z} ) \ ,
\end{equation}
where $\rho_n$ are small compared to $\rho_0$. In the presence of a
non-BPS perturbation the solution turns into a sum over GL modes
of the form
\begin{equation}
h_{\mu\nu} = \sum_{n \neq 0} h^{(n)}_{\mu\nu} (u) \cos( q_n
\hat{z} ) \exp \left( \frac{\gamma
u_0^{3-\frac{p}{2}}}{\sqrt{K}\sqrt{1+\gamma^2}} q_n t \right) \ ,
\end{equation}
with $u_0$ small. In this way we can find the non-BPS
continuation of any extremal perturbation of the extremal smeared
branes, since we can Fourier decompose the perturbation and put it
into the form \eqref{rhorho}.

\subsection{GL-instability of D-brane bound states}
\label{sec:GLbound}

So far we have considered only the  case of
singly-charged D-brane configurations that arise as thermal
excitations of half BPS branes in type IIA/B string theory and
M-theory. Another interesting class of examples in type II string
theory and M-theory are non-extremal D-brane bound states.
In this subsection we review known facts about the stability properties of
a class of D-brane bound states in type II theory and mention some interesting
possible generalizations.

General spinning D-brane bound state solutions of type II supergravity
appear in \cite{Harmark:1999xt}, \cite{Harmark:1999rb}, \cite{Harmark:2000wv}.
For completeness, we summarize here the general solution and then proceed to
discuss some special cases in more detail. The general solution represents a
spinning bound state of D$(p-2k)$-branes with $k=0,...,m$ and non-zero
$B$-field. The background depends on the non-extremality parameter
$r_0$, the charge parameter $\alpha$, the angular momenta $\ell_i$
$(i=1,2,...,n=\frac{9-p}{2})$ and the angles $\theta_k$, $k=1,2,...,m$.
The metric in the string frame is
\begin{eqnarray}
\label{boundaa}
ds^2&=&H^{-1/2}\left(-f dt^2+\sum_{k=1}^m D_k\left[ (dy^{2k-1})^2+(dy^{2k})^2\right]
+\sum_{i=2m+1}^p (dy^i)^2\right)
\nonumber\\
&&+H^{1/2}\left( \bar f^{-1} K_{9-p} dr^2+\Lambda_{\alpha \beta}d\eta^\alpha d\eta^\beta\right)
\nonumber\\
&&+H^{-1/2}\frac{1}{W_p}\frac{r_0^{7-p}}{r^{7-p}}
\left( \sum_{i,j=1}^n \ell_i \ell_j \mu_i^2 \mu_j^2 d\phi_i d\phi_j-2\cosh \alpha
\sum_{i=1}^n \ell_i \mu_i^2 dt d\phi_i\right)
\end{eqnarray}
with a dilaton of the form
\begin{equation}
\label{boundab}
e^{2\phi}=H^{(3-p)/2}\prod_{k=1}^m D_k
~.
\end{equation}
The NSNS $B$-field has rank $2m\leq p$ and is given by the relation
\begin{equation}
\label{boundac}
B_{2k-1,2k}=\tan \theta_k \left( H^{-1} D_k-1\right)~, ~ ~ k=1,...,m
~.
\end{equation}
The non-zero RR gauge potentials, whose explicit form can be found
in Appendix B of \cite{Harmark:1999rb}, are $A_{p-2k+1}$ with $k=0,...,m$.

The functions entering in this background are
\begin{equation}
\label{boundad}
H=1+\frac{1}{W_p}\frac{r_0^{7-p} \sinh^2 \alpha}{r^{7-p}}~, ~ ~
D_k=\left(\sin^2\theta_k H^{-1}+\cos^2\theta_k\right)^{-1}
~,
\end{equation}
\begin{equation}
\label{boundae}
f=1-\frac{1}{W_p} \frac{r_0^{7-p}}{r^{7-p}}~, ~ ~
\bar f=1-\frac{1}{L_{9-p}}\frac{r_0^{7-p}}{r^{7-p}}~, ~~
L_{9-p}=\prod_{i=1}^n \left(1+\frac{\ell_i^2}{r^2} \right)
\end{equation}
with
\begin{equation}
\label{boundaf}
W_p=K_{9-p}L_{9-p}
~.
\end{equation}
$K_{9-p}$ and $\Lambda_{\alpha \beta}$ are functions
appearing in the flat transverse space metric
\begin{equation}
\label{boundag}
\sum_{a=1}^{9-p}(dx^a)^2=K_{9-p}dr^2+\Lambda_{\alpha \beta}
d\eta^\alpha d\eta^\beta
~.
\end{equation}
Explicit expressions for $K_{9-p}$, $\Lambda_{\alpha \beta}$ and $\mu_i$
can be found in Appendix B of Ref.\ \cite{Harmark:1999xt}.
It should be pointed out that the following condition holds as a
consequence of charge quantization of the D$p$-brane
\begin{equation}
\label{boundai}
r_0^{7-p}\cosh \alpha \sinh \alpha=
\frac{(2\pi)^{7-p}Ng_s l_s^{7-p}}{(7-p)\Omega_{8-p}}\prod_{k=1}^m
(\cos\theta_k)^{-1}
~.
\end{equation}

\subsubsection{The D$(p-2)$-D$p$ system}

A direct analysis of the classical stability properties of the non-rotating
D0-D2 bound state has been performed by Gubser
in \cite{Gubser:2004dr}. The non-extremal D0-D2 solution
is a special case of the above general background, whose explicit
form is
\begin{eqnarray}
\label{boundaj}
ds^2&=&H^{-1/2}(-fdt^2+D((dx^1)^2+(dx^2)^2))
+H^{1/2}\left(\frac{1}{f} dr^2+r^2 d\Omega_6^2 \right)
~,\nonumber\\
A_1&=&\coth \alpha ~\sin \theta \left(1-\frac{1}{H}\right)dt
~,\nonumber\\
A_3&=&\coth \alpha~ \sec \theta \left(1-\frac{D}{H}\right)dt \wedge dx^1 \wedge dx^2
~,\\
e^{2\phi}&=&H^{1/2}D~, ~ ~ B_2=\tan \theta \left(1-\frac{D}{H}\right) dx^1 \wedge dx^2
~,\nonumber\\
H&=&1+\frac{r_0^5 \sinh^2\alpha}{r^5}~, ~ ~ D=\frac{1}{H^{-1}\sin^2\theta+\cos^2\theta}
~, ~ ~ f=1-\frac{r_0^5}{r^5}
~.\nonumber
\end{eqnarray}
The thermodynamic quantities of the solution will be given in Section \ref{sec:examples}
 when we discuss thermodynamic stability and the correlated stability conjecture.

The above static solution describes a configuration of uniformly smeared
D0 charge inside the worldvolume of a D2-brane. As in the
case of non-extremal D-branes smeared uniformly on a transverse
circle one might expect that this configuration exhibits a GL instability
towards a non-uniform redistribution of the D0 charge. Gubser analyzed
an important aspect of the perturbative stability of \eqref{boundaj} by looking
for an $s$-wave marginal threshold mode that depends only on $x^1$ and
$r$. Reducing the full ten-dimensional action on the $S^6$, $t$ and $x^2$
directions he obtained a complicated two dimensional system whose
dynamics he was able to analyze with a combination of analytical and numerical
techniques. The results exhibit the presence of a boundary of stability
in the phase diagram of $(Q_0/M,Q_2/M)$ charges, which separates
an unstable from a stable phase. Moreover, it was shown that the presence
of a GL instability is in very good agreement with expectations based on
a local thermodynamic stability and the correlated stability conjecture.
The precise form of the boundary of stability and its connection with thermodynamics
will be discussed in more detail in Section \ref{sec:CSC}.

Once a marginal threshold mode is obtained for the D0-D2 system
a corresponding threshold mode can be deduced for the more general
non-rotating D$(p-2)$-D$p$ bound state with T-duality. It would be interesting
to proceed further analyzing the full phase diagram of the D$(p-2)$-D$p$ bound
state in analogy with what has been achieved for the singly-charged
non-extremal D$p$-branes on a transverse circle.

\subsubsection{Other bound states and further generalizations}

One can generalize the above discussion to other non-extremal
bound states of D-branes. The classical stability of D$(p-4)$-D$p$ and
F1-D$p$ bound states has been considered with general arguments based
on boost/U-duality transformations in \cite{Ross:2005vh} and
from the point of view of thermodynamics and the correlated stability
conjecture in \cite{Ross:2005vh,Friess:2005tz} (see Section \ref{sec:CSC}
for further details on this aspect).

We have seen the presence of a GL instability in a variety of black systems.
In general, the task of determining the existence of an instability and the
full phase structure of the system requires a very complicated analysis
in classical gravity coupled with a set of matter fields. The
number of charges and parameters of the system has a non-trivial
effect on the properties of the GL instability, which one would like to explore
in as many different cases as possible. Some interesting generalizations
include the above-mentioned bound states with rotation,
black objects with multiple (more than two) charges, non-extremal
intersecting branes \cite{Aref'eva:1997nz} etc.

\section{Thermodynamics and the correlated stability conjecture}
\label{sec:CSC}

In this section we review the connection between classical stability
and thermodynamic stability of black branes in supergravity. This connection
is known as the correlated stability conjecture, which we first present
in some detail. As an illustration, we consider the conjecture in two examples, namely that
of smeared D$p$-branes and D-brane bound states, which were the subject of
the previous section. Subsequently we discuss the support of the conjecture based
on a general semiclassical argument. We end by commenting on known counter
examples to the conjecture and how the conjecture may be refined or restricted
to avoid violation.

\subsection{Connecting thermodynamic and classical stability}
\label{subsec:csc1}

We have seen that the Gregory-Laflamme instability is a classical
instability of black brane solutions in gravity where the mass
tends to clump together non-uniformly. It is natural to ask if
there is a simple intuitive understanding for the existence of
such instabilities. Using the entropy to say something about the
dynamical properties of classical gravity is natural from the
point of view of the second law of thermodynamics. Already in
\cite{Gregory:1993vy}, a global thermodynamic argument was
presented for the instability, namely that for small masses the
entropy of an array of black holes is higher than that of the
black brane with the same mass. This argument suggests that the
black branes might be unstable.  Nevertheless, a global
thermodynamic argument like the one above can sometimes be
misleading. Given the existence of a higher entropy configuration
the second law permits a classical instability, but does not
require it. For a black string, in particular, this means that we
cannot deduce the endpoint of the Gregory-Laflamme instability
solely on the basis of global thermodynamic arguments like the
above entropy argument.\footnote{In fact, the viewpoint that an
unstable uniform black string decays to a localized black hole has
been challenged in \cite{Horowitz:2001cz} where it was argued that
the black string horizon cannot pinch off in finite ``horizon
time'' and that the endpoint of the Gregory-Laflamme decay should
be some type of non-uniform black string. See Sections \ref{sec:GL}, \ref{sec:end}
for a discussion of the endpoint of the GL instability and Section
\ref{sec:open} for some further comments.} It is clear that one
should look for a principle that relates two local properties on
the configuration space of gravity.

As a refinement of the entropy argument Gubser and Mitra proposed
in \cite{Gubser:2000ec,Gubser:2000mm} a natural conjecture that
relates classical and local thermodynamic stability. The
conjecture, which from now on will be referred to as the
Correlated Stability Conjecture (CSC), postulates that:

\vspace{2mm}
\noindent
{\it Gravitational systems with translational symmetry and infinite extent exhibit
a Gregory-Laflamme instability if and only if they have a local thermodynamic
instability.}

\vspace{2mm} \noindent Local thermodynamic stability is defined
here as positive-definiteness of the Hessian matrix of second
derivatives of the mass with respect to the entropy and any other
charges that can be redistributed over the direction in which a
Gregory-Laflamme instability can in principle occur. The
additional assumption of translational symmetry and infinite
extent has been added to insure that finite size effects do not
spoil the thermodynamic nature of the argument and to exclude
well-known cases like the Schwarzschild black hole which is
classically stable yet has negative specific heat and is therefore
locally thermodynamically unstable.

The CSC provides an exciting link between perturbative classical
stability and local thermodynamic stability bypassing the obvious
problems with using global entropic arguments. The importance of
such a connection on a practical level is obvious given the usual
complexity of the classical stability analysis. In what follows,
we outline the main aspects of the CSC, review its validity in a
series of examples and discuss its limitations, known
counterexamples and previous attempts to prove it.

For a system with $n$ conserved charges $Q_i$ the criterion of local
thermodynamic stability translates (after making the appropriate choice of ensemble)
into a positivity criterion for the $(n+1)\times (n+1)$ Hessian matrix
$H_M=\big(\frac{\d ^2 M}{\d q_\alpha \d q_\beta}\big)$, which
involves the second derivatives of the mass $M$ with respect to
the entropy $S$ and the charges $Q_i$. The charges $Q_i$ entering this definition
are determined by the choice of thermodynamic ensemble, which is crucial,
because it affects the discussion of local thermodynamic
stability and ultimately the validity of the conjecture as such.
The usual practice is to consider gravity in the canonical
ensemble, where the temperature $T$ is kept fixed
and the partition function of the system is expressed
as a function of $T$. With more degrees of freedom
the choice of ensemble for a general system can become less obvious.
To illuminate this point it is instructive to consider the
class of smeared or unsmeared, magnetically or
electrically charged black brane solutions of the previous sections.

Consider then a general (non-extremal) black D$p$-brane background
of type II string theory. Let $\{z_i\}$ denote the set of
non-compact directions along which the brane background exhibits
translational symmetry. The brane may be charged electrically or
magnetically under a $(p+1)$-form gauge field, with a
corresponding charge $Q_p$. We say that an electrically charged
brane is smeared along a direction $z_i$ when it is not charged
along this direction. For magnetically charged branes, we follow
the opposite convention and say that the brane is smeared along
$z_i$, whenever it is charged along $z_i$.\footnote{This
definition is also very natural for D3-branes which are self-dual
under electric-magnetic duality.}

We can now formulate for the CSC in a precise manner which
thermodynamical ensemble we should work in. If a brane is not
smeared in any of the non-compact directions with respect to a
charge $Q_p$ then we should work in the {\it canonical} ensemble
with respect to that charge, $i.e.$ we should keep $Q_p$ fixed. On
the other hand, if a brane is smeared in a particular direction
$z_i$ with respect to the charge $Q_p$ we should work in the {\it
grand canonical} ensemble with respect to that charge, $i.e.$ we
should instead keep the chemical potential $\nu_p$ corresponding
to $Q_p$ fixed. This rule for the ensemble used for CSC applies
equally well to electrically or magnetically charged branes as
expected by any sensible formulation of the CSC that is invariant
under the electric-magnetic duality.

As a simple illustration, consider a non-extremal
D0-brane solution smeared on a transverse direction $z$
in type IIA theory. The corresponding supergravity solution in the string
frame takes the form
\begin{equation}
\begin{array}{c}
\displaystyle
\label{aac}
ds^2=-H^{-1/2} f dt^2+H^{1/2}\left(
f^{-1} dr^2 +r^2 d\Omega_7^2 + dz^2\right) \spa e^{2\phi}=H^{3/2}\
,
\\[2mm]
\displaystyle  A_0=\coth \alpha (H^{-1}-1) ~,
f(r)=1-\frac{r_0^6}{r^6}~, ~ H(r)=1+\sinh^2 \alpha (1-f)\ .
\end{array}
\end{equation}
In this example the D0-brane charge $Q$ is smeared along the
$z$-direction and can be redistributed there freely. Hence, in the
context of the CSC it is appropriate to consider local thermodynamic stability in
the grand canonical ensemble, where $Q$ is allowed to
vary.

With a T-duality transformation along $z$
the background (\ref{aac}) turns into the D1-brane solution
\begin{eqnarray}
\label{aad} ds^2&=&H^{-1/2} \left(-f dt^2+dz^2\right)+H^{1/2}\left(
f^{-1} dr^2 +r^2 d\Omega_7^2\right)~,
\\ \nonumber
e^{2\phi}&=&H~, ~ ~ A_{0z}=\coth \alpha (H^{-1}-1) ~.
\end{eqnarray}
The D1-brane is now charged along the $z$-direction and the
corresponding charge is a fixed quantity. Accordingly, we should
now consider local thermodynamic stability in the canonical ensemble.
The application of the CSC with this choice of thermodynamics
yields a picture of stability that fits very nicely with the
classical stability properties of these solutions as we see
explicitly below.

As a final comment, it is well-known in thermodynamics that one can
move back and forth between the canonical and grand canonical
ensembles with a simple Legendre transform. The above example
suggests that within the context of the CSC it is natural to
associate a Legendre transform with a T-duality transformation in
supergravity. It should be noted, however, that the appearance of
the Legendre transform is an essential feature of supergravity that
treats momentum and winding modes asymmetrically. In the full string
theory, where momentum and winding are exchanged by T-duality the
Legendre transform would be unnecessary. For instance, a momentum
instability for a smeared brane transforms under T-duality into a
winding instability for a wrapping brane which is invisible in
supergravity.

\subsection{Examples \label{sec:examples}}

The first evidence for the validity of the CSC was given in the
original papers \cite{Gubser:2000ec,Gubser:2000mm} from the
perspective of the AdS/CFT correspondence. These papers studied a
set of $AdS_4$-Reissner-Nordstrom black hole solutions in $\NN=8$
supergravity in the large mass limit where the black holes are
expected to exhibit similar behavior to black branes. Up to a small
discrepancy arising from a numerical error they found that a
classical instability appears precisely when a local thermodynamic
instability sets in.

Further progress was made in the paper by Reall \cite{Reall:2001ag}
where a semiclassical proof was found of the CSC (see Section
\ref{sec:count}) in the case of singly-charged black branes. In this
case the local thermodynamics in the canonical ensemble is stable
for large charges and unstable for small charges. This fits
precisely with the results of the numerical study of the classical
stability for these branes
\cite{Gregory:1994bj,Gregory:1994tw,Hirayama:2002hn,Gubser:2002yi,Kang:2004hm,Miyamoto:2006nd}.

Below we consider instead another class of brane backgrounds, in
which the charges can be smeared in one or more directions.

\subsubsection{Non-extremal smeared black branes}

In Section \ref{sec:sugra} we saw that the non-extremal smeared
D$p$-branes exhibit a Gregory-Laflamme instability below a critical
mass. According to the CSC a corresponding instability should be
present in thermodynamics in the grand canonical ensemble
\cite{Harmark:2005jk}. In verifying this statement it is instructive
to consider both the canonical and grand canonical ensembles.

In the canonical ensemble, the temperature and charge are kept fixed and the
appropriate thermodynamic potential is the Helmholtz free
energy
\begin{equation}
F (T,Q) = M - T S \spa d F = -S d T  + \nu d Q \ .
\end{equation}
The condition for thermodynamic stability requires that
the specific heat $C_Q$ is positive, $i.e.$
\begin{equation}
\label{specheat} C_{Q} \equiv \left( \frac{\partial M}{\partial T}
\right)_Q = T \left( \frac{\partial S}{\partial T} \right)_Q > 0 \
.
\end{equation}

On the other hand, in the grand canonical ensemble the temperature
and chemical potential are kept fixed so the appropriate
thermodynamic potential is the Gibbs free energy
\begin{equation}
G (T,\nu)= M - TS - \nu Q \spa dG = - S dT - Q d\nu \ .
\end{equation}
The condition for thermodynamic stability is in this case
\begin{equation}
\label{grand}
C_Q \equiv \left( \frac{\partial M}{\partial T} \right)_Q > 0
\spa c \equiv \left( \frac{\partial \nu}{\partial Q} \right)_T > 0
\ ,
\end{equation}
where $C_Q$ is again the specific heat and $c$ is the inverse isothermal electric
permittivity. The condition \eqref{grand} follows from demanding that the
Hessian $H_G$ of the Gibbs free energy is negative definite.
Because of the matrix relation $H_G= - H_M^{-1}$,
where $H_M$ is the Hessian of the mass $M(S,Q)$, this
is equivalent to demanding that $H_M$ is positive
definite, as required by local thermodynamic equilibrium.

The thermodynamics of the non-extremal D$p$-branes \eqref{uni1}-\eqref{uni3}
smeared on a transverse circle of circumference $L$ is given by
\begin{equation}
\label{thermon1}
\begin{array}{c}
\ds \frac{M}{L} =\frac{\Omega_{7-p}V_p}{16 \pi G} \frac{L^{6-p}}{(2\pi)^{6-p}}
R_0^{6-p}[7-p+(6-p)\sinh^2 \alpha] \spa
 \frac{S}{L}= \frac{\Omega_{7-p} V_p}{4 G} \frac{L^{7-p}}{(2\pi)^{7-p}}
R_0^{7-p} \cosh \alpha \ ,
\\[4mm] \ds
 \frac{Q}{L}=\frac{\Omega_{7-p} V_p}{16 \pi G} \frac{L^{6-p}}{(2\pi)^{6-p}}
R_0^{6-p} (6-p)
\sinh \alpha \cosh \alpha \spa
 T=\frac{6-p}{2 L R_0 \cosh \alpha} \spa
 \nu=\tanh \alpha  \ ,
\end{array}
\end{equation}
where $V_p$ is the world-volume of the brane and the extensive quantities, $M/L$, $Q/L$ and $S/L$ correspond
respectively to the mass density, charge density and entropy density
along the transverse $z$ direction. $T$ is the temperature and $\nu$ the
chemical potential. These quantities satisfy the usual first law of
thermodynamics $dM = T d S  + \nu d Q$.

With the use of the  definitions \eqref{grand} and the explicit thermodynamic
quantities above, we find for the non-extremal smeared branes\footnote{These
expressions are most easily computed using the
identity that $\frac{ \partial y(r,s)}{\partial x(r,s)} \vert_{z(r,s)}
= \frac{ \partial(y,z)} {\partial(r,s)}
[\frac{ \partial(x,z)}{\partial (r,s)}]^{-1}.
$}
\begin{equation}
\label{cqnex}
C_Q= \left[ \frac{ 7-p  + (8-p) \sinh^2 \alpha}{-1 + (4-p) \sinh^2
\alpha} \right] S \spa c =   \frac{1}{ \cosh^2 \alpha [1 -(4-p)
\sinh^2 \alpha]}\frac{1}{TS} \ .
\end{equation}
We see for $p\leq 3$ that the requirement of positive specific heat
translates to
\begin{equation}
\label{cq2}
C_Q > 0 \qquad \Leftrightarrow \qquad \alpha>
{\rm arcsinh}  (1/\sqrt{4-p}) \ ,
\end{equation}
showing that there is a lower bound on the charge of the branes in order
to be thermodynamically stable in the canonical ensemble.
On the other hand, we have
\begin{equation}
\label{cq3} c > 0 \qquad \Leftrightarrow \qquad 0 < \alpha <  {\rm
arcsinh}  (1/\sqrt{4-p}) \ ,
\end{equation}
which is incommensurate with the condition in \eqref{cq2}. For
$p=4,5$ we have $C_Q < 0$ and $c>0$. Hence for any
$p=0,1,...,5$ it is impossible to satisfy the requirement
\eqref{grand} of thermodynamic stability in the grand canonical
ensemble. The presence of a local thermodynamic instability
in the grand canonical ensemble verifies the validity of the CSC
in this case and fits nicely with the general criteria for the choice of
thermodynamic ensemble presented in the previous subsection.
Note, however, that it is not possible to use the CSC to determine the exact value
of the Gregory-Laflamme threshold unstable mode.

It is also worth pointing out that the above thermodynamic quantities are invariant under T-duality.
For example, the thermodynamics of a D$p$-brane smeared on a transverse circle
is identical to that of a D$(p+1)$-brane wrapped around the T-dual circle.
The wrapped non-extremal D$(p+1)$-brane solution (for $1\leq p+1\leq 4$) is known to
be stable in supergravity, which is reflected nicely in the thermodynamics
if we choose to consider the CSC in the canonical ensemble.

\subsubsection{The near extremal limit}

The near-extremal limit of smeared D$p$-branes presents an
interesting subtlety. Since we send the charge to infinity, it seems
that in the near-extremal limit there is only one relevant ensemble,
namely the canonical ensemble. The corresponding Helmholtz potential
is given by
\begin{equation}
F (T) = E - TS  \spa d F = -S d T \ ,
\end{equation}
and thermodynamic stability requires positivity of the specific heat
$C \equiv \partial E(T)/\partial T$.

The thermodynamic quantities of near-extremal D$p$-branes  \eqref{NEbrane}
smeared on a transverse circle are given by%
\footnote{In terms of dimensionless quantities this thermodynamics was given
in \eqref{unise}.}
\begin{equation}
\label{thermon3}
E =  \frac{1}{{\cal{G}}} u_0^{6-p} \frac{8-p}{2} \spa
S=  \frac{4\pi}{{\cal{G}}} u_0 ( u_0^{6-p} K)^{1/2} \spa T  =
\frac{6-p}{4\pi u_0} \left( \frac{u_0^{6-p}}{K} \right)^{1/2} \ .
\end{equation}
Here, $E$ is the energy above extremality defined by $E = \lim (M- Q)$
in the near-extremal limit \eqref{nelim}, whereas $K$ (defined in \eqref{Kdef})
and ${\cal{G}}$ given by
\begin{equation}
\label{Gdef}
\frac{L V_{p} \Omega_{7-p} }{16 \pi G} =
\frac{1}{{\cal{G}}} l_s^{-2(6-p)} \spa
\frac{1}{{\cal{G}}} \equiv
(2\pi)^{2p-9} \hat{L} V_p \Omega_{7-p} \frac{N^2}{\lambda^2} \ ,
\end{equation}
are both finite in the limit.
The quantities in \eqref{thermon3} satisfy the usual first law of
thermodynamics $dE = T d S$.

From \eqref{thermon3} one obtains for the specific heat
the simple relation\footnote{An alternative way to obtain the same
result is by taking the near-extremal limit
($\alpha \rightarrow \infty$) of the expression \eqref{cqnex} for $C_Q$.}
\begin{equation}
\label{cqneex} C = \frac{8-p}{4-p} S \ .
\end{equation}
Hence, for any non-zero temperature, the specific heat is positive for
near-extremal smeared D$p$-branes when $p\leq 3$, which is a
well-known result. Since the charge goes to infinity in the near-extremal limit and
the chemical potential goes to one, it seems that these quantities
cannot be varied anymore. This suggests that it does not make
sense to consider the grand canonical ensemble for near-extremal
branes.

To put it differently, one could consider what happens to the quantity
$c$ in \eqref{cqnex} when we take the near-extremal limit. As a consequence of
the limit $\alpha \rightarrow \infty $ we deduce that $c \rightarrow 0$. This would seem
to imply that, infinitesimally close to the near-extremal limit,
non-extremal branes are marginally thermodynamically stable in the
grand canonical ensemble in clear contradiction with the classical
stability analysis of Section \ref{sec:nonGL}, where we saw explicitly
the presence of a Gregory-Laflamme instability in the near-extremal limit.
It would appear that the near-extremal limit is an example where the CSC is manifestly violated.

Fortunately, there is a simple resolution to this discrepancy \cite{Harmark:2005jk}.%
\footnote{Note that this also resolves the contradiction found in
the numerical investigations of Ref.~\cite{Kang:2005is}.}
It turns out that in the near-extremal limit
the right way to compute the quantity $c$ in the grand canonical ensemble
is not by using the non-extremal chemical potential $\nu$, but by using a new
chemical potential $\hat \nu$, which can be understood
as the ``chemical potential above extremality".
This is in analogy to the fact that we do not compute $C_Q$ using the mass
$M$ (which goes to infinity in the near-extremal limit), but using
the energy $E$ above extremality, which is finite.

To derive an expression for $\hat \nu$, we note that the first law of
thermodynamics for non-extremal branes takes the form
\begin{equation}
\label{law1} d E = T d S + (\nu -1) d Q \ ,
\end{equation}
when written in terms of $E = M -Q$. Moreover, one can easily see
from \eqref{thermon1}, that the chemical potential $\nu$
approaches 1 in the near-extremal limit $\alpha \rightarrow
\infty$, and is not a free parameter anymore. It is therefore
natural to define the rescaled chemical potential $\hat \nu$ and
the corresponding rescaled charge ${\hat Q}$ as
\begin{equation}
\label{nuhatdef}
\hat \nu \equiv \frac{1}{g^2_{\rm YM}} \lim_{l_s \rightarrow 0}
\frac{1}{ l_s^4} (\nu -1 )
\spa \hat Q \equiv g^2_{\rm YM}\lim_{l_s \rightarrow 0}  l_s^4 Q \ ,
\end{equation}
where the factor of $l_s$ has been chosen such that $\hat \nu$ is
finite in the near-extremal limit. The finite factor $g_{YM}^2$ has
been inserted for later convenience. With the definitions
\eqref{nuhatdef}, we then find that \eqref{law1} can be written as
\begin{equation}
\label{law1p} d E = T d S + \hat \nu d \hat Q \ ,
\end{equation}
which is a well-defined differential relation in the near-extremal
limit. We have thus obtained  a modified version of the first law of
thermodynamics for near-extremal branes involving an extra term
containing the rescaled chemical potential $\hat \nu$ and rescaled
charge $\hat Q$.

Consequently, by using the non-extremal quantities \eqref{thermon1}
 and the near-extremal limit \eqref{nelim} in the
definitions \eqref{nuhatdef} we find for near-extremal smeared
D$p$-branes the expressions
\begin{equation}
\label{thermon5} \hat Q = (6-p)  \frac{K}{{\cal{G}}} \spa \hat \nu
= - \frac{u_0^{6-p}}{2 K} \ ,
\end{equation}
where $K$, ${\cal{G}}$ are defined in \eqref{Kdef}, \eqref{Gdef} respectively.
The near-extremal Gibbs free energy is then naturally defined by
$G(T,\hat \nu) = E - TS - \hat \nu \hat Q$, and
the condition for thermodynamic stability of near-extremal
smeared branes in the grand canonical ensemble follows immediately
by analogy with \eqref{grand}, namely
\begin{equation}
C_{\hat Q}=\left(\frac{\partial E}{\partial T}\right)_{\hat Q} >0
\spa \hat c=\left( \frac{\partial \hat \nu}{\partial \hat
Q}\right)_T > 0 \ .
\end{equation}
Computing these quantities for near-extremal smeared D$p$-branes,
we find that the quantity $C_{\hat Q}$ is identical to the one computed in
\eqref{cqneex}. On the other hand, for $\hat c$ we find
using \eqref{thermon5} and \eqref{unise} that%
\footnote{One can compute $\hat c$ directly from the near-extremal
thermodynamics using $\hat c = \frac{ \partial ( \hat \nu,T)
}{\partial (u_0,K)} [\frac{ \partial(\hat Q,T)}{\partial
(u_0,K)}]^{-1}$. Alternatively, one can show from \eqref{nuhatdef}
that $\hat c = \lim l_s^{-8} g_{\rm YM}^{-4} c $. We obtain the same result
by using in this expression
the non-extremal relation \eqref{cqnex} for $c$ and then taking the
near-extremal limit.}
\begin{equation}
\hat c=- \frac{{\cal{G}}}{g_{\rm YM}^4 K^2}
\frac{u_0^{6-p}}{(4-p)(6-p)} \ ,
\end{equation}
which is finite and negative for $p \leq 3$.
Hence we conclude that near-extremal smeared D$p$-branes with $p\leq 3$
are thermodynamically unstable in the near-extremal
grand canonical ensemble defined above,
in full agreement with the classical stability analysis and the CSC.

\subsubsection{The D$(p-2)$-D$p$ system}

Another interesting class of examples in type II string theory involves
systems of D-branes, for example D-brane bound states, which preserve
various amounts of supersymmetry in the extremal case. For instance,
consider the bound state of a D$(p-2)$- and a D$p$-brane, where the
D$(p-2)$-brane is embedded in the D$p$-brane and smeared along its two
transverse directions. In the extremal case, this system is a half BPS state.
The corresponding supergravity solution appears in
Section \ref{sec:GLbound}. The classical stability properties for the D0-D2 bound
state were discussed in \cite{Gubser:2004dr,Ross:2005vh}, where it was
found that the system exhibits a Gregory-Laflamme instability in accordance with the local thermodynamic stability analysis and the CSC. Here we review the stability pattern
that emerges from thermodynamics.

The thermodynamic quantities of the D$(p-2)$-D$p$ bound state can be
found for instance in \cite{Harmark:1999rb}. For zero angular momentum
we have
\begin{equation}
\label{boundthermo2}
\begin{array}{c}
M=\frac{V_p \Omega_{8-p}}{16\pi G} r_0^{7-p}(8-p+(7-p)\sinh^2 \alpha)~, ~ ~
T=\frac{7-p}{4\pi r_0 \cosh \alpha}~,
\\[4mm]
S=\frac{V_p \Omega_{8-p}}{4G}r_0^{8-p}\cosh \alpha~, ~ ~
Q=\frac{(7-p)V_p \Omega_{8-p}}{16\pi G} r_0^{7-p}\sinh \alpha \cosh\alpha~, ~ ~
\nu=\tanh \alpha~,
\\[4mm]
\nu_p=\nu \cos\theta~, ~ ~ \nu_{p-2}=\nu \sin \theta~, ~ ~
Q_p=Q\cos\theta~, ~ ~ Q_{p-2}=Q\sin\theta
~.
\end{array}
\end{equation}
To determine the local thermodynamic stability of this system, we
analyze the thermodynamics in a mixed canonical/grand canonical ensemble,
where we allow $Q_{p-2}$ to vary while keeping $Q_p$ fixed.  In other words,
we consider the smeared D$(p-2)$-brane in the grand canonical ensemble
and the D$p$-brane in the canonical ensemble adhering to the general criteria
of the previous subsection.

\begin{figure}
\begin{center}
\includegraphics[width=8cm,height=5cm]{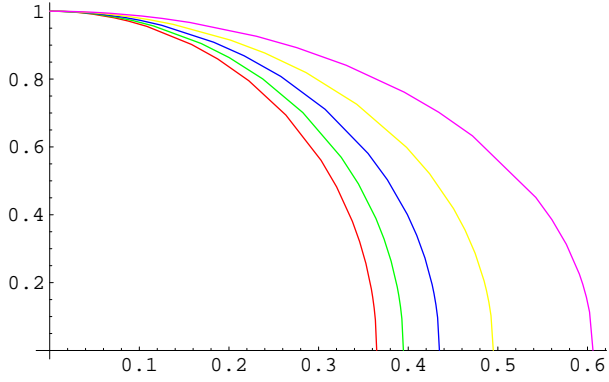}
\caption{Sketch of the $(Q_{p-2}/M,Q_p/M)$ boundary of stability for $p=0$ (red), $p=1$ (green), $p=2$ (blue), $p=3$ (yellow), $p=4$ (pink). The bound state is unstable in the region below the boundary of stability.}
\label{d0d2}
\end{center}
\end{figure}

Much like in the previous example of the non-extremal smeared branes one can
show that the determinant of the Hessian matrix
$H_M=\left( \frac{\d^2 M}{\d q_{\alpha} \d q_{\beta}}\right)$ (with
$q_{\alpha} \in \{ S, Q_{p-2} \} )$ is
\begin{equation}
\label{boundhessian2}
\det H_M=\frac{\d (T,\nu_{p-2})}{\d(S,Q_{p-2})}=
\frac{16 ~G ~{\rm sech}^4 \alpha}{V_p^2\Omega^2_{8-p} r_0^{16-2p}\left((9-p)\cosh^2\alpha-1\right)}
\left((5-p)\sinh^2 \alpha \cos^2\theta-1\right)
~.
\end{equation}
Hence, the condition of thermodynamic stability translates to
\begin{equation}
\label{stabbound} \det H_M >0 ~ ~ \Leftrightarrow ~~ {\rm csch}^2
\alpha \leq (5-p) \cos^2\theta ~,
\end{equation}
which can be satisfied only if $p\leq 4$.
The boundary of stability, where $\det H_M=0$, can be written more compactly as
\begin{equation}
\nu_{p-2}^2+(6-p)\nu_p^2=1
~.
\end{equation}
The stability pattern is summarized in Fig.\ \ref{d0d2}.

\subsubsection{The D$(p-4)$-D$p$ system}

Similar statements apply to the D$p$-D$(p-4)$ bound
states, which are quarter BPS in the extremal case.
The supergravity solution for these states appears already in
Section \ref{sec:GLbound}. Again, one can apply the CSC to
find the stability pattern for these states with a simple calculation
in thermodynamics. As above, we are instructed to work
in a mixed canonical/grand canonical ensemble, where we allow
$Q_{p-4}$ to vary holding $Q_p$ fixed.

In a suitable parametrization (see \cite{Friess:2005tz}) the thermodynamic
variables of the D$(p-4)$-D$p$ system take the form
\begin{equation}
\label{boundthermo4}
\begin{array}{c}
M=\frac{V_p \Omega_{8-p}}{16\pi G} \frac{r_0^{7-p}}{2}
(2+(7-p)(\cosh 2 \alpha_p+\cosh 2\alpha_{p-4}))~, ~ ~
T=\frac{7-p}{4\pi r_0 \cosh \alpha_p \cosh \alpha_{p-4}}~,
\\[4mm]
S=\frac{V_p \Omega_{8-p}}{4G}r_0^{8-p}\cosh \alpha_p \cosh \alpha_{p-4}~, ~ ~
\nu_p=\tanh\alpha_p~, ~ ~ \nu_{p-4}= \tanh \alpha_{p-4}~, ~ ~
\\[4mm]
Q_p=\frac{(7-p)V_p \Omega_{8-p}}{32 \pi G}r_0^{7-p} \sinh 2\alpha_p~, ~ ~
Q_{p-4}=\frac{(7-p)V_p \Omega_{8-p}}{32 \pi G}r_0^{7-p} \sinh 2\alpha_{p-4}
~.
\end{array}
\end{equation}
Consequently, the determinant of the Hessian matrix becomes
\begin{eqnarray}
\label{boundhessian4}
&&\det H_M=\frac{\d(T,\nu_{p-4})}{\d(S,Q_{p-4})}=
\\
&&=-\frac{8G^2}{V_p^2\Omega^2_{8-p}}
\frac{r_0^{2(p-8)}{\rm sech}^2\alpha_p {\rm sech}^4 \alpha_{p-4}}
{\cosh 2\alpha_{p-4}(7-p+2\cosh 2\alpha_p)+(7-p)\cosh2\alpha_p}
(7-p-(5-p)\cosh2\alpha_p)
~.\nonumber
\end{eqnarray}
In this expression all the factors have a positive definite sign except
for the last.

There are three values of interest for $p$. $p=4$ represents
the D0-D4 bound state, for which we find an instability when
\begin{equation}
\cosh2\alpha_4<3
~.
\end{equation}
The existence of a corresponding threshold unstable mode
in the classical stability analysis of this bound state
was argued in \cite{Ross:2005vh}. For $p=5$ and $p=6$
there is no stability boundary and the bound state is unstable
all the way up to extremality.

\subsubsection{Other examples}

Other examples of bound states can be obtained from the above
in the following manner. For instance, by S-dualizing the D1-D3 bound
state we can obtain the F1-D3 bound state. This state exhibits the same
thermodynamics as the D1-D3 system and therefore also the same stability pattern.
We can also lift the D2-D4 bound state to the M2-M5 system. By further
reducing the latter on a different circle we obtain the F1-D4 bound state.
All three of these bound states have the same thermodynamics and
the same stability pattern. Unfortunately, we cannot use
T-duality to extend this list even further. As already pointed out, T-duality
acts non-trivially on the CSC. On the level of the classical stability analysis
the translation invariance is broken explicitly by the threshold unstable mode
along the direction of the Gregory-Laflamme instability thus making it impossible
to perform a T-duality transformation along these
directions. On the level of thermodynamics, the same effect reappears in some cases
as a change of the ensemble.

One can also consider independently the general
F1-D$p$ bound state or three-charge bound states like the D0-D4-F1 system.
The former was considered in \cite{Friess:2005tz} and the latter in
\cite{Ross:2005vh}.

Beyond these cases it would be interesting to consider the CSC in a wider class
of examples, for instance in cases that exhibit rotation, such as spinning
bound states.

\subsection{Towards a proof of the conjecture and known counterexamples \label{sec:count}}

The correlated stability conjecture has proven to be a rather robust statement
whose validity has been verified in a large class of examples. A semi-classical proof of the
conjecture for magnetically charged black brane solutions was given in
\cite{Reall:2001ag} by Reall (see also \cite{Gregory:2001bd}). Elements of this proof for more complicated systems,
like that of electrically charged smeared black branes, were discussed in
\cite{Friess:2005tz} and \cite{Harmark:2005jk} (see also \cite{Buchel:2005nt}
for another approach to a proof based on holography). Recently \cite{Friess:2005zp} presented
a set of counterexamples in solutions with scalar hair, which suggest
that the conjecture cannot survive in its current form as a statement
that applies to the most general case. To see what may go wrong, it is instructive
to review the key elements of the existing arguments in favor of the CSC, the
weak points of these arguments and the known counterexamples.

To expedite the discussion let us denote again by $\{ z^i \}$ the
infinite extent directions along which a Gregory-Laflamme
instability can occur and by $x^{\mu}$ all the other space-time
directions. For compactness, it is also convenient to denote the
field content of the theory collectively by the fields $\Psi^I$.
$I$ is a multi-index label that summarizes all the metric
components and any other fields including scalar fields ($e.g.$
the dilaton) and gauge fields. The central object in the
semi-classical argument of Reall \cite{Reall:2001ag} is the
Euclidean path integral dimensionally reduced along $z^i$. The
Euclidean path integral can be written compactly as
\begin{equation}
Z=\int d[\Psi^J] e^{-I[ \Psi^J ]}
~,
\end{equation}
where $I[ \Psi^J ]$ is the Euclidean action and the path integral is taken
over the fields $\Psi^I$ in the appropriate thermodynamic ensemble specified
(for a class of examples) in Section \ref{subsec:csc1}. For gravity
this is always the canonical ensemble and the path integral is taken over
all the smooth Riemannian geometries that are asymptotically $S^1 \times \MM$.
$S^1$ is the Euclidean time circle with circumference $\beta=\frac{1}{T}$ and
$\MM$ the asymptotic spatial manifold.

The Euclidean path integral is well-defined in the semi-classical approximation,
where one looks for saddle points of the Euclidean action. We can expand the
fields $\Psi^I$ around the saddle points\footnote{In what follows, a subindex $0$
will denote the saddle point expressions.}
\begin{equation}
\Psi^I=\Psi^I_{0}+\delta \Psi^I
\end{equation}
and write the perturbed action as
\begin{equation}
I[\Psi^I]=I_0[\Psi^I]+\delta_2 I[\Psi^I]
~.
\end{equation}
The first order perturbation vanishes by the equations of motion and
the second order perturbation can be written in terms of the Euclidean
Lichnerowicz operator $\Delta_L$ as
\begin{equation}
\begin{array}{c}
\label{order2pert}
\delta_2 I[\Psi^I]=\frac{1}{2}\int d^d x d^d x' \sqrt{g_0}
\delta \Psi^I(x)\frac{\delta^2 I}{\delta \Psi^I(x) \delta \Psi^J(x')} \delta \Psi^J(x')=
\\[4mm]
=\int d^d x \sqrt{g_0} \delta \Psi^I(x) (\Delta_L)_{IJ} \delta \Psi^J(x')
~.
\end{array}
\end{equation}

The Euclidean path integral $Z$ receives two contributions: one
from the saddle point configuration and another from the determinant
of the Lichnerowicz operator, so
\begin{equation}
Z \propto (\det \Delta_L)^{-1/2} e^{-I_0[\Psi^I_0]}
~.
\end{equation}
When the Lichnerowicz operator has a negative eigenvalue, in other words
when the system of differential equations
\begin{equation}
\label{lichne}
(\Delta_L)_{IJ} \delta \Psi^J=\lambda \GG_{IJ} \delta \Psi^J
\end{equation}
has a solution with $\lambda<0$, the Euclidean path
integral $Z$ receives an imaginary contribution, which can be seen
as a signal of thermodynamic instability. $\GG_{IJ}$ is a natural
metric on the space of perturbations defined implicitly by \eqref{lichne}.

The existence of a negative eigenvalue for the dimensionally reduced
Euclidean Lichnerowicz operator provides a natural bridge
between classical and local thermodynamic instability.
The connection to a classical instability of the Lorentzian theory
arises in the following way. As we found in previous sections,
a clear signal of a dynamical instability is the existence of a threshold
unstable mode. Let us denote this threshold mode as $\delta\psi^I$.
We can separate the $z^i$ dependence of this mode by setting
\begin{equation}
\delta \psi^I={\rm Re}\left( e^{ik_j z^j} \delta \Psi^I \right)
~.
\end{equation}
Plugging this ansatz into the second order perturbation of the
equations of motion of the theory leads to a system of differential
equations of the form
\begin{equation}
\label{lichne1}
(\hat{\Delta}_L)_{IJ} \delta \Psi^J=-k^2 \GG_{IJ} \delta  \Psi^J
~,
\end{equation}
where $(\hat{\Delta}_L)_{IJ}$ is the dimensionally reduced Lorentzian
Lichnerowicz operator (see \eqref{Goncrith}).
After a trivial Wick rotation equation \eqref{lichne1}
becomes \eqref{lichne} with $\lambda=-k^2$. Hence, we see that there is
a close relation between the negative modes appearing in the dimensionally
reduced Euclidean path integral
and the momentum of the threshold unstable modes of the classical stability
analysis.

On the other hand, one can argue for a natural relation between the semi-classical
path integral and local thermodynamic stability. To make this relation more
precise Reall suggested an argument, based on an earlier construction given in
\cite{Whiting:1988qr,Prestidge:1999uq}. The goal of the argument is to relate the Hessian
$\frac{\delta^2 I}{\delta \Psi^I \delta \Psi^J}$
of the action $I$ with the Hessian $H_M$ of the local thermodynamic stability analysis.
For that purpose, consider a system with temperature $T$, entropy $S$ and
$n$ additional charges $Q_i$. The $n$ corresponding chemical potentials
will be denoted as $\nu_i$. In terms of these variables the first law of thermodynamics
takes the form
\begin{equation}
dM=TdS+\sum_i \nu_i dQ_i
~.
\end{equation}
In the grand canonical ensemble, the idea is to construct an $n+1$-parameter
family of configurations with Euclidean action
\begin{equation}
I(x_0,...,x_{n};\beta,\nu_1,....,\nu_n)=
\beta M(x_0,...,x_{n})-S(x_0,...,x_{n})-\beta \sum_{j=1}^n\nu_j Q_j(x_0,...,x_{n})
~.
\end{equation}
The generic point in this family is an off-shell background, which does not
satisfy the Einstein equations of motion, but satisfies the appropriate Hamiltonian
constraints. For a particular value of the $n+1$ parameters
$(x_0,....,x_{n})=(x_0(\beta,\nu_1,...,\nu_n),...,x_{n}(\beta,\nu_1,....,\nu_n))$
the background becomes an exact solution of the equations of motion.
At this point the functions $M,S$ and $Q_i$ are precisely the energy, entropy
and charges of the corresponding gravitational system. For other values of
the parameters $(x_0,...,x_{n})$ the interpretation of the functions
$M,S$ and $Q_i$ is not important.

Given the existence of such a family of configurations, we can immediately
draw a number of interesting conclusions. First of all, since the configurations
become an exact solution of the equations of motion at the special point
\begin{equation}
P=(x_0,....,x_{n})=(x_0(\beta,\nu_1,...,\nu_n),...,x_{n}(\beta,\nu_1,....,\nu_n))
\nonumber
\end{equation}
the action $I$ is extremized there yielding
\begin{equation}
\label{extrem}
\left( \frac{\d I}{\d x_j}\right)_{\beta,\nu_1,...,\nu_n}=0~, ~ ~ j=0,...,n
~
\end{equation}
or equivalently the standard thermodynamic relations
\begin{equation}
\frac{1}{\beta}=\left( \frac{\d M}{\d S} \right)\bigg|_{(x_0,...,x_n)=P}~, ~ ~
\nu_i=\left( \frac{\d M}{\d Q_i} \right)\bigg|_{(x_0,...,x_n)=P}~, ~ ~
i=1,...,n~.
\end{equation}

The quadratic perturbation of the action $I$ around the special point $P$
involves the Hessian matrix $I_{(2)}=\frac{\d^2 I}{\d x_j \d x_k}$, which
can be recast in terms of the Hessian matrix $H_M$ as
\begin{equation}
\label{I2IM}
\left( I_{(2)} \right)_{ab}=M_a^{~ \alpha} \left( H_M^{-1} \right)_{\alpha \beta}
M^{\beta}_{~b}
~.
\end{equation}
In this relation, which can be deduced easily with the use of the first order
equations \eqref{extrem}, $M_a^{~ \alpha}$ is the Jacobian matrix of the transformation
from the variables $(x_0,...,x_n)$ to the thermodynamic variables
$(\beta,\nu_1,...,\nu_n)$. Therefore, assuming $M$ to be an invertible matrix,
we can use equation \eqref{I2IM} to achieve the desired relation between the Hessian
matrix $I_{(2)}$ of classical stability and the Hessian matrix $H_M$ of
local thermodynamic stability.

The construction of the $n+1$ parameter family of configurations is a non-trivial
exercise that has not been performed for the most general system. So far the construction
has been performed explicitly for magnetically charged black brane solutions in
\cite{Reall:2001ag} and for electrically charged smeared black branes in \cite{Harmark:2005jk}.

To complete the proof one has to verify an additional set of properties.
For instance, one should verify that the norm of the on-shell perturbations
on the space of theories is positive definite. This ensures that the analysis
is restricted to normalizable on-shell perturbations excluding for example any non-physical
negative-norm conformal perturbations. The positivity of the norm
\begin{equation}
||\delta \Psi ||^2=\int d^d x ~\delta \Psi^I \GG_{IJ} \delta \Psi^J
\nonumber
\end{equation}
also ensures that the action $I$ decreases, and therefore an
instability exists, precisely when the Hessian $\frac{\delta^2
I}{\delta \Psi^I \delta \Psi^J}$ fails to be positive definite
(see equations \eqref{order2pert}, \eqref{lichne}).

Finally, one should demonstrate that there is sufficient overlap between the off-shell
deformations $(x_0,...,x_n)$ with the actual on-shell perturbations $\delta \psi^I$.
In \cite{Reall:2001ag} it was pointed out that a path in the family of off-shell
configurations is not related directly to an eigenfunction of the Lichnerowicz
operator, but rather some linear combination of eigenfunctions. This suggests
that when the action decreases along this path, at least one of the eigenvalues
of the Lichnerowicz operator must be negative and therefore an actual on-shell
instability should exist. It is not immediately obvious, however, that the converse
is also true. To complete the proof one should demonstrate that the off-shell
deformations and the actual on-shell perturbations cover the same linear space.
This is a rather subtle point that has not been demonstrated fully in any known example.
It is natural to expect in general that there will be sufficient overlap between the
eigenfunctions of the Lichnerowicz operator and the family of off-shell deformations
for systems that are specified uniquely by a complete set of conserved charges.
Systems with hair do not fall into this category and are therefore the prime candidates
as systems that violate the CSC in its current formulation. In such situations the full set of conserved
charges is not enough to determine uniquely the classical solution. As we saw
the conserved quantities are the only input of the local thermodynamic stability;
a dynamical instability, however, is a property of the entire solution.

This crucial point was put forward in \cite{Friess:2005zp}, which presented
a set of theories where the exact black brane solutions are specified by a set of conserved
charges and the asymptotic values of a set of scalar fields. It was found that
these systems can develop a classical instability associated with the dynamics of the
scalar fields despite the presence of local thermodynamic stability.

The violation of the CSC observed in \cite{Friess:2005zp} is reminiscent of
an earlier counterexample proposal in \cite{Marolf:2004fy}. The central objects
in that paper are three-charge D1-D5-P spinning black brane solutions in type
IIB supergravity compactified on $T^4$. This class of solutions contains gyrating
black strings \cite{Horowitz:1996cj} carrying the same charges as ordinary
spinning black strings, but a different value of an extra parameter, which measures the angular momentum carried by gyrations of the string (as opposed to spin).
\cite{Marolf:2004fy} argued that a small perturbation
of a non-gyrating black string could grow to become large for sufficiently large angular
momentum. This new instability was proposed as a possible counterexample
to the CSC.

Given the explicit violation of the CSC in the above examples, is it still possible to find
a suitable refinement or restriction of the conjecture that can survive
as a correct statement? The above counterexamples suggest
two natural possibilities. If we choose to leave the definition of local thermodynamic
stability unchanged in the formulation of the conjecture, then it is natural
to add a further assumption and conjecture that the CSC works provided
that there is a unique solution with a spatially uniform horizon and specified
conserved charges \cite{Friess:2005zp}. No counterexamples of this restricted
version of the CSC are known. An alternative, more ambitious, refinement
of the conjecture has been suggested by B.\ Kol (see \cite{Friess:2005zp}).
We can redefine the meaning of
local thermodynamic stability by enlarging the Hessian matrix to include
derivatives with respect to those quantities that characterize the asymptotics
of the scalar fields or any other quantity that determines uniquely a solution.
In that case, local thermodynamic stability translates into a positivity criterion
for the new enlarged Hessian matrix. It seems quite likely that with this refinement
the CSC survives all tests known up to date.

\section{Holographic implications \label{sec:holo}}

Holography provides a non-trivial relation between a gravitational
theory in the bulk
and a non-gravitational theory ($e.g.$ a Yang-Mills theory) on the boundary.
A well-known example of this relation is the AdS/CFT
correspondence between the $\CN = 4$ Super-Yang-Mills theory and
type IIB string theory on $AdS_5 \times S^5$
\cite{Maldacena:1997re,Gubser:1998bc,Witten:1998qj,Witten:1998zw}.
In general, holography relates gravity in the near-horizon background of a
brane configuration in string/M-theory and the
non-gravitational world-volume theory living on the branes
in a low-energy decoupling limit \cite{Itzhaki:1998dd}.%
\footnote{See also the review \cite{Aharony:1999ti} and references therein.}

In this section we discuss gauge/gravity duals in which the gravity
side exhibits a Gregory-Laflamme instability. The presence of this
classical gravitational instability in the bulk has important
consequences for the gauge theory on the boundary, which we review
following \cite{Aharony:2004ig,Harmark:2004ws,
Harmark:2005pq,Harmark:2005jk,Harmark:2005dt}. Focusing on a set of
examples that are closely connected to the material of this review
we will consider first D-branes on a circle, where the relevant
gauge theory is super-Yang-Mills theory on a circle. This example
can be generalized with D-branes on higher-dimensional torii, which
we discuss briefly at the end of the first subsection. Other
interesting examples include the M2- and M5-brane on a circle, where
the relevant non-gravitional theory is (2+1)-dimensional
super-Yang-Mills and Little String theory. We review the basic
features of both cases giving more emphasis to the second example,
which exhibits new properties not present in the other cases.
We conclude with a brief discussion of
the implications of Gregory-Laflamme instabilities for non-commutative
Yang-Mills and non-commutative open string theories, which arise as
part of the worldvolume dynamics of D-brane bound states.

\subsection{D-branes on a circle}
\label{sec:SYMcirc}

Consider the supergravity solution of a large number of coincident
D$(p-1)$-branes on a transverse circle.
In Section \ref{sec:nonphases} we found that there is a map between
the phases of
neutral Kaluza-Klein black holes (see Section \ref{sec:KKphases}) and
the phases of non- and near-extremal D-branes on a transverse circle.
Furthermore, we saw in Section \ref{sec:nonGL} that we can map the
Gregory-Laflamme mode of the neutral black string to a
Gregory-Laflamme mode of non- and near-extremal D-branes smeared on
a transverse circle. In this section we will discuss the consequences of
this unstable mode for the holographically dual theories living on the D-branes.

The low-energy theory on $N$ coincident D$p$-branes in type II string
theory in flat space is given by the $(p+1)$-dimensional $U(N)$ super-Yang-Mills
theory with sixteen supercharges (on $\R^{p,1}$). According to holography
\cite{Maldacena:1997re,Itzhaki:1998dd}, the large $N$, finite temperature,
$SU(N)$ version of this  theory\footnote{The $U(1)$ part of the gauge group
decouples in holography \cite{Aharony:1999ti}.} is equivalent to the
near-extremal limit of the
supergravity solution of $N$ coincident D$p$-branes.
The perturbative Yang-Mills description is valid at small 't Hooft coupling
and the supergravity description at large 't Hooft coupling.

Similarly, since a near-extremal D$(p-1)$-brane on a transverse circle
is T-dual to a D$p$-brane wrapped on a circle we conclude
that the gauge theory dual of $N$ coincident near-extremal D$(p-1)$-branes on a
transverse circle is $(p+1)$-dimensional SYM on
$\R^{p-1,1}\times S^1$ with gauge group $SU(N)$ and
sixteen supercharges \cite{Susskind:1997dr,Barbon:1998cr,Li:1998jy,%
Martinec:1998ja,Martinec:1999bf,Fidkowski:2004fc}.

For example, this implies a correspondence between the
phases of near-extremal D2-branes on a transverse
circle and the phases of ${\cal{N}}=4$ SYM theory
on $\R^{2,1} \times S^1$ . By Wick rotating to Euclidean space,
where the non-zero temperature translates to a compact direction,
we get $\NN=4$ SYM on $\R^2 \times \T^2$, the
two-torus being $\T^2 \equiv S_\beta^1 \times S^1$. Two other interesting
examples are near-extremal D0- and D1-branes on a transverse circle.
These are respectively dual to $(1+1)$-dimensional SYM theory on
$\T^2=S_\beta^1 \times S^1$
and to $(2+1)$-dimensional SYM on $\R \times \T^2$ (in Euclidean space).
In what follows we will restrict the discussion to the cases of
D0-, D1-, D2- and D3-branes on a circle, but one should keep in mind that
the results of this section can be easily extended to include the
F-strings and the
remaining D-branes.

The main point of the discussion here is that we can use the gravity results of
Sections \ref{sec:nonphases} and \ref{sec:nonGL} to obtain predictions for
the thermodynamics of the dual strongly coupled gauge theories described
in the previous paragraphs \cite{Aharony:2004ig,Harmark:2004ws,
Harmark:2005jk,Harmark:2005dt}. We will see, for example, that the
localized phase of the supergravity phase diagram
corresponds to the low temperature/low energy regime of gauge theory,
while the non-uniform phase corresponds to a new phase that emerges from the
uniform phase. The latter describes the high temperature/high energy regime
of gauge theory. The Gregory-Laflamme instability in supergravity has
a natural counterpart
in gauge theory as an instability of the high energy/temperature phase when the
energy/temperature becomes sufficiently low.

The dictionary between the supergravity parameters of the near-extremal
D$p$-branes on a (transverse) circle and the parameters of SYM theory
has been established in Section \ref{sec:sugra}. We summarize it here
for the convenience of the reader. On the gravity side, $L$ and $\hat L$
denote respectively the circumference of the transverse circle and its T-dual.
$g_s$ and $\hat g_s$ denote the string coupling in the theory before and
after T-duality. These quantities are related to each other via the relations
\eqref{tdual}. On the gauge theory side, $N$ is the rank of the $SU(N)$ gauge
group and $\lambda=g_{\rm YM}^2 N=(2\pi )^{p-2} \hat g_s l_s^{p-3}$
the 't Hooft coupling. The energy and temperature of the field theory will
be denoted by $E$ and $\hat T$ respectively. The definition of the near-extremal
limit \eqref{nelimit} on the gravity side involves two parameters $g$ and $l$,
which are kept fixed. These parameters can be re-expressed in terms of
the SYM variables in the following way\footnote{We use the notation and
conventions of Ref.~\cite{Harmark:1999xt}.}
\begin{equation}
\label{dictdp}
g = \frac{1}{(2\pi)^3 N^2 } \frac{\hat L^{p}}{V_{p-1}} \left( \lambda \hat
L^{3-p} \right)^2 \spa l  = \frac{\hat L }{\sqrt{2\pi}}\sqrt{
\lambda \hat L^{3-p} } \ .
\end{equation}
From $g$, $l$ and $\hat T$ we can form two
independent dimensionless parameters, $l/g$ and $l \hat T$, which
will play an important role below. These parameters are functions of
the dimensionless temperature and coupling $t'=\hat L \hat T$,
$\lambda'=\lambda \hat L^{3-p}$.

It should be noted that stringy corrections to the supergravity description
of coincident D$p$-branes wrapping a circle\footnote{The metric is presented
here in the string frame. Compare with Eqs.\
\eqref{uninear1}-\eqref{uninear3} in Section \ref{sec:nonphases},
where the T-dual solution is written in the Einstein frame.
The full background also includes a non-trivial dilaton and a
$p$-form RR potential which can be deduced easily
from the corresponding expressions in Section \ref{sec:sugra}.}
\begin{eqnarray}
\label{p+1}
l_s^{-2}ds^2=\frac{R^{\frac{7-p}{2}}}{\sqrt {\hat h_d}} \left[
-\left(1-\frac{R_0^{7-p}}{R^{7-p}}\right) dt^2+
\frac{d\hat z^2}{(2\pi)^2}\right]+
\frac{\sqrt {\hat h_d}}{R^{\frac{7-p}{2}} \left( 1-\frac{R_0^{7-p}}{R^{7-p}}
\right)}dR^2+ \sqrt {\hat h_d} R^{\frac{p-3}{2}} d \Omega^2_{8-p}
\nonumber
\end{eqnarray}
are negligible when the string scale is small compared to the length scale
generated at the horizon
$\ell_H \sim \sqrt{\alpha'} \hat h^{1/4} R_0^{\frac{p-3}{4}}L^{-\frac{1}{2}}$, $i.e.$
when $\ell_H \gg \sqrt{\alpha'}$.\footnote{The extra factor
$L^{-1/2}$ in the definition of $\ell_H$ follows from the extra rescaling that we
do when we take the near-extremal limit (see sentence below \eqref{Hhat}).}
This is equivalent (up to a constant factor with powers of $\pi$ that we drop)
to the condition $\hat L \hat T \ll (\lambda \hat L^{3-p})^{\frac{1}{3-p}}$.\footnote{To
derive this result we use the definition of $\hat h_d$ in \eqref{uninear3} together
with \eqref{dictdp} and the relation
\begin{equation}
R_0^{\frac{5-p}{2}}=\frac{2 (2\pi)^{\frac{5-p}{2}} l \hat T}{(6-p) \sqrt{(6-p) \Omega_{8-p}}}
~.
\end{equation}
}
On the other hand, the typical mass of the winding modes is
$M_w \sim \frac{R_0^{(7-p)/2}L^{1/2}}{\sqrt{\alpha'} \hat h^{1/4}}$, so the effects of
the winding modes can be neglected when $\ell_H M_w \gg 1$, $i.e$ when
$R_0 \gg 1$. This is equivalent to $\hat L \hat T \gg (\lambda \hat L^{3-p})^{-\frac{1}{2}}$.
Hence, when $p=1,2$ and $\lambda'=\lambda \hat L^{3-p} \gg 1$
the supergravity solution \ref{p+1} is valid for a large range of temperatures $t'$.
For $p=3$ the supergravity solution is valid when
${\lambda'}^{-\frac{1}{2}}\ll t' \ll 1$ and for
$p=4$ when ${\lambda'}^{-\frac{1}{2}}\ll  t' \ll {\lambda'}^{-1}$.
The latter condition implies $\lambda' \ll 1$.

In a similar fashion, the T-dual background of smeared D$(p-1)$ branes
(see Eqs.\ \eqref{uninear1}-\eqref{uninear3})
is valid ($i.e.$ stringy and winding mode corrections can be ignored)
when $t' \ll {\lambda'}^{\frac{1}{3-p}}$, $t' \ll 1$.

Having said this, we are now ready to use the supergravity results
of Section \ref{sec:sugra} to make definite predictions for
the phase diagram of the dual gauge theory at strong
coupling. For that purpose we recast the thermodynamics of Section
\ref{sec:sugra} in terms of gauge theory variables. In \eqref{epsrdef}
we defined and computed the rescaled energy, temperature, entropy
and free energy quantities $\epsilon$, $\hmt$, $\hms$ and
$\hmf \equiv \epsilon - \hmt \hms$. In terms of these variables
we obtain the dual gauge theory entropy and free energy $\hat S$ and
$\hat F$ by reinstating the dimensions
\begin{equation}
\label{physsf} \hat S (E) = \frac{l}{g} \hms ( g E) \spa
 \hat{F} (\hat T) = \frac{1}{g} \hmf ( l \hat T)
~.
\end{equation}
In the ensuing we concentrate mostly on the canonical ensemble
and summarize the phases of ($p+1$)-dimensional SYM theory on
$\R^{p-1,1} \times S^1$ that follow from the uniform, localized and
non-uniform phases of near-extremal D$(p-1)$-branes in supergravity.

\subsubsection{Uniform phase: High temperature phase of SYM on a circle}

The uniform phase of near-extremal D$(p-1)$-branes on a circle
corresponds to the high temperature phase of ($p+1$)-dimensional
SYM theory on $\R^{p-1,1} \times S^1$. This is physically sensible
because high temperatures correspond to short distances. In this regime
the details of the compact direction in $\R^{p-1,1} \times S^1$ are invisible
and the circumference $\hat{L}$ appears only
as a trivial proportionality factor in the free energy. This
is precisely what we find in the uniform phase that corresponds
to a uniformly smeared near-extremal D$(p-1)$-brane (recall that this
solution has the
same thermodynamics as a near-extremal D$p$-brane wrapping the T-dual circle).
We can check this statement explicitly using the thermodynamics
\eqref{unifr} and the dictionary \eqref{physsf}. The
resulting free energy is
\begin{equation}
\label{unidp}
\hat{F}^{\rm u}_{{\rm SYM}(p+1)} (\hat T) =
\hat{F}^{\rm u}_{{\rm D}(p-1)}(\hat T) = - k_{p}V_{p-1} \hat L  N^2
\lambda^{- \frac{3-p}{5-p}} \hat T^{2 \frac{7-p}{5-p}} \ ,
\end{equation}
where the coefficients $k_p$ are given in Table \ref{tabcoef}. The
expression \eqref{unidp} is indeed the familiar result for the free
energy of the D$p$-brane theory, with $V_{p} = V_{p-1} \hat L$.

\subsubsection{Localized phase: Low temperature phase of SYM on a circle}

The localized phase corresponds to the low
temperature phase of ($p+1$)-dimensional SYM theory on
$\R^{p-1,1} \times S^1$. Low temperatures correspond to large distances
in the SYM theory, so in this regime the presence of the circle
enters in a non-trivial manner and the whole tower of Kaluza-Klein
states can contribute. We will see that as we send the temperature
to zero, the leading order thermodynamics is captured by the thermodynamics
of the near-extremal D$(p-1)$-brane, which is dual to the strongly coupled
$p$-dimensional SYM theory on $\R^{p-1,1}$. The next-to-leading order
corrections to the thermodynamics of the localized phase
(see Section \ref{sec:nonphases}) will lead to gauge theory corrections
that capture the contribution of the Kaluza-Klein modes on the circle.

Using the expressions for the corrected free energy in
\eqref{freene} together with the dictionary \eqref{dictdp}, \eqref{physsf}
we deduce that the free energy of the localized phase of thermal
($p+1$)-dimensional SYM theory on $\R^{p-1,1} \times S^1$ is
\begin{eqnarray}
\label{locdp}
 & & \hat{F}^{\rm loc}_{{\rm SYM}(p+1)} (\hat T) =
\hat{F}^{\rm loc}_{{\rm D}(p-1)}(\hat T) \\ \nn && \simeq -
k_{p-1}V_{p-1} N^2 \left( \frac{\lambda}{\hat L} \right)^{-
\frac{4-p}{6-p}} \hat T^{2 \frac{8-p}{6-p}} \left\{ 1 + \frac{2
(9-p) \zeta (8-p) k_{p-1} }{(6-p)^2 (2\pi)^3 \Omega_{9-p}} \left[
\hat L \hat T  \sqrt{ \lambda\hat L^{3-p}}
\right]^{2\frac{8-p}{6-p}}  \right\} \ .
\end{eqnarray}
The coefficients $k_p$ are the same as above and are listed in Table
\ref{tabcoef}.%
\footnote{These coefficients are related to $\hat K_1^{(d)}$ defined
in \eqref{K1def} via the relation $\hat K_1^{(9-p)}
(2\pi)^{2\frac{4-p}{5-p}} = k_p$.}

\begin{table}
\begin{center}
\begin{tabular}{|c||c|c|c|c|c|}
\hline $p$ & 0 & 1 & 2 & 3 & 4 \\ \hline $k_p$ & $(2^{21} 3^2 5^7
7^{-19} \pi^{14})^{1/5} $ & $2^4 3^{-4} \pi^{5/2}  $ & $ (2^{13} 3^5
5^{-13} \pi^8)^{1/3}$ & $2^{-3} \pi^2 $ & $2^5 3^{-7} \pi^2$ \\
\hline
\end{tabular}
\end{center}
\caption{Coefficients for the free energy of D$p$-branes.
\label{tabcoef} }
\end{table}

Notice that the leading term in \eqref{locdp} is in
perfect agreement with the expected result for the strong coupling
limit of the $p$-dimensional SYM theory. In particular,
the 't Hooft coupling appears through the combination $\lambda/\hat L$, which
is precisely what we expect to find by
compactifying the $(p+1)$-dimensional theory with 't Hooft coupling $\lambda$.
The second term in \eqref{locdp} gives a quantitative prediction for the first
correction in terms of the dimensionless parameter
\begin{equation}
\delta = \hat L \hat T \sqrt{\lambda \hat L^{3-p}} \ .
\end{equation}
The next corrections will be of order
$(\delta^{2\frac{8-p}{6-p}})^2$. It would be very interesting
to reproduce these corrections from the field theory side.
We refer the reader to \cite{Harmark:2004ws}
for the explicit expressions of \eqref{locdp} in the cases
$p=1,2,3$ and 4.

Similar results in the microcanonical
ensemble can be easily obtained using the corrected entropy
\eqref{entne} of the localized near-extremal branes and
\eqref{dictdp}, \eqref{physsf}.

\subsubsection{Non-uniform phase: New phase in SYM on a circle}

The non-uniform phase of near-extremal $D(p-1)$-branes on
a circle (see Section \ref{sec:nonphases}) predicts
the existence of a new phase of ($p+1$)-dimensional
SYM theory on $\R^{p-1,1} \times S^1$ at intermediate temperatures.
With the use of \eqref{freeen} and the dictionary
\eqref{dictdp}, \eqref{physsf} we can also deduce
the first correction to the thermodynamics around the
point where the non-uniform phase connects to the uniform phase.
One finds the free energy
\begin{eqnarray}
\label{nudp} & & F^{\rm nu}_{{\rm SYM}(p+1)} (\hat T) = F^{\rm
nu}_{{\rm D}(p-1)}(\hat T) \\ \nn & & \simeq -(2\pi)^3 V_{p-1} N^2
\lambda \hat{L}^{3-2p}
 \left[
\frac{5-p}{9-p} \epsilon_c + \hmt_c \hms_c \left( \frac{\hat T}{\hat
T_c} -1\right) + \frac{\hmt_c c}{2} \left( \frac{\hat T}{\hat T_c}
-1\right)^2  \right] \ .
\end{eqnarray}
The critical temperature in this expression is given by
\begin{equation}
\hat T_c =\frac{\hmt_c}{\hat L} \sqrt{\frac{2\pi}{\lambda  \hat
L^{3-p}}}  \ .
\end{equation}
and the values for $\hmt_c$, $\epsilon_c$, $\hms_c$ and $c$ can be
read off from Table \ref{tabnear} by using $d=10-p$.
Notice that the first term in the expression \eqref{nudp} corresponds to  $\hat
F^{\rm u}_{{\rm SYM}(p+1)}(\hat T_c)$, $i.e.$ the free energy
\eqref{unidp} of the uniform phase evaluated at the critical
temperature. The second term is $-\hat S^{\rm u}_{{\rm
SYM}(p+1)}(\hat T_c) (\hat T - \hat T_c)$ and therefore involves the
entropy of the uniform phase evaluated at the critical temperature.
The third term contains the departure of the free energy from the
uniform phase into the non-uniform phase.
It would be very interesting to understand the results
\eqref{locdp} and  \eqref{nudp} for thermal $(p+1)$-dimensional SYM
theory on $\R \times \R^{p-1} \times S^1$ from the gauge theory
side.

Again, similar results for the non-uniform phase in the microcanonical
ensemble can be obtained using the corrected entropy \eqref{ssu} of
the localized
near-extremal branes and \eqref{physsf}, \eqref{dictdp}.

Finally, since both the localized and non-uniform phase of near-extremal
branes have copies, we also have copies of the thermal
$(p+1)$-dimensional SYM phases obtained above.
See Ref.~\cite{Harmark:2004ws} for further details.

\subsubsection{The expected T(E) diagram for D-branes on a circle}

As we discussed above, the thermodynamics of near-extremal D$p$-branes on a
transverse circle, is related to the thermodynamics of
phases of Kaluza-Klein black holes in $10-p$ dimensions. In particular,
for the cases of interest in this section, namely D0, D1, D2 and D3-branes,
we need the phases of Kaluza-Klein black holes in 10, 9, 8 and 7
dimensions. However, at present we do not have complete analytical or
numerical data (see, however \cite{Sorkin:2006wp}) for the
localized and non-uniform phases in those dimensions.
For this reason, our results so far ($c.f.$ the free energies
\eqref{locdp} and \eqref{nudp}
for the localized and non-uniform phase respectively) have been leading order
results around some specific point in the phase diagram. For the localized phase
this point is the origin (large radius limit) and for the non-uniform
phase it is the
GL point. For the uniform phase we have of course the exact result
\eqref{unidp}.

Nevertheless, based on the expectation that the $(\mu,n)$ phase diagram
for neutral Kaluza-Klein black holes in those dimensions
has a similar form as those in 5 and 6 dimensions (see Fig.~\ref{fig1}) we
can use the map \eqref{nemap} to make some general observations
\cite{Harmark:2005dt}.
For example, the map implies that the $(E,r)$ phase diagram for the
near-extremal D$p$-branes has the same qualitative form as that in
Fig.~\ref{fig1}.
From this we get a prediction for the qualitative form of the $T(E)$
diagram, because
given the curves in the $(E,r)$ phase diagram the entire
thermodynamics is determined.

For instance, consider the case of $N$ near-extremal D2-branes on a
transverse circle, which is dual to finite temperature $\NN=4$
supersymmetric $SU(N)$ Yang-Mills on $\R^2\times \T^2$.
A similar qualitative picture is expected also for the other D$p$-brane
cases. In the uniform phase, which consists of near-extremal D2-branes
uniformly smeared on a transverse circle, the relation between temperature
and energy is
\begin{equation}
 T(E) \propto E^{\frac{1}{4}} \ ,
\end{equation}
We argued that the uniform phase dominates at high energies.
In the localized branch which dominates at low energies
we know \cite{Harmark:2004ws} that the temperature/energy relation is
\begin{equation}
T(E) \propto E^{\frac{3}{10}} ( 1 + \alpha E )  \spa E \ll 1 \ ,
\end{equation}
with $\alpha < 0$ a known constant. Finally, there is a non-uniform
phase which emanates from the uniform phase at the
critical energy $E_{\rm c}=  0.71 \cdot (2\pi)^3 V_2 N^2/(\lambda
\hat L^3)  $. At the critical point, the ratio of the
specific heat of the non-uniform phase to the specific heat of the
uniform phase is \cite{Harmark:2004ws}
\begin{equation}
\label{cnu} \frac{c_{\rm nu}}{c_{\rm u}} = \frac{1}{1 + \frac{6}{7}
\hat \gamma \epsilon_c} = 0.41 \  .
\end{equation}
The ratio is less than one, a feature that holds for any of the
D$p$-brane cases mentioned above.

\begin{figure}[ht]
\centerline{ \epsfig{file=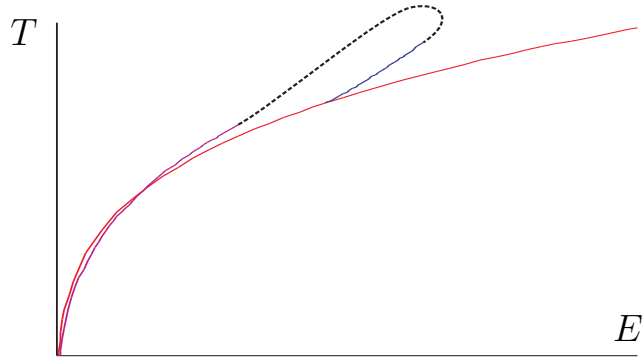,width=8 cm,height=5cm} }
\caption{The qualitative picture of energy versus temperature for the
phases of near-extremal D$p$-branes on a transverse circle. One can see
the uniform (red), non-uniform (blue) and localized (magenta)
phases. The dashed curve connects the localized and non-uniform
phase. \label{figdp}}
\begin{picture}(0,0)(0,0)
\put(87,215){\Large $T$} \put(315,103){\Large $E$}
\end{picture}
\end{figure}

The salient features of the emerging picture in the $T(E)$ diagram
are the following
\begin{itemize}
\item Both the uniform and localized phases begin at the origin (with
infinite slope)
and at very low energies the uniform phase has higher temperature
than the localized phase (at a given energy). As the energy is increased
the localized phase is expected to cross the uniform phase at
relatively low energies below the critical energy where
the non-uniform phase emerges.
\item The slope in the $T(E)$ diagram is the inverse of the specific heat.
Therefore we deduce from \eqref{cnu} that the slope of the non-uniform phase
at the critical point $(E,T)=(E_{\rm c}, T_{\rm c})$ is larger than
that of the uniform phase.
\item It is natural to assume that the localized and non-uniform branches
meet at a merger point in a qualitatively similar way to the one
observed explicitly
in the NS5-brane case in \cite{Harmark:2005dt} (see Fig.~\ref{figm5te}).
\end{itemize}
The emerging picture is summarized in Fig.~\ref{figdp}.

We notice that as we increase the energy above extremality in the localized
phase we first reach a point where the specific heat becomes infinite.
This is the point where the localized phase attains its maximal
temperature. Increasing the energy further we reach a point
where the specific heat vanishes. This is the point where the localized
phase attains its maximal energy.%
\footnote{The conditions for infinite and zero specific heat are
in this case $E r'(E)= (5-p)/2 - r(E)$ and $1/r'(E) =0 $ respectively.}
Despite the presence of a maximal temperature in the localized phase,
the uniform phase exists for any temperature and therefore no limiting
temperature exists for the system of near-extremal branes on a circle.
Hence, for sufficiently high temperatures the system will necessarily
find itself in the uniform phase. At low temperatures the localized
phase will be favored.

The above behavior is the same in both the canonical and
microcanonical ensembles.
In Section \ref{sec:LST} we will see that the case of NS5-branes is
very different
in this respect. In that case the uniform phase has a constant
Hagedorn temperature
and there is a striking difference between the canonical and
microcanonical ensembles.

\subsubsection{From strong to weak gauge coupling}

So far we have obtained a set of predictions for strongly coupled
$(p+1)$-dimensional
SYM theory on a circle by analyzing the supergravity phases of
near-extremal black hole
solutions on a transverse circle. It is interesting to ask what happens
to all these phases in gauge theory when we continue the theory from
strong to weak 't Hooft coupling $\lambda$. If there is a line of
phase transitions
at strong coupling, this line is not expected to end abruptly in the
$(\lambda, T)$ parameter
space as we change the coupling $\lambda$. It may either go to zero or infinite
temperature at a finite value of the coupling, or it may continue smoothly
to weak coupling, where one can analyze it with standard perturbative
methods directly in gauge theory. An example of the latter possibility is
the Hawking-Page transition \cite{Hawking:1983dh,Witten:1998zw}
in $AdS_5$, which is dual to a thermal deconfinement
transition in $(3+1)$-dimensional $\NN=4$ SYM theory on $S^3$. This transition
can indeed be continued to weak coupling where it can be analyzed explicitly
\cite{Sundborg:1999ue,Aharony:2003sx,Aharony:2005bq}. A similar
continuation to weak coupling also occurs for the case of interest in
this section,
namely the Gregory-Laflamme instability%
\footnote{In another recent example, Ref.~\cite{Hollowood:2006xb} found
the weak coupling version of the Gregory-Laflamme localization instability of the
small $AdS_5$ black hole.}
 and the associated transition in the dual
SYM theories on a circle. This has been shown for the $(1+1)$-dimensional
maximally supersymmetric YM theory in \cite{Aharony:2004ig}.
We review the main features of this example in the ensuing.

The Euclidean $(1+1)$-dimensional $SU(N)$ SYM theory on the torus
$S^1_\beta \times S^1$ has two non-contractible Wilson loops, the loops
running around the time $\tau$ and circle $z$. The spatial Wilson loop
\begin{equation}
\label{Wilsonz}
\PP_z = Pe^{i \oint dz A_z}
\end{equation}
is of special interest here, because it serves as an order parameter for the
weak coupling manifestation of the Gregory-Laflamme transition.
Indeed, the eigenvalues of $\PP_z$ represent the positions of T-dual
D0-branes on a circle and their distribution can distinguish between
the three different phases. The uniform phase corresponds to a uniform
eigenvalue distribution, the non-uniform phase corresponds to a non-uniform
eigenvalue distribution that breaks the translational invariance on
the transverse
circle and the localized phase maps to an eigenvalue distribution,
which is localized
on the transverse circle. \cite{Aharony:2004ig} showed that the eigenvalue
distribution of the Wilson loop $\PP_z$ is sharply localized at small
temperatures
and weak coupling. As we continue the theory to higher temperatures
the eigenvalues spread out and fill out the whole circle. There is a critical
temperature $T_c \simeq 1.4/(\lambda \hat L^3)$ where the theory undergoes
a Gross-Witten like phase transition \cite{Gross:1980he} from the localized
distribution to the uniform distribution. For more details we refer the reader
to the original discussion of the Refs.~\cite{Aharony:2004ig,Aharony:2005ew}.

\subsubsection{Implications of the GL instability for higher tori}

Above we considered the implications for Yang-Mills theory on a
circle due to the Gregory-Laflamme instability of D-branes smeared
on a transverse circle. Similar results can be obtained
for Yang-Mills theory on a torus $\T^k$ by
analyzing the Gregory-Laflamme instability of D-branes
smeared on a transverse torus $\T^k$.

As a special case, consider the $\CN=4$ SYM theory on $\R \times \T^3$.
The holographic dual of this theory arises from
D0-branes smeared on the three-torus $\T^3$. This brane
configuration has Gregory-Laflamme instabilities in the periodic
directions when the radii are sufficiently large. This implies an
interesting phase structure which generalizes the phase structure of
non-extremal D-branes on a circle presented in Section
\ref{sec:nonphases}. First, for $\T^3 = S^1 \times \T^2$
we can obtain several phases for the D0-branes on $\T^3$
from the phases of D2-branes on a circle with T-duality, imposing that
there is no dependence on the $\T^2$ directions. Thus, we
inherit all the phases of the D2-brane on a circle. Moreover, it is
conceivable that there are new localized phases, for example, phases in
which the D0-brane horizon is topologically $S^3 \times S^5$
corresponding to having the D0-branes localized in all three torus
directions, or one could have that the D0-brane is localized in two
directions giving an $S^1 \times S^2 \times S^5$ horizon. In
addition, one can well imagine new non-uniform phases, for example
in which the D0-branes are not localized in any of the three
directions but nevertheless are non-uniform in all three directions.
This would correspond to a $\T^3 \times S^5$ topology of the
horizon. Therefore, we see that the phase structure of D0-branes on
$\T^3$ is even richer than that of D-branes on a circle. This is in
accordance with the phase structure of neutral $p$-branes on tori as
discussed in Section \ref{sec:highdspaces}.
In particular, via the boost/U-duality map
discussed in Section \ref{sec:nonphases} one can directly relate the phases
of neutral Kaluza-Klein black holes on $\CM^{1,6} \times \T^3$
to those of non- and near-extremal D0-branes smeared on $\T^3$.
Through holography, the richness of the phase structure for D0-branes on
$\T^3$ implies an equally rich phase structure for $\CN=4$ SYM theory
on $\R \times \T^3$. We refer to \cite{Aharony:2005ew} for more
considerations on the phase structure of gauge theories on tori.

\subsection{M2- and M5-branes on a circle}
\label{sec:M2M5}

In this section we consider separately the cases of near-extremal
M2- and M5-branes on a transverse circle. By IIA/M S-duality, the former
is dual to the near-extremal D2-brane and the latter is dual to
the near-extremal NS5-brane. Hence, it is conjectured that
$N \gg 1$ coincident near-extremal M2-branes on a circle are dual
to (2+1)-dimensional SYM theory with sixteen supercharges and
gauge group $SU(N)$ \cite{Maldacena:1997re,Itzhaki:1998dd}.
Similarly, the system of $N \gg 1$ concident near-extremal M5-branes
on a transverse circle is dual to the six-dimensional $(2,0)$ LST.
In both cases, the transverse circle is the M-theory circle and the
dual non-gravitational theory lives on the non-compact space
$\R^{p,1}$. $p=2$ for the case of the M2-brane and $p=5$ for the
case of the M5-brane.

\subsubsection{$(2+1)$-dimensional SYM theory}
\label{sec:3dimSYM}

In this case we place $N$ near-extremal M2-branes transverse
to the M-theory circle, which we take to have circumference
$L = 2\pi g_s l_s$. Using the quantization condition that comes from
the M2-branes together with \eqref{nelimit} and \eqref{Hhat} we find
that we can recast the supergravity parameters $g$, $l$ as
\begin{equation}
\label{m2gl} g = \frac{(2\pi)^2}{V_2} \frac{N^3}{\lambda^3} \spa l =
2\pi \frac{N^{3/2}}{\lambda} \ ,
\end{equation}
where $\lambda = \gym^2 N = g_s l_s^{-1} N$. As in the
cases of the previous subsection, these relations provide a
dictionary between the supergravity results of the near-extremal
M2-branes on a circle and the dual $(2+1)$-dimensional SYM theory.
Since there are many similarities between this case and the case
of D$p$-branes on a transverse circle of the previous subsection,
we will provide here only a brief summary of the main results
and refer the reader to \cite{Harmark:2004ws} for a more detailed
discussion. Qualitatively similar results can be obtained in both
the microcanonical and canonical ensembles.

The uniform phase of near-extremal M2-branes on a circle corresponds
to the high temperature/high energy phase of thermal
$(2+1)$-dimensional SYM. The localized phase corresponds to the low
temperature/low energy phase. At leading order this phase is captured
by the near-extremal M2-brane solution and is dual to the infrared
fixed point of
$(2+1)$-dimensional SYM which is a superconformal field theory with
$SO(8)$ R-symmetry.
Using the results of Section \ref{sec:sugra} and the dictionary of
eqs.\ \eqref{physsf}
and \eqref{m2gl} we can compute the free energy for both of these phases.
In the uniform phase the result reproduces the free energy of the near-extremal
D2-brane theory. In the localized phase the leading order result reproduces
the free energy of $N$ near-extremal M2-branes. The small black hole
corrections
capture the non-trivial effects of the theory as we move away from the infrared
fixed point.

The phase diagram of the near-extremal M2-brane on a circle also
predicts the existence of a new non-uniform phase of uncompactified
($2+1$)-dimensional SYM theory on $\R^{2,1}$ at intermediate
temperatures. This phase emerges from the uniform phase at a
critical temperature and presumably connects to the localized phase
at a merger point. The first correction to the thermodynamics
around the critical point of the uniform phase can be computed
from the available supergravity data as above. It would be interesting
to understand these results from the gauge theory side.
Finally, since both the localized and the non-uniform phase of near-extremal
branes have copies, there are similar copies of the thermal
uncompactified ($2+1$)-dimensional SYM phases discussed above.

\subsubsection{Little String Theory}
\label{sec:LST}

In the decoupling limit of $N$ coincident NS5-branes in type IIA string
theory the string length $l_s$ is kept fixed and the string coupling $g_s$ is
sent to zero \cite{Maldacena:1997cg,Aharony:1998ub}. In this limit
the dynamics of the theory is believed to reduce to a string theory
without gravity called Little String Theory (LST), or more precisely
$(2,0)$ LST of type $A_{N-1}$ \cite{Berkooz:1997cq,Seiberg:1997zk,
Dijkgraaf:1997ku,Aharony:1999ks,Kutasov:2001uf}.\footnote{A similar
decoupling limit of NS5-branes in type IIB string theory gives
the $(1,1)$ LST of type $A_{N-1}$. In this review we will focus on the
type IIA NS5-branes, which are related more directly to the transverse
M5-branes of M-theory.}
In this subsection (see also \cite{Harmark:2005pq,Harmark:2005dt})
we will consider the thermodynamics
of the near-extremal NS5-brane in type IIA string theory using as a starting
point the S-dual description as a near-extremal M5-brane on a transverse circle
and the input of the phases obtained from six-dimensional neutral
Kaluza-Klein black holes in Section \ref{sec:KKphases}.
Other interesting work on the thermodynamics of LST can be found
in \cite{Maldacena:1996ya,Maldacena:1997cg,Harmark:2000hw,
Berkooz:2000mz,Kutasov:2000jp,Rangamani:2001ir,Buchel:2001dg,
Narayan:2001dr,DeBoer:2003dd,Aharony:2004xn,Parnachev:2005hh}.

The thermodynamics of the NS5-brane exhibit a variety of new features
compared to the other cases analyzed earlier in this section. The
usual near-extremal NS5-background (dual to M5-branes uniformly
smeared on a transverse circle) has entropy as function of energy
$S(E)=T_{\rm hg}^{-1} E$, where $T_{\rm hg}$ is a fixed Hagedorn
temperature. In \cite{Harmark:2000hw,Berkooz:2000mz}
the idea was put forward that the singular thermodynamics of the
NS5-brane background could be resolved by computing a string one-loop
correction to the background. This was done in \cite{Kutasov:2000jp},
where it was shown that the corrected background has $T>T_{\rm hg}$
and negative specific heat. In a subsequent analysis the results of
\cite{Kutasov:2000jp} were interpreted to indicate that one does not
have a Hagedorn phase transition and that the Hagedorn temperature is
a limiting temperature \cite{Aharony:2004xn}.

Here we will see, following \cite{Harmark:2005dt}, that in the canonical
ensemble the usual near-extremal M5-brane background is thermodynamically
subdominant at tree-level to a background of near-extremal M5-branes localized
on the transverse circle. The new localized phase exhibits a maximal
temperature and it remains unclear what happens if one tries to heat up
the type IIA NS5-brane to a temperature higher than this maximal temperature.
A related additional new feature of the NS5-brane is the fact that the
canonical and microcanonical ensembles seem to be inequivalent at large
energies. In what follows we will review separately the basic features of each
ensemble.

\subsubsection*{\sl Canonical ensemble}

\begin{figure}[ht]
\centerline{\epsfig{file=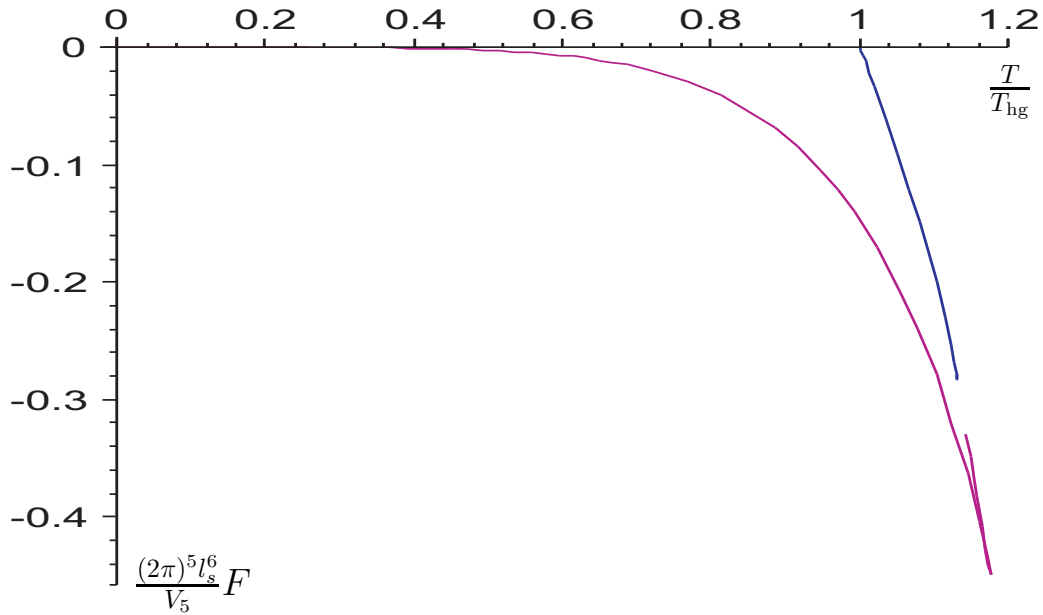,width=14cm,height=8cm} }
\caption{Free energy versus temperature for the near-extremal
NS5-branes. Displayed are the non-uniform (blue) and localized
(magenta) phases. The uniform phase is located at the point
$(T,F)=(T_{\rm hg},0)$. \label{figm5temp} }
\begin{picture}(0,0)(0,0)
\put(388,263){\Large $\frac{T}{T_{\rm hg}}$} \put(65,75){\Large
$\frac{(2\pi)^5 l_s^6}{V_5} F$}
\end{picture}
\end{figure}

We begin with the canonical ensemble, since this is the ensemble that exhibits
the most interesting results. The free energy versus temperature
is displayed in Fig.~\ref{figm5temp}.

Before considering the new data for the localized phase, we
list the main properties of the three different phases
(already stated in \cite{Harmark:2004ws}) that arise from the
corresponding phases of neutral Kaluza-Klein black holes:%
\footnote{Note that the non-uniform and localized phase have copies
with free energy  $F' = F/k^2$ and temperature $T' = T$
($k=2,3,...$). We do not consider the copied phases here since they
are subdominant in this ensemble.}
\begin{itemize}
\item {\sl Uniform phase.} This phase has zero free
energy and a fixed temperature $T = T_{\rm hg}$ and captures
the well-known Hagedorn-like behavior of LST.
It corresponds to a single point in the free energy versus temperature
diagram in Figure \ref{figm5temp}.
\item {\sl Non-uniform phase.} This phase emerges out of the point
$(T,F) = (T_{\rm hg},0)$ and exhibits increasing temperature and
decreasing free energy. It has positive specific heat. For
temperatures near the Hagedorn temperature $0 \leq T-T_{\rm hg} \ll
T_{\rm hg}$, one finds
\begin{equation}
F_{\rm nu} (T ) \simeq - 2 \pi V_5 N^3  T_{\rm hg}^6 \left[ 1.54
\cdot \left( \frac{ T}{ T_{\rm hg}} - 1 \right) + 3.23 \cdot \left(
\frac{ T}{ T_{\rm hg}} - 1 \right)^2 \right] \ .
\end{equation}
This expression has been computed in \cite{Harmark:2004ws}
using results of \cite{Sorkin:2004qq}.
\item
{\sl Localized phase.} The localized phase begins at the point
$(T,F)=(0,0)$, which corresponds to the $(2,0)$ SCFT, the
infrared fixed point of the type IIA near-extremal NS5-brane. The
first correction to the thermodynamics as we raise the temperature and
move away from the infrared fixed point is
 \cite{Harmark:2004ws}
\begin{equation}
F_{\rm loc} (T) = - \frac{2^6 \pi^3}{3^7} V_5 N^{3}  T^6 \left[ 1 +
 \frac{2^5 \zeta (3)}{3^6}
 \frac{T^6}{T_{\rm hg}^6} + {\cal{O}}\left( \frac{ T^{12}}{T_{\rm hg}^{12}}
 \right) \right] \ .
\end{equation}
This has been computed using the results of \cite{Harmark:2003yz}.
\end{itemize}

One can say more about the localized phase by using the numerical data of
\cite{Kudoh:2004hs}. The localized phase begins at the point
$(T,F)=(0,0)$ and has decreasing free energy until the maximum
temperature $T_{\rm \star} = 1.177 \cdot T_{\rm hg}$.
At this point the curve reverses direction and has decreasing
temperature and increasing free energy.

\begin{figure}[ht]
\centerline{\epsfig{file=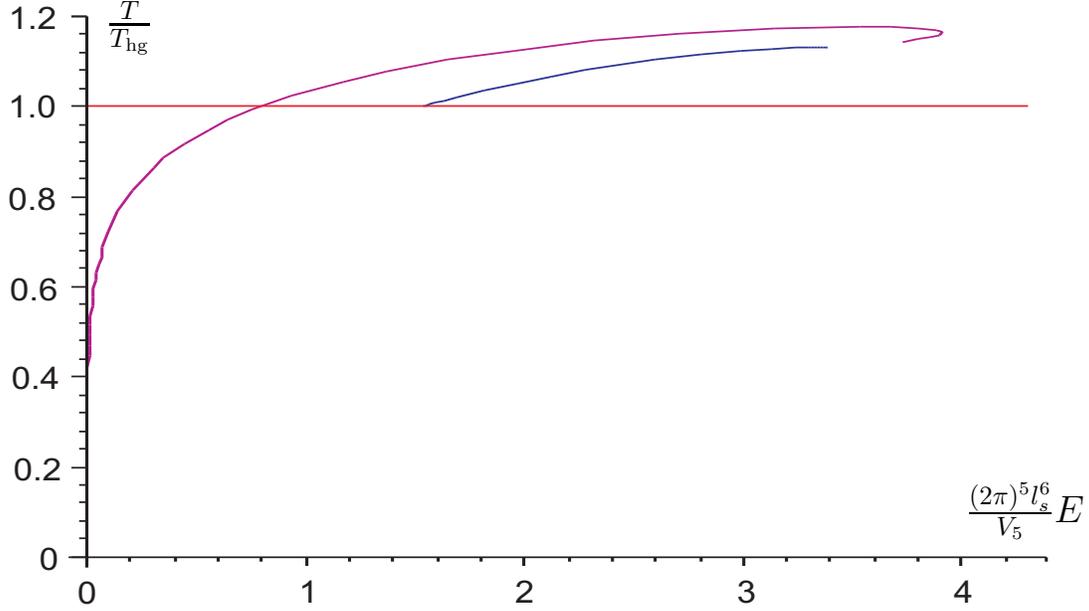,width=14cm,height=8cm} }
\caption{Temperature versus energy for the near-extremal NS5-branes.
Displayed are the uniform (red), non-uniform (blue) and localized
(magenta) phases. \label{figm5te} }
\begin{picture}(0,0)(0,0)
\put(50,275){\Large $\frac{T}{T_{\rm hg}}$} \put(375,90){\Large
$\frac{(2\pi)^5 l_s^6}{V_5} E$}
\end{picture}
\end{figure}

The most intriguing part of the localized phase is evidently the
behavior near the maximal temperature $T_{\rm \star}$. One way to
understand the behavior around this point is to consider the
temperature versus energy diagram in Fig.~\ref{figm5te}.
In this diagram the localized phase is apparently a smooth curve.
Hence, the entropy is continuous at  $T \simeq T_{\rm \star}$.
However, since the temperature is maximal at $T= T_{\rm \star}$ the
$(E,T)$ curve has a horizontal tangent at that point.
This implies that the localized phase has positive
specific heat as $T$ increases from zero to $T_{\rm \star}$ which
becomes infinite at the maximal temperature $T_{\rm \star}$.
Continuing along the same curve in the $(E,T)$ diagram we begin
with minus infinite specific heat and continue with negative specific heat
until the curve develops a vertical tangent. At that point the
specific heat becomes
zero. For the remaining part of the curve the specific heat is
positive.%
\footnote{In the $(E,r)$ phase diagram the point with
infinite specific heat is characterized by the value of $r$ where
$E r'(E) = 1/2-r(E)$. Zero specific heat occurs when $r'(E)$
is infinite. It is not difficult to see from the phase diagram in
Fig.~1 in \cite{Harmark:2005dt} that for each of these two equations
there exists a particular  solution of $r$ on the localized branch.}

Let us consider more closely the thermodynamics around
the critical temperature $T_{\star}$. From the available data we get
\begin{equation}
T \simeq T_{\star} - c (E - E_{\star} )^2  \spa c = 0.038 \spa
\frac{E_{\star}}{V_5} = 3.60 \cdot (2\pi )^{-5} l_s^{-6} \ .
\end{equation}
From this expression we can obtain the free energy as a function of
temperature near the singularity
\begin{equation}
\label{Fnearmax}
\begin{array}{c} \ds
F (T) \simeq F_{\star} - S_{\star} (T - T_{\star}) \pm \frac{2}{3
\sqrt{c} T_{\star}} (T- T_{\star})^{3/2} \ ,
\\[3mm] \ds
\frac{S_{\star}}{V_5} = 3.5 \cdot (2\pi)^{-4} \sqrt{N} \spa
\frac{F_{\star}}{V_5} \equiv \frac{1}{V_5}(E_{\star} - S_{\star} T_{\star}) =
- 0.53 \cdot (2\pi)^{-5} l_s^{-6} \ .
\end{array}
\end{equation}
In the first equation the $\pm$ sign corresponds to the two branches
on either side of the maximum temperature point. The minus
branch is the one that has positive specific heat (for this branch
$E \lesssim E_{\star}$) and the plus branch is the one that has
negative specific heat (for this branch $E \gtrsim E_{\star}$).
It is immediately clear from \eqref{Fnearmax} that the entropy is indeed
finite and continuous at the critical point, whereas the heat capacity
diverges and changes sign.

Also it is clear from Fig.~\ref{figm5temp} that there are temperatures
where more than one phases are available. The dominant phase
will be the one with the lowest free energy.\footnote{Indeed,
for two phases 1 and 2 at the same temperature with free energies
$F_1<F_2$ the Euclidean path integral $Z$ will be dominated
by phase 1, since $Z \simeq e^{ -\beta F_1} + e^{ -\beta F_2} =
e^{-\beta F_1} ( 1 +e^{-(F_2 - F_1)}) \simeq e^{-\beta F_1}$ at
the large $N$ limit.} It corresponds to the localized phase, which lies
within the energy range $0 \leq E \leq E_{\star}$.
Hence, in the canonical ensemble we conclude that the system is
(globally) thermodynamically unstable in the uniform phase
(at zero free energy and temperature $T=T_{\rm hg}$)
and will decay to the localized solution at the same temperature.
In the dominant, localized phase the heat capacity will increase
as we increase the temperature and eventually will become
infinite at the maximal temparature $T=T_{\rm \star}$.
This implies that $T_{\star}$ is a limiting temperature. Note
that this limiting temperature exceeds the temperature $T_{\rm hg}$
associated with the uniform phase.

\subsubsection*{\sl Microcanonical ensemble}

\begin{figure}[ht]
\centerline{\epsfig{file=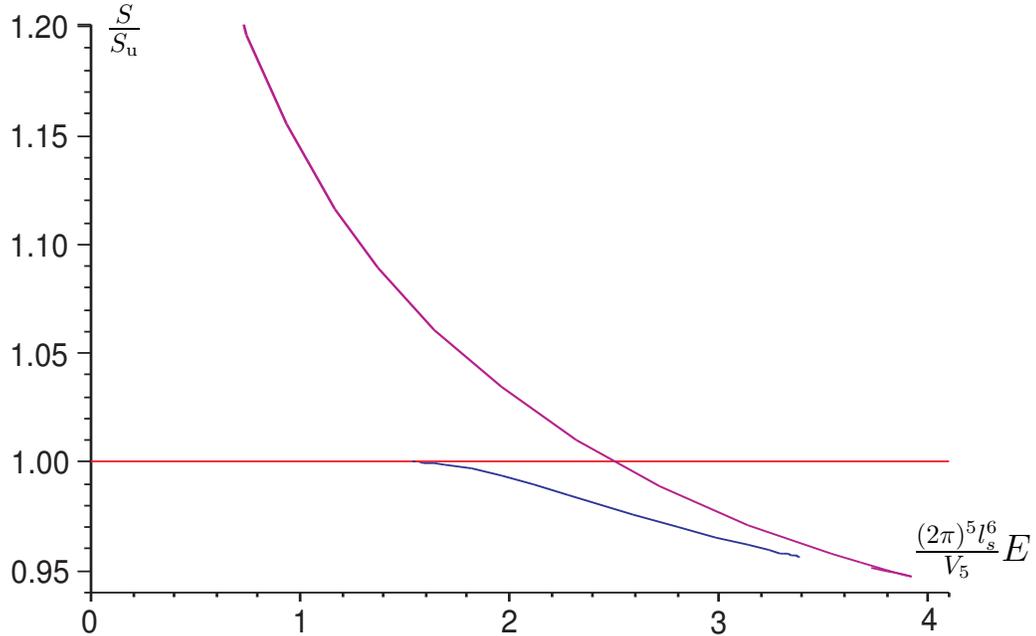,width=13cm,height=9cm} }
\caption{Entropy versus energy for the near-extremal NS5-branes. On
the vertical axis we plot the ratio $S(E)/S_{\rm u}(E)$ and on the
horizontal axis $E/E_{\rm c}$. $S_{\rm u}(E)$ denotes the entropy
function of the uniform phase of near-extremal NS5-branes.
Displayed are the uniform (red), non-uniform (blue) and localized
(magenta) phases.  \label{figm5ent} }
\begin{picture}(0,0)(0,0)
\put(73,328){\Large $\frac{S}{S_{\rm u}}$} \put(378,130){\Large
$\frac{(2\pi)^5 l_s^6}{V_5} E$}
\end{picture}
\end{figure}

We now turn to the microcanonical ensemble. The entropy
of the three different phases as a function of energy
is displayed in Fig.~\ref{figm5ent}.
The main properties of these phases are%
\footnote{The non-uniform and localized phase have copies
with entropy  $S' = S/k^2$ and energy $E' = E/k^2$ ($k=2,3,...$). We
will not consider the copied phases here since they are subdominant in
this ensemble.}

\begin{itemize}
\item {\sl Uniform phase.}
This phase corresponds to the horizontal line in
Fig.~\ref{figm5ent} and exhibits Hagedorn behavior.
The entropy as a function of energy (see \eqref{unise}) is
\begin{equation}
S_{\rm u} (E) = \frac{E}{T_{\rm hg}} \ .
\end{equation}
%
\item {\sl Non-uniform phase.}
This phase emerges at the point $(E,S)=(E_{\rm c},S_{\rm c})$,
and for energies close to the critical energy $E_{\rm c}$
we have \cite{Harmark:2004ws}
\begin{equation}
\frac{S (E)}{S_u (E)} \simeq 1 - 0.05 \cdot \left[ \frac{(2\pi)^5
l_s^6}{V_5} (E- E_{\rm c}) \right]^2 \spa \frac{E_{\rm c}}{V_5} =
1.54 \cdot (2\pi)^{-5} l_s^{-6}  \spa 0 \leq \frac{l_s^6
(E-E_c)}{V_5} \ll 1 \ .
\end{equation}
This expression has been computed using the results of \cite{Sorkin:2004qq}.
The entropy function for the entire non-uniform phase, as displayed in
Fig.~\ref{figm5ent}, was computed in \cite{Harmark:2004ws} using the
data of \cite{Wiseman:2002zc}.
\item {\sl Localized phase.} The localized phase begins at the
origin $(E,S)=(0,0)$. At very low energies we obtain, using the
results of \cite{Harmark:2003yz}, the analytical result
\cite{Harmark:2004ws}
\begin{equation}
S_{\rm loc} (E) \simeq \frac{2 \pi^{1/3}}{3} (\frac{6}{5})^{5/6}
\frac{ \sqrt{N}}{(2\pi)^4} \frac{V_5}{l_s^5} \left( (2\pi)^5
\frac{l_s^6 E}{V_5} \right) ^{5/6} \left(1 + \frac{\zeta(3)}{10
\pi^2} (2\pi)^5 \frac{l_s^6 E}{V_5} \right) \spa
 \frac{l_s^6 E}{V_5}
\ll 1  \ .
\end{equation}
The entropy function for the entire localized phase, as displayed in
Fig.~\ref{figm5ent}, has been computed using the data of
\cite{Kudoh:2004hs} and the map of \cite{Harmark:2004ws}.
\end{itemize}

In the microcanonical ensemble, the phase with the highest entropy
dominates. Hence, from Fig.~\ref{figm5ent} we conclude that for
energies below $E_{\#}/V_5 = 2.51 \cdot (2\pi)^{-5} l_s^{-6}$,
the localized phase is preferred in the canonical ensemble,
whereas for energies above that value the
system is in the uniform phase. In the range $E_{\#} < E \leq E_{\rm max}$,
with $E_{\rm max}/V_5 = 3.92 \cdot (2\pi)^{-5} l_s^{-6}$,
the entropy of the uniform phase is greater than that of
the localized phase. For $E > E_{\rm max}$ only the uniform phase is
available. Note that $E_{\#} < E_{\star} < E_{\rm max}$. The
non-uniform phase has always lower entropy than the other two
phases.

The point of the localized phase where $E = E_{\rm max}$ is special.
The heat capacity is zero there, which corresponds
to a singular behavior in the microcanonical ensemble and
prompts the interpretation of the energy $E_{\rm max}$ as a limiting
energy for the localized phase.
It is also interesting to consider what happens to the system with
energy $E < E_{\#}$ when we throw in a finite amount of energy
$\Delta E$. There are three possibilities.
If $E + \Delta E < E_{\#}$ or $E_{\#} < E + \Delta E < E_{\rm max} $
the system will remain in the localized phase.
In the second case, however, the entropy of the uniform phase
at the new energy $E + \Delta E$ will be greater so the system will
probably go through a phase transition to end up
in the uniform phase. Finally, if $E + \Delta E > E_{\rm max}$ the
only available phase is the uniform phase and the system presumably
has to end up in that phase.

\subsubsection*{\sl Comparing the ensembles}

Comparing the canonical and microcanonical ensembles, we see
significant differences in the qualitative behavior of the
near-extremal NS5-brane thermodynamics. In the canonical ensemble,
the localized phase is always the dominant phase in the interval
$E \leq E_{\star}$, and has a positive specific heat and a
limiting temperature $T_{\star}$. In the microcanonical ensemble,
the localized phase is again the dominant
phase for energies $E \leq E_{\#}$. However, the qualitative
behavior is very different for energies $E_{\#}< E < E_{\rm max}$.
In this energy range the uniform phase is dominant. Hence, we
find different preferred phases in the two ensembles. For
energies $E > E_{\rm max}$ there is an even more dramatic difference
between the two ensembles since the only available phase is
the uniform phase in the canonical ensemble. Consequently,
it seems that the two ensembles are completely inequivalent at
large energies.

Since we discuss a system without gravity one would expect that it is always
possible to bring it in contact with a heat bath of a
certain temperature. Therefore, the inequivalence of ensembles could
suggest a subtlety in the definition of the canonical
ensemble. Alternatively, it could point towards the existence of a new
hitherto unknown phase of near-extremal NS5-branes. This
could potentially resolve the issue arising from the absence of
any stable phases with temperatures greater than $T_{\star}$. See
the conclusions of Ref.~\cite{Harmark:2005dt} for further discussion on
these issues.

\subsection{D-brane bound states}

In the above discussion we considered the implications of the
Gregory-Laflamme instability when there is a singly-charged brane with
a compact direction. It is also interesting to consider brane bound
state backgrounds without compact directions. In this case the CSC
conjecture still applies, as reviewed in Section \ref{sec:CSC}, thus
giving a connection between the Gregory-Laflamme instability, the
local thermodynamical stability and the stability of the dual
non-gravitational theories living on the brane bound states.

In \cite{Gubser:2004dr} the implications of the Gregory-Laflamme
instability for the phase-structure of the non-commutative
Yang-Mills (NCYM) theories were described. The dynamics of open strings
ending on a D$p$-brane are described at low energies by super-Yang-Mills
theory in $p+1$ dimensions. For a D-brane in the presence of an NS-NS
B-field the same dynamics is captured by a non-commutative gauge theory
\cite{Connes:1997cr,Douglas:1997fm,Cheung:1998nr,Seiberg:1999vs}.
One can then obtain the holographic dual of the NCYM theory by taking
the near-horizon limit of $N$ parallel D-branes in the presence of an NS-NS
B-field. This implies, in particular, that the near-extremal limit of
the thermal
D1-D3 brane bound state is holographically dual to a
four-dimensional NCYM theory with a non-zero temperature
\cite{Maldacena:1999mh,Alishahiha:1999ci,Harmark:1999rb,Cai:1999aw,Cai:2000hn}.
In \cite{Gubser:2004dr} it was argued that the D$(p-2)$-D$p$ brane
bound state is unstable in the near-extremal limit for all
temperatures above zero. This can be seen using the condition
\eqref{stabbound} for thermodynamic stability of the D$(p-2)$-D$p$
brane bound state derived in Section \ref{sec:examples}.

Ref.\ \cite{Friess:2005tz} discussed similar implications for the
non-commutative
open string (NCOS) theories living on the brane bound states.
The effective low energy dynamics of open strings ending on
an F1-D$p$ brane bound state are described by a Non-Commutative
Open String (NCOS) theory, which is a theory of open strings on a
target space in which space and time does not commute
\cite{Seiberg:2000ms,Gopakumar:2000na}. Taking the near-extremal limit
of the non-extremal F1-D$p$ bound state we obtain a gravity dual for the
$(p+1)$-dimensional NCOS theory \cite{Harmark:2000wv}. In
\cite{Friess:2005tz} it was argued that the near-extremal F1-D$p$
bound state is unstable above a certain critical temperature.

Similar considerations also hold for other non-gravitational
theories on brane bound states, such as the OM theory on the M2-M5
brane bound state \cite{Gopakumar:2000ep,Bergshoeff:2000ai} and the
ODp-theories on the D$p$-NS5 brane bound state
\cite{Gopakumar:2000ep,Harmark:2000ff}.

\section{Discussion and outlook \label{sec:conc}}

In this review we summarized the current state of knowledge on the
classical stability of black branes in pure gravity and
supergravity. We focused mainly on the case of static black hole
solutions that asymptote to a Kaluza-Klein space of the form $\MM^d
\times S^1$. Although this is the simplest possible case in a large
class of solutions that asymptote to Kaluza-Klein spaces of the form
$\MM^d \times \NN$, where $\NN$ is any compact Ricci flat manifold,
we have seen that it exhibits a profoundly rich structure with many
features that are expected to be part of the more general story. We
would like to conclude with a summary of other known classical
instabilities in gravity and a list of important questions that
deserve further study.

\subsection{Other classical instabilities \label{sec:otherins}}

\subsubsection*{\it Ultra-spin instability}

Rotating black holes in higher dimensions with arbitrary angular
momentum were discovered twenty years ago by Myers and Perry
\cite{Myers:1986un}. The absence of an upper bound on the angular
momentum (in six and higher dimensions) should be contrasted with
the situation in four dimensions, where the Kerr black hole obeys
the famous Kerr bound $J < GM^2$, and with the similar bound $J^2 <
32 G M^3/(27\pi)$ for the five dimensional Myers-Perry black hole.
Nevertheless, it was argued in \cite{Emparan:2003sy} that in six or
higher dimensions the Myers-Perry black hole becomes unstable above
some critical angular momentum thus recovering a dynamical Kerr
bound. The instability was identified as a Gregory-Laflamme
instability by showing that in a large angular momentum limit the
black hole geometry becomes that of an unstable black membrane. This
result is also an indication of the existence of new rotating black
holes with spherical topology, where the horizon is distorted by
ripples along the polar direction. Furthermore it suggests that a
possible endpoint of the decay process in ultra-spinning black holes
is the fragmentation of the original black hole into multiple black
holes. Ref.~\cite{Emparan:2003sy} verified that this possibility is
consistent with a global thermodynamic argument. An alternative
conjecture for the evolution of the ultra-spin instability is that
the decay proceeds via emission of gravitational radiation. In the
process, the black hole sheds a fraction of both its spin and mass
and settles down in a range of parameters within the stable regime.
Comments on this alternative possibility can also be found in
\cite{Emparan:2003sy}.

It should be emphasized that up to date the complete analytic
theory of perturbations around the higher-dimensional rotating
black holes has not been developed. Progress in this direction has
been achieved in special cases (odd number of space-time
dimensions, equal angular momenta) in \cite{Kunduri:2006qa}.

Although the arguments presented in \cite{Emparan:2003sy} apply only
to six or higher dimensions an instability in five dimensions is
also expected \cite{Emparan:2001wn}. Entropy arguments suggest that
in this case the endpoint of the instability is the five dimensional
black ring.

\subsubsection*{\it Gyration instability}

In the context of D1-D5-P spinning black brane solutions in type IIB
supergravity compactified on $\T^4$ it has been argued in
\cite{Marolf:2004fy} that a small perturbation of a spinning black
string could grow to become large for sufficiently large angular
momentum. The endpoint of this instability is expected to be a
gyrating string where part of the original angular momentum is
carried by gyrations of the string (as opposed to spin). This type
of instability has been proposed as a possible counterexample to the
CSC (see also Section \ref{sec:count}).

\subsubsection*{\it Superradiance instability}

Superradiance is a phenomenon that occurs in a scattering process
when the incident wave is reflected back amplified. The first
classic example of superradiant scattering, which later led to the
notion of superradiant scattering in black hole space-times, was
given by Zel'dovich in \cite{Zeldovich1,Zeldovich2}. By examining
what happens when scalar waves of the form $e^{-i\omega t+im\phi}$
scatter off a rotating absorbing cylinder, Zel'dovich concluded
that the scattered wave is amplified provided that the frequency
$\omega$ of the incident wave satisfies the inequality
\begin{equation}
\label{discaaa} \omega<m\Omega ~.
\end{equation}
In this relation $\Omega$ is the angular velocity of the rotating
cylinder. Similar superradiant scattering occurs in the background
of Kerr black holes
\cite{Zeldovich1,Zeldovich2,Bardeen:1972fi,Starobinsky1,Starobinsky2}
for frequencies $\omega$ still satisfying \eqref{discaaa} with
$\Omega$ the angular velocity of the black hole.

An instability can occur if the black hole is surrounded by a
reflecting wall that scatters the returning wave back towards the
horizon. In that case the wave will bounce back and forth from the
horizon extracting an exponentially growing amount of energy from
the black hole until the pressure destroys the mirror. The combined
system of the black hole and the mirror is known as Press and
Teukolsky's black hole bomb \cite{Press,Cardoso:2004nk}. A natural
reflecting wall occurs in a variety of situations: a massive scalar
field propagating in a Kerr background
\cite{Damour:1976kh,Detweiler:1980uk, Furuhashi:2004jk}, a wave
propagating around spinning black strings
\cite{Marolf:2004fy,Cardoso:2004zz,Cardoso:2005vk},
Kerr-anti-deSitter black holes \cite{Cardoso:2004hs}, large radius
doubly spinning black rings \cite{Dias:2006zv}. In general, the
unstable modes can be scalar, electromagnetic or gravitational.

\subsubsection*{\it Ergoregion instability}

Space-times with ergospheres and no horizon are unstable under
perturbations of scalar and electromagnetic fields or
gravitational waves \cite{Friedman}. The instability also occurs
in the case of rotating stars with an ergoregion
\cite{Comins,Yoshida} and in smooth non-supersymmetric D1-D5-P
geometries \cite{Cardoso:2005gj}. In the latter case the endpoint
of the instability has been argued to be a smooth supersymmetric
configuration.

\subsubsection*{\it A detour on the stability of black rings}

The recent discovery of the black ring solution in five dimensions
\cite{Emparan:2001wn} demonstrates clearly that in higher dimensions
new qualitative features can appear which are not present in four
dimensions. Most notably, one finds that there are black holes with
non-spherical horizon topologies\footnote{ The horizon topology of
black rings is $S^2\times S^1$. In general, topological arguments
allow for a large class of horizon topologies in higher dimensions.
For recent discussions of the relevant theory see
\cite{Helfgott:2005jn,Galloway:2005mf}.} and that conventional
notions of black hole uniqueness do not apply. This realization has
made the five dimensional black ring a protagonist in the arena of
higher dimensional black hole physics. Some important partial
results on the classical stability of the five-dimensional black
ring have been obtained. Here we review the emerging picture
beginning with a quick reminder of the major features of black
rings. We will see that black rings can exhibit a combination of the
above-mentioned instabilities and more. We will focus mostly on
neutral black rings.

\begin{figure}[ht]
\centerline{\epsfig{file=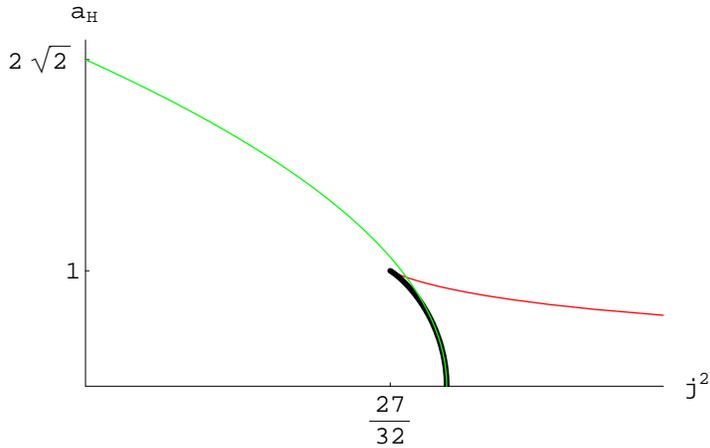,width=10 cm,height=6cm}}
\caption{Rescaled horizon area $a_H$ versus angular momentum squared
$j^2$ for five-dimensional Myers-Perry black holes and black rings.
The green curve represents the Myers-Perry black hole, the red curve
(for $\nu>\frac{1}{2}$)
the branch of fat black rings and the solid black curve (for
$0<\nu<\frac{1}{2}$)
the branch of thin black rings.}
\label{BSphase}
\end{figure}

The neutral black ring metric has been discussed in a series of
papers \cite{Emparan:2001wn,Elvang:2003mj,Emparan:2004wy} (for a
recent review of the subject with a complete list of references see
\cite{Emparan:2006mm}). The resulting one-parameter family of
solutions is conveniently parameterized by a single variable $\nu$
in the range $0<\nu\leq 1$. In terms of $\nu$ the dimensionless
angular momentum and horizon area are given by the expressions
\begin{equation}
\label{discbaa}
j^2=\frac{27 \pi}{32
G}\frac{J^2}{M^3}=\frac{(1+\nu)^3}{8\nu}~, ~ ~
a_H=\frac{3}{16}\sqrt{\frac{3}{\pi}} \frac{\AA_H}{(GM)^{3/2}}=
2\sqrt{\nu(1-\nu)} ~.
\end{equation}
Compare these with the corresponding expressions for the
five dimensional Myers-Perry black hole
\begin{equation}
\label{discbab}
j^2
=\frac{2\nu}{1+\nu}
~, ~ ~
a_H
=2\sqrt{2\frac{1-\nu}{1+\nu}}
~.
\end{equation}
The situation is summarized for fixed mass in Fig.\ \ref{BSphase} along
with the corresponding curve for the Myers-Perry
black hole. This diagram has two striking features. First, there is
a range of $j$ where three objects with the same mass and spin
coexist (one Myers-Perry black hole and two black rings). Secondly,
the black ring curve exhibits a cusp at $\nu=\frac{1}{2}$ which is
the meeting point of two branches of black rings. One branch has
$0<\nu<1/2$ and has been called the branch of {\it thin black
rings}, because it includes the regime $\nu \sim 0$ of very thin
(highly spinning) rings. The other branch with $1/2<\nu<1$ has been
called the branch of {\it fat black rings} and terminates at the
singular solution with $\nu=1$.

The classical stability of these solutions has been studied with a
combination of methods (qualitative and semi-quantitative) in
\cite{Emparan:2001wn,Arcioni:2004ww,Arcioni:2005fm,Nozawa:2005eu,
Cardoso:2005sj,Hovdebo:2006jy,Astefanesei:2005ad,Elvang:2006dd}
and has led to the following picture (this picture has been recently
summarized and further elaborated upon in \cite{Elvang:2006dd} to which
we refer the interested reader for more details):
\begin{itemize}
\item[$\bullet$] At large angular momentum the thin black ring becomes
identical
to a highly boosted black string (see Section \ref{sec:booststr}) which is known
to exhibit a Gregory-Laflamme instability. This suggests
\cite{Emparan:2001wn,Hovdebo:2006jy} that the thin black ring branch
suffers from a Gregory-Laflamme instability quite possibly driving
the system towards a configuration of two or more spinning black
holes. There are some similarities here with the case of
ultra-spinning Myers-Perry black holes in six or higher dimensions.
\item[$\bullet$] Black rings can become unstable under radial perturbations.
The analysis of \cite{Hovdebo:2006jy} indicates that fat black rings
are radially unstable while thin black rings are radially stable.
The black ring with minimal spin is also argued to be unstable. This
type of instability appears to mesh nicely with the topological
arguments of \cite{Arcioni:2004ww,Arcioni:2005fm} which are based on
the Poincar\'e method.
\item[$\bullet$] The angular momentum of black rings is bounded from below.
By absorbing a counter-spinning null-particle the black ring can
decrease its angular momentum $j$. For the lowest-spinning black
ring this process indicates an instability. On the other hand, it is
impossible for a Myers-Perry black hole or a fat black ring to
overspin by absorbing a co-rotating null particle or to
spontaneously decay by emitting null particles without violating the
area law \cite{Elvang:2006dd}.
\item[$\bullet$] Doubly spinning large radius ($i.e.$ $\nu\rightarrow 0$) thin
black rings have been argued to suffer from superradiance
instabilities \cite{Dias:2006zv}.
\end{itemize}
The addition of charges can improve the stability of black rings.
For instance, supersymmetric black rings \cite{Elvang:2004rt} are
expected to be stable. Black rings with dipole charges
\cite{Emparan:2004wy} are expected to suffer from similar
instabilities as the neutral ones \cite{Elvang:2006dd}.

\subsection{Future directions and open problems \label{sec:open}}

We conclude with a number of important issues and questions for
future research.

\subsubsection*{\it Pure gravity: static solutions asymptoting to
$\MM^d\times S^1$}

\begin{itemize}
\item[$\bullet$] At present the complete non-uniform branch is
known numerically for $d=4$ and $d=5$
\cite{Wiseman:2002zc,Kleihaus:2006ee} and recently also higher
dimensions $6 \leq d \leq 10$ have been studied
\cite{Sorkin:2006wp}. It would be interesting to compute it above
the critical dimension, $i.e.$ $d \geq 13$. For the black hole
branch, the present status is that we know numerically the entire
branch only for $d=4$ and $d=5$
\cite{Sorkin:2003ka,Kudoh:2003ki,Kudoh:2004hs}. Having the numerical
data for other dimensions would be interesting as well.
\item[$\bullet$] If possible, it would be interesting to find the analytic form of the
non-uniform and black hole branches, for example using the ansatz
\eqref{ansatz}. A first step in this direction would be to extend
the known first order analytical results to second order. For the
$d=4$ black hole this has been done in Ref.\
\cite{Karasik:2004ds}, while \cite{Chu:2006ce} gives the corrected thermodynamics
for all $d$, but not the metric.
\item[$\bullet$] For the neutral black string the study of the
classical stability revealed the presence of the GL instability. It
would be highly interesting if one could study the classical
stability of other branches of solutions, such as the non-uniform
black string, the localized black hole, or a black hole attached to
a Kaluza-Klein bubble.
\item[$\bullet$]
The bubble-black hole sequences are only known in five and six
dimensions. It is interesting to ask if solutions for bubble-black
hole sequences exist for $D\geq 7$, and if so, whether they can be
related to the solutions of \cite{Elvang:2004iz} by a map similar
to the one mapping five- to six-dimensional solutions. For spaces
with more than one Kaluza-Klein circle, $i.e.$ asymptotics $\MM^d
\times \T^{k\geq 2}$, it is easy to construct solutions describing
regular bubble-black hole sequences \cite{Elvang:2004iz}. On the
other hand, finding bubble-black hole sequences in $\MM^d \times
S^1$ may be difficult since one cannot use the generalized Weyl
ansatz for such solutions, but we do not see any physical
obstructions to their existence. Should these higher dimensional
bubble-black hole sequences exist, it would be interesting to see
what type of horizon topologies they exhibit.
\item[$\bullet$] In
connection with the above points, also the general question of the
$d$-dependence of the phase diagram is an important one. For
example, the non-uniform branch is known to have a critical
dimension beyond which the slope changes sign. Likewise one could
imagine new phases appearing (or other phases disappearing) as the
number of dimensions increases.
\item[$\bullet$] The instability
of the uniform black string implies that a naked singularity may be
formed when the horizon of the black string pinches off. As
originally pointed out by Gregory and Laflamme
\cite{Gregory:1993vy,Gregory:1994bj}, this would entail a violation
of the Cosmic Censorship Hypothesis. The results of
\cite{Wiseman:2002zc,Kudoh:2004hs,Kleihaus:2006ee,Sorkin:2006wp}
suggest that, for certain dimensions, the localized black hole is
the only solution with higher entropy than that of the uniform black
string (for masses where the black string is unstable). Thus, the
endpoint of the instability seems to be the localized black hole
which means that the horizon should pinch off in the classical
evolution. On the other hand, in \cite{Horowitz:2001cz} it was
argued that the horizon cannot pinch off in finite affine parameter,
thus suggesting that it is impossible for the black string horizon
to pinch off. A way to reconcile these two results is if the horizon
pinches off in infinite affine parameter (see also
\cite{Marolf:2005vn}). Recently, the numerical analysis of
\cite{Choptuik:2003qd,Garfinkle:2004em,Anderson:2005zi} indicates
that this indeed is the case. If this is correct  it would be
interesting to examine the implications for the Cosmic Censorship
Hypothesis.
\item[$\bullet$] A similar formation of a
naked singularity (now for static configurations) has been
predicted at the merger point of the non-uniform and localized
black hole branch in the phase diagram of black strings. Direct
evidence for the existence of the merger point has been found with
a numerical study in \cite{Kol:2003ja} (see also
\cite{Kol:2002xz,Kol:2004ww}). It would be very interesting to
explore the physics of the merger point transition further.
\end{itemize}

\subsubsection*{\it Pure gravity: static solutions asymptoting to
$\MM^d\times \NN$}

Static solutions in pure gravity asymptoting to $\MM^d\times S^1$
exhibit a rich structure. An even richer structure is expected for
static solutions asymptoting to the more general manifold $\MM^d
\times \NN$ where $\NN$ is any compact Ricci-flat manifold. It would
be interesting to explore the $\NN$ dependence of the phase
structure of solutions in certain examples, $e.g.$ with $\NN$ being
an $n$-torus $\T^n$ or $\NN$ being a Calabi-Yau manifold. Uniformly
smeared solutions are still expected on general grounds to exhibit
Gregory-Laflamme instabilities, but it would be further interesting
to see if the phase diagram of the general case exhibits new
qualitative features which are not present in the simplest $\NN=S^1$
case. A recent discussion of the Gregory-Laflamme instability for
$\NN=\T^n$ appeared in \cite{Kol:2006vu}. For some preliminary
comments on the case of a general compact Ricci flat manifold $\NN$
see Section \ref{sec:highdspaces}.

\subsubsection*{\it Static black brane solutions in supergravity}

Many of the above pure gravity neutral static solutions can be
easily lifted to corresponding solutions in type IIA or type IIB
ten-dimensional supergravity where one can also make contact with
D-brane configurations and gauge theories. It would be interesting
to explore the full phase diagram of static solutions in
ten-dimensional supergravity independently and see whether it admits
new phases which cannot be obtained through a map from the neutral
pure gravity ones. In this connection, we note that in Ref.~\cite{Miyamoto:2006nd}
a new stable phase of non-uniform black strings was found.

\subsubsection*{\it Black hole microscopics}

Another interesting direction is to make contact with microscopic calculations
of the entropy of black holes. While this is difficult for neutral Kaluza-Klein black holes,
charging up the solutions by the boost/U-duality map
and taking the near-extremal limit opens up the possibility of applying microscopic conjectures.
This was successfully applied in Ref.~\cite{Harmark:2006df} where five-dimensional three-charge
black holes on a a circle were obtained by extending the map of Ref.~\cite{Harmark:2004ws} to
include three charges. As a result a rich phase structure  of three-charge black holes,
including a new phase of three-charge
black holes that are non-uniformly distributed on the circle, was found
via a map that relates them to the phases of five-dimensional
neutral Kaluza-Klein black holes. Moreover, for three-charge black holes localized
on the transverse circle, it was shown that in a partial extremal limit with
two charges sent to infinity and one finite, the first correction to the finite
entropy is in agreement with the microscopic entropy. This was achieved
by taking into account that the number of branes shift as a consequence of the
interactions across the transverse circle. This matching was extended in
Ref.~\cite{Chowdhury:2006qn} to second order. Furthermore, in Ref.~\cite{Chowdhury:2006qn}
a simple microscopic model was proposed that reproduces most of the features
of the phase diagram, including the new non-uniform phase. It would be interesting
to pursue this microscopic description further (in this connection
see e.g. the reviews \cite{Mathur:2005zp,Mathur:2005ai}).

\subsubsection*{\it Connection to thermodynamics}

A natural and intuitive connection between classical and
thermodynamic instabilities of black brane solutions is expected to
hold in general. The conjecture underlying this relation (Correlated
Stability Conjecture) has passed a large number of tests, but is
known to fail in certain cases as we saw in section \ref{sec:CSC}.
It would be interesting to formulate a refinement of the conjecture
that survives all known tests. Then one could hope to find a general
proof for its validity.

\subsubsection*{\it Other compactified solutions}

It would also be interesting to examine the existence of other
classes of solutions with a compactified direction. For example in
Ref.~\cite{Maeda:2006hd} a supersymmetric rotating black hole in a
compactified space-time was found and charged black holes in
compactified space-times are considered in Ref.~\cite{Karlovini:2005cn}.
In another direction, new solutions with Kaluza-Klein boundary conditions
for anti-de-Sitter space times have recently been constructed in
Refs.~\cite{Copsey:2006br,Mann:2006yi}.

\subsubsection*{\it Black rings}

For black rings in five dimensions (and for that matter also for
other higher dimensional black holes like the rotating Myers-Perry
solutions) it is important to develop further the explicit linear
stability analysis. The main obstacle in this enterprize is of
course the complexity of the equations in higher dimensions. The
ultimate goal in that respect is a better understanding of the
full phase diagram. Beyond that it would be very exciting to see
if there are black ring solutions in $D>5$ dimensions or other
solutions with the more complex horizon topology allowed by
topological arguments in higher dimensions
\cite{Helfgott:2005jn,Galloway:2005mf}.

\subsubsection*{\it Braneworld black holes}

The higher dimensional black holes and branes described in this
review appear naturally in the discussion of the braneworld model
of large extra dimensions \cite{Arkani-Hamed:1998rs,
Antoniadis:1998ig}. In other braneworld models such as the one
proposed by Randall and Sundrum \cite{Randall:1999ee,
Randall:1999vf} the geometry is warped in the extra direction and
the discovery of black hole solutions in this context has proven
more difficult. A black string solution is also allowed in this
model and is expected to have a Gregory-Laflamme instability as in
the case of uniform black strings in KK space-times. Localized
black hole solutions are anticipated as end-products of the
Gregory-Laflamme instability of black strings, but have not been
constructed explicitly so far. A partial list
of references includes \cite{Wiseman:2001xt, Karasik:2003tx,
Kudoh:2003xz, Kudoh:2003vg} (see also the recent work in
\cite{Creek:2006je, Fitzpatrick:2006cd} and references therein).
It should be noted that exact black hole solutions localized on
the four-dimensional braneworld have been constructed by the
authors of \cite{Emparan:1999wa, Emparan:1999fd}.

\section*{Acknowledgments}

We thank Ofer Aharony, Micha Berkooz, Jan de Boer, Poul Henrik Damgaard, Oscar
Dias, Roberto Emparan, Dan Gorbonos, Shinji Hirano, Gary Horowitz, Veronika Hubeny, Barak Kol,
Kristjan Kristjansson, Hideaki Kudoh, Finn Larsen, Luis Lehner, Ernesto Lozano-Tellechea,
Don Marolf, Rob Myers, Mukund Rangamani, Simon Ross, Peter R{\o}nne, Evgeny Sorkin
and Toby Wiseman for useful
discussions related to the material presented in this work. We thank
Henriette Elvang for useful discussions and for collaboration on one
of the topics presented in this work. We thank Hideaki Kudoh and
Toby Wiseman for providing us with the numerical results of
Ref.~\cite{Kudoh:2004hs}, and Burkhard Kleihaus, Jutta Kunz and
Eugen Radu for kindly providing the numerical results of
Ref.\cite{Kleihaus:2006ee}. We thank Ruth Gregory for permission to
reprint Figure \ref{figGL}. TH and NO would also like to thank the
KITP for hospitality during the program ``Scanning new horizons: GR
beyond 4 dimensions'', while part of this work was completed. The
work of TH and NO is partially supported by the European Community's
Human Potential Programme under contract MRTN-CT-2004-005104
`Constituents, fundamental forces and symmetries of the universe'.
TH would like to thank the Carlsberg Foundation for support. VN
acknowledges partial financial support by the EU under the contracts
MEXT-CT-2003-509661, MRTN-CT-2004-005104 and MRTN-CT-2004-503369.

\begin{appendix}
\section{The Gregory-Laflamme mode}
\label{app:mode}

In this appendix we consider in more detail the Gregory-Laflamme mode
\cite{Gregory:1993vy,Gregory:1994bj} for the metric \eqref{ublstr}
that describes a uniform black string in $D=d+1$ dimensions.%
\footnote{See also \cite{Gross:1982cv,Prestidge:1999uq,Kol:2004pn}
and the more recent \cite{Kol:2006ga} for
related work on this.} The instability mode is given by a metric
perturbation $h_{\mu\nu}$ that solves the perturbed Einstein equations
of motion
around the black string metric \eqref{ublstr}. The precise form of the
perturbation $h_{\mu\nu}$ appears in the main text in eqs.\
\eqref{GLmode1} and \eqref{GLmode2}. In
\eqref{GLmode2} $\psi$, $\eta$, $\chi$ and $\kappa$ are all
functions of the variable
\begin{equation}
x = \frac{rk}{r_0} \ ,
\end{equation}
and this will be the variable with respect to which we take derivatives.
The function $f$ is given in terms of $x$ as
\begin{equation}
\label{fkx} f = 1 - \frac{k^{d-3}}{x^{d-3}} \ .
\end{equation}

Following \cite{Gross:1982cv,Prestidge:1999uq,Kol:2004pn} we
work here in a transverse $(\nabla ^{\mu} h_{\mu \nu}=0)$, traceless
$(g^{\mu \nu} h_{\mu \nu}=0)$ gauge.
The tracelessness condition is
\begin{equation}
\label{traceeq}
 (d-2) \kappa + \chi + \psi = 0 \ .
\end{equation}
The transversality conditions are
\begin{equation}
\label{transeqs}
\begin{array}{c} \ds
\frac{\Omega}{k} x \psi + (d-3)(1-f) \eta + f \left((d-2) \eta + x
\eta' \right)=0 \ ,
\\[3mm] \ds
-2 \frac{\Omega}{k} x  \eta+(d-3)(1-f)(\chi-\psi)+2 f \left((d-1)
\chi+\psi+ x \chi'\right)=0 \ .
\end{array}
\end{equation}
The four independent Einstein equations take the form
\begin{eqnarray}
& \ds \label{E1}
\begin{array}{l} \ds x \frac{\Omega}{k} \psi +
(d-3)(1-f) \eta +\frac{k}{\Omega} f^2 \big(
(d-2)\psi'+x\psi''\big)
\\[2mm] \ds
+f \left[ 2(d-2) \eta - x \frac{k}{\Omega}\psi + 2 x \eta'
+\frac{k}{\Omega}(d-3)(1-f)(-\chi'+\psi')\right] =0~, \end{array}
& \\[3mm] & \ds
\label{E2}
-(d-3)(1-f)x \frac{\Omega}{k}(\chi+\psi)+ 2f\left[
x\frac{\Omega}{k}((d-1)\chi+\psi+x(\chi'+\psi'))+\eta x^2\right]=0
~,~~
& \\[3mm] & \ds
\label{E3}
\begin{array}{l} \ds
x^2\bigg(\frac{\Omega^2}{k^2}+f\bigg)\chi-(d-3)(1-f)\frac{\Omega}{k}x\eta
\\[4mm] \ds
+f\left[ -2 x^2 \frac{\Omega}{k} \eta'+(d-3)(1-f)x(\chi'-\psi')
+xf\big(d \chi'+2\psi'+x\chi''\big)\right]=0~, \end{array}
& \\[3mm] & \ds
\label{E4}
\begin{array}{l} f\bigg[
\psi\big(-x^2+2(d-3)(1-f)\big)+\chi\big(-x^2+2(d-1)(d-3)(1-f)\big)
\\ \ds
 -2(d-2)x\frac{\Omega}{k}\eta +(d-3)(1-f)x(\chi'+\psi')\bigg]
+f^2\bigg[2(d-1)(d-3)\chi
\\ \ds
+2(d-3)\psi+x\big(3(d-2)\chi'+(d-2)\psi'+x(\chi''+\psi'')\big)\bigg]
-x^2\frac{\Omega^2}{k^2}(\chi+\psi)=0 ~. \end{array} &
\end{eqnarray}
Combining the gauge conditions \eqref{traceeq}-\eqref{transeqs}
with the Einstein equations \eqref{E1}-\eqref{E4} one can derive
Eq.~\eqref{psieq} which is a second order differential equation
for $\psi$ of the form
\begin{equation}
\label{E5}
\psi ''(x)+{\cal Q}_d(x) \psi'(x)+{\cal P}_d(x)\psi(x)=0 ~.
\end{equation}
${\cal Q}_d$ and ${\cal P}_d$ are $d$-dependent rational functions
of $x$, which we summarize here for completeness
\begin{eqnarray}
{\cal Q}_d(x)&=&f^{-1} x^{-1} \Big[f^3 k^4
(39+10d+3d^2)+\Omega^2\big(k^2(3-d)^2-4\Omega^2\big)
\nonumber\\
& &+k^2 f^2 \big(-2 k^2(41-16d+d^2)-3(d+5)(5-3d)\Omega^2\big)
\nonumber\\
& &-k^2 f \big(k^2(3-d)^2+2(-5-12d+5d^2)\Omega^2\big)\Big]^{-1}
\nonumber\\
& & \times \Big[k^4 f^4
(36+11d+6d^2-d^3)-3(3-d)\Omega^2\big(k^2(3-d)^2-4\Omega^2\big)
\nonumber\\
& &+k^2 f^2
\big(k^2(3-d)(64-23d+d^2)+\Omega^2(-431+405d-181d^2+31d^3)\big)
\nonumber\\
& &+k^2 f^3 \big(k^2
(-178+28d-2d^2+4d^3)+\Omega^2(135-94d+51d^2-8d^3)\big)
\nonumber\\
& &f \big(2k^4(3-d)^3+k^2
\Omega^2(3-d)(163-127d+26d^2)+4(-7+2d)\Omega^4\big)\Big] ~,
\end{eqnarray}
\begin{eqnarray}
{\cal P}_d(x)&=&k^{-2} f^{-2} x^{-2} \Big[f^3 k^4
(39+10d+3d^2)+\Omega^2\big(k^2(3-d)^2-4\Omega^2\big)
\nonumber\\
& &+k^2 f^2 \big(-2 k^2(41-16d+d^2)-3(5+d)(5-3d)\Omega^2\big)
\nonumber\\
& &-k^2 f \big(k^2(3-d)^2+2(-5-12d+5d^2)\Omega^2\big)\Big]^{-1}
 \times \Big[-2 k^6 f^5(1-d)^2(3-d)^2
\nonumber\\
& &+ k^4 f^4 \big( 2 k^2
(-3+d)(-16+ 21d-11d^2+2d^3)+(-39-10 d-3d^2) x^2)
\nonumber\\
& &-(3-d)^2(1-d)(37-7d)\Omega^2\big)
\nonumber\\
& &-2k^4 f^3\big( k^2(
(-3+d)(-25+21d-7d^2+d^3)+(-41+16d-d^2)x^2)
\nonumber\\
&
&+((3-d)(-239+ 273d-91d^2+9d^3)+
(-21+18d+7d^2)x^2)\Omega^2\big)
\nonumber\\
& &+2 fk^2\Omega^2
\big(k^2(3-d)^2(35-14d+d^2+x^2)
\nonumber\\
& & +(
2(3-d)(d-2)+(57-56d+13d^2)x^2)\Omega^2\big)
\nonumber\\
& &+\Omega^2\big(k^4(3-d)^4-k^2(3-d)^2(-2+7x^2)\Omega^2+4x^2\Omega^4\big)
\nonumber\\
& &+k^2f^2\big(k^4(3-d)^2(8-2d+x^2)-2k^2(
(-3+d)(-302+279d-80d^2+7d^3)
\nonumber \\
& &
-6(d-2)^2 x^2)\Omega^2+(2(d-3)(d-1)+(9+22d-19d^2)x^2)\Omega^4\big)\Big] ~.
\end{eqnarray}

The threshold unstable mode obeys the differential equation
\eqref{E5} with $\Omega=0$, which is equivalent
to the second order equation for the scalar $\chi$ that appears
in (2.10) of \cite{Kol:2004pn}
\begin{eqnarray}
-f \chi''&+&\left[ \frac{2 x^2 (f f''-f'^2)-x(d-2)f f'+2d
f^2}{x(xf'-2f)}\right] \chi'
\nonumber\\
&+&\left[ \frac{x^2 f' f''+x\left[ 2(d-1) f f''-(d+2){f'}^2\right]+4 f
f'}{x(x f'-2f)}\right]
\chi=-r_0^2 \chi
~.
\end{eqnarray}

The differential equations governing the threshold unstable mode
have appeared in the literature in various forms depending on the
choice of gauge. A complete list of the gauge choices that have been
employed so far in the literature can be found in appendix A of
\cite{Kol:2006ga}. Differential equations for the general time-dependent
Gregory-Laflamme mode have appeared also in \cite{Gregory:1993vy}
(see eq.\ 10, which is a second order differential equation for $h^{tr}$)
and in \cite{Aharony:2004ig} (see eq.\ 12).

\end{appendix}




\providecommand{\href}[2]{#2}\begingroup\raggedright\endgroup

\end{document}